# Orpiment under compression: metavalent bonding at high pressure

V.P. Cuenca-Gotor,[1] J.A. Sans,[1,*] O. Gomis,[2] A. Mujica,[3] S. Radescu,[3] A. Muñoz,[3] P. Rodríguez-Hernández,[3] E. Lora da Silva,[1] C. Popescu,[4] J. Ibañez,[5] R. Vilaplana,[2] and F.J. Manjón[1,*]

[1] Instituto de Diseño para la Fabricación y Producción Automatizada, Universitat Politècnica de València, 46022 Valencia (Spain)
[2] Centro de Tecnologías Físicas, Universitat Politècnica de València, 46022 Valencia (Spain)
[3] Departamento de Física, Instituto de Materiales y Nanotecnología, MALTA Consolider Team, Universidad de La Laguna, 38200 San Cristóbal de La Laguna (Spain)
[4] ALBA-CELLS, 08290 Cerdanyola, Barcelona (Spain)
[5] Institute of Earth Sciences Jaume Almera, CSIC, 08028 Barcelona (Spain)

ABSTRACT: We report a joint experimental and theoretical study of the structural, vibrational, and electronic properties of layered monoclinic arsenic sulfide ($\alpha$-As$_2$S$_3$), aka mineral orpiment, under compression. X-ray diffraction and Raman scattering measurements performed in orpiment samples at high pressure and combined with *ab initio* calculations have allowed us to determine the equation of state and the tentative assignment of the symmetry of many Raman-active modes of orpiment. From our results, we conclude that no first-order phase transition occurs up to 25 GPa at room temperature; however, compression leads to an isostructural phase transition above 20 GPa. In fact, As coordination increases from threefold at room pressure to more than fivefold above 20 GPa. This increase in coordination can be understood as the formation of metavalent bonding at high pressure, which results in a progressive decrease of the electronic and optical bandgap, an increase of the dielectric tensor and Born effective charges, and a considerable softening of many high-frequency optical modes with pressure. The formation of metavalent bonding may also explain the behavior of other group-15 sesquichalcogenides under compression. Moreover, our results suggest that group-15 sesquichalcogenides either show metavalent bonding at room pressure or undergo a transition from p-type covalent bonding at room pressure towards metavalent bonding at high pressure, as a preceding phase towards metallic bonding at very high pressure.

## 1.- INTRODUCTION

Arsenic sulfide ($As_2S_3$) and in particular the monoclinic polymorph ($\alpha$-$As_2S_3$), aka mineral orpiment, is one of the ores of As together with minerals realgar ($\alpha$-$As_4S_4$) and arsenopyrite (FeAsS) [1]. In particular, orpiment and realgar have been known since ancient times, with realgar being used as an orange-red pigment and orpiment as a gold-like pigment, hence his mineral name orpiment (*Aurum pigmentum* in latin) [2]. Due to the high chemical stability of $\alpha$-$As_2S_3$, the canary yellow or King's yellow pigment was obtained from molten orpiment and has been extensively used in papyrus and objects of ancient Egypt that date back to 3.1 millennia BC [3-5]. Additionally, $As_2S_3$ crystals and glasses have been used as drugs to treat different illnesses, such as prophylactic diseases, asthma, tuberculosis or diabetes, and were also prescribed as antiseptics and sedative by Aristotle and Hippocrates (IV century BC). Moreover, orpiment has been used since ancient times in traditional Chinese medicine for the treatment of inflammation, ulcers, convulsions, and schistosomiasis [6]. Nowadays, arsenic trisulfide has proved to exhibit an excellent activity and positive effects in cancer therapy [7]. Finally, it must be stressed that in the last decades arsenic sulfides have been investigated for different applications in photonics and non-linear optics since they possess high values of the refractive index, high IR transparency, high optical non-linearity, stability to crystallization, the possibility to be modified by a femtosecond laser irradiation, and an inertness to environment [8,9].

$As_2S_3$ usually crystallizes in the monoclinic $\alpha$ polymorph [space group (SG) $P2_1/n$ or $P2_1/c$, No. 14] [10,11]; although a triclinic dimorph of orpiment, called anorpiment [SG $P\bar{1}$, No. 2] was recently discovered and approved as a new mineral by the International Mineralogical Association [12]. The crystalline structure of the monoclinic $\alpha$ phase (see **Figure 1**) is composed by corrugated or zigzag layers piled up mainly along the *b* axis (see **Figure 1(a)**), which are linked by weak van der Waals (vdW) forces **[13]**.

The properties of arsenic sulfides have been extensively studied, mainly for melts and glasses, because $As_2S_3$ ranks first among chalcogenides from the viewpoint of the production of amorphous and glass-like industrial materials. However, many properties of their crystalline phases, orpiment, and anorpiment, are not yet well known, likely due to the complex monoclinic and triclinic crystalline structures of these compounds. For instance, a definitive symmetry assignment of the vibrational modes of orpiment is still under debate **[14]** despite a number of papers reporting Raman scattering (RS) measurements, as well as infrared (IR) measurements, since the advent of laser-based Raman spectroscopy **[14-29]**.

In this context, the pressure is a valuable tool to study the properties of materials and several high-pressure (HP) studies have been performed in glassy $As_2S_3$ and $\alpha$-$As_2S_3$ in order to help us to understand their properties. In particular, results obtained from HP optical absorption, HP x-ray absorption spectroscopy (XAS), HP Raman scattering (RS) measurements, HP electrical measurements and from a few *ab initio* calculations have been reported for $\alpha$-$As_2S_3$ **[23,26,27,30-35]**. In particular, a HP and high-temperature (HT) polymorph of orpiment ($\gamma$-$As_2S_3$), whose structure has not been resolved yet, has been found **[34]**. This is not surprising because little is also known about the structure of orpiment and its behavior under compression. It is only known that the remarkable high packing of layers in orpiment is responsible for its high stability under pressure as suggested by previous experiments. Besides, HP-XAS measurements suggested some structural changes above 10 GPa and an increase in As coordination above 30 GPa. On the other hand, recent electrical measurements and calculations suggest an isostructural phase transition (IPT) around 25 GPa **[33]**.

To verify these hypotheses, it would be desirable to perform HP x-ray diffraction (XRD) measurements on $\alpha$-$As_2S_3$ at room temperature in order to obtain the equation of state of orpiment and explore the origin of the changes proposed in the literature. Such measurements will allow us to investigate the possible appearance of HP phases and its relationship mainly with those of other sesquioxides and sesquichalcogenides. On the other hand, it will be important to perform new HP-RS measurements in order to correlate the structural and vibrational changes in orpiment and to aid in the assignment of the symmetries and the origin of the different vibrational modes with the help of *ab initio* calculations. All these measurements and calculations would allow us to understand the behavior of orpiment under compression.

For all the mentioned reasons, we report in this work a joint experimental and theoretical study of the structural, vibrational, and electronic properties of $\alpha$-$As_2S_3$ under compression at room temperature by means of powder HP-XRD and HP-RS measurements in orpiment. Experimental results are complemented with theoretical *ab initio* electronic and lattice-dynamics calculations. Our results show a continuous compression of orpiment up to 25 GPa with a loss of the vdW character of the interlayer forces above 10 GPa. Moreover, our calculations show that there is a progressive increase of As coordination with pressure, changing from threefold coordination at 0 GPa to more than fivefold coordination above 20 GPa, where a pressure-induced IPT occurs. Additionally, we provide a tentative symmetry assignment of the Raman-active modes in orpiment and explain their behavior under compression.

Our electronic band structure calculations suggest that there is a progressive closing of the bandgap of $\alpha$-$As_2S_3$ with increasing pressure leading from a trivial semiconductor to a trivial metal well above 40 GPa, which is in rather good agreement with previous optical absorption and resistivity measurements [26,27,31,33]. Finally, we show that all the changes observed in orpiment under compression, such as the strong decrease of the bandgap; the softening of many optical vibrational modes; the evolution of the electron localization function (ELF) along the As-S bonds; and the strong increase of the dielectric tensor and Born effective charges, can been explained in the framework of the formation of metavalent bonding at HP. This behavior has been recently proposed for other chalcogenides [36-41]. In summary, we show in the present paper that orpiment under compression is an example, together with GeSe [39], of a compound that develops a pressure-induced transition from covalent to metavalent bonding without the need of undergoing a first-order structural phase transition. This feature contrasts to Se and Te, which undergo this transition after being subject of a first-order phase transition. Moreover, we propose that the formation of metavalent bonding may explain the behavior under compression of other group-15 chalcogenides, so our results pave the way for extending the study of the transformation from covalent to metavalent bonding to other group-15 chalcogenides and related compounds.

## 2.- EXPERIMENTAL DETAILS

Almost pure mineral orpiment from Hunan province (China) was used in the present study. Angle-dispersive powder HP-XRD experiments at room temperature up to 25 GPa were conducted in a membrane-type diamond anvil cell (DAC) at the BL04-MSPD beamline of ALBA synchrotron. Incident monochromatic beam with wavelength of 0.4246 Å was focused to 20 x 20 μm using a pinhole of 50 μm to cut the x-ray beam tail [42]. Images covering a 2θ range up to 18° were collected using a SX165 CCD located at 240 mm from the sample. One-dimensional diffraction profiles of intensity as a function of 2θ were obtained by integration of the observed intensities with the Fit2D software [43]. Le Bail refinements were carried out with GSAS package software for synchrotron measurements [44,45]. Interatomic distances were obtained with the Visualization for Electronic and Structural Analysis (VESTA) software [46]. The equation of state (EoS) of copper [47], which was mixed with the sample powder, was used for pressure calibration.

Room-temperature unpolarized HP-RS measurements up to 14.5 GPa were carried out with a Horiba Jobin Yvon LabRAM HR spectrometer equipped with a thermoelectrically cooled multichannel CCD detector. RS measurements with a spectral resolution better than 2 cm$^{-1}$ were excited using the 632.8 nm line of a He:Ne laser. The use of an edge filter limited the observation of Raman-active modes below 40 cm$^{-1}$. Pressure was determined with the ruby fluorescence method [48]. In both HP-XRD and HP-RS measurements, a methanol-ethanol (4:1 ratio) mixture was used as a pressure-transmitting medium with quasi-hydrostatic conditions up to 10 GPa and deviatoric stresses are within the experimental uncertainty of this soft material up to 25 GPa [49,50].

## 3.- THEORETICAL DETAILS

*Ab initio* calculations within the density functional theory (DFT) [51] were performed to compute the electronic and structural properties of α-$As_2S_3$ by using the plane-wave Vienna Ab initio Simulation Package (VASP) [52,53]. The projector-augmented wave scheme (PAW) [54] was considered to treat the valence and the semi-core states, more explicitly by including the 4s and 4p outermost electrons of As and the 3s and 3p electrons of S as valence electrons, with the remaining of the electrons being considered frozen at the core. In this work, the generalized gradient approximation with the Perdew-Burke-Ernzerhof (PBE) [55] parametrization was used for the exchange and correlation energy, after evaluating calculations performed also with the PBEsol modified version [56] in Ref. **35**. The plane-wave kinetic-energy cutoff was defined with 360 eV, which together with the use of a dense Monkhorst-Pack grid [57] with a 6×4×2 k-point reciprocal space sampling mesh, ensured a convergence of the total energy around 1 meV with deviations of the stress tensor from a diagonal hydrostatic form of less than 1 kbar (0.1 GPa). The vdW corrections to the total energy were taken into account within DFT by using the D2 method [58]. Electronic band-structure calculations were carried out at different pressures along selected high-symmetry k-points on the first Brillouin-zone (BZ). Additionally, analysis of the electron density topology and of the electron localization function (ELF) were accomplished with the VESTA program employing data from the VASP code [59,60].

Lattice-dynamics calculations were performed at the zone center ($\Gamma$-point) and along high-symmetry segments of the BZ as a function of pressure, by using the direct-force constant approach [61,62]. The separate calculations of the forces, which result from a fixed displacement away from equilibrium of the atoms in the primitive cell, necessary to the construction of the dynamical matrix at the $\Gamma$-point of the BZ were achieved with VASP. The diagonalization of the dynamical matrix provides the normal-mode frequencies and allows identifying the irreducible representations and the character of the vibrational phonon modes at the $\Gamma$-point.

## 4.- RESULTS AND DISCUSSION

### 4.1.- Structural and vibrational characterization of $\alpha$-As$_2$S$_3$ at room conditions

As already commented, $\alpha$-As$_2$S$_3$, with monoclinic $P2_1/c$ space group at room conditions, is a layered material with corrugated (or zigzag) layers extended along the $ac$-plane that are linked by vdW interlayer interactions along the $b$-axis (**Figure 1(a)**). Each corrugated layer is formed by heart-shaped six-member rings of corner-sharing AsS$_3$ pyramids (**Figure 1(b)**) Traditionally, the zigzag layers have been understood as formed by spiral AsS chains extended along the $a$-axis and held together by inter-chained S atoms, which allow the link between the chains along the $c$-axis (see **Figure 1(c)**). On the other hand, zigzag layers can also be seen as a W-shape formation of molecules (S1-As2-S3-As1-S2) where each molecule is linked with four adjacent inverted W-shape molecules, mostly along the $c$-axis due to covalent bonding (see **Figure 1(d)**).

At room conditions, each As1 and As2 atom is linked by covalent bonds to one S1, one S2 and one S3 atom at distances around 2.3 Å, leading to an As threefold-coordination (see pyramids in **Figure 1(a)**). Additionally, each As1 and As2 atom has two farther S neighbors on the same layer at distances above 3.0 Å (see dashed lines inside the layers in **Figures 1(a)** and **1(b)**) and other two S neighbors on an adjacent layer at distances above 3.5 Å (see dashed lines between the layers at the top part of **Figures 1(a)** and **1(d)**). In the following section, we will discuss the importance of these long As-S intra- and inter-layer distances (above 3 Å at room pressure) in orpiment at high pressure.

According to the difference of the As-S distances, we can consider orpiment as an arrangement of $AsS_3$ pyramids forming $AsS_3E$ tetrahedral units, where E refers to the lone electron pair (LEP) of the As atoms. Therefore, each layer can be described alternatively by an arrangement of two tetrahedral $AsS_3E$ units; i.e. those formed by the two inequivalent As atoms, each surrounded by three S atoms and the corresponding cation LEP that points towards the interlayer space. The strong As LEP distorts the electronic distribution and the geometry of the polyhedral units leading to the layered structure of orpiment.

The layered structure of orpiment bears similarities with the chain-like structure of the valentinite mineral ($\beta$-$Sb_2O_3$), where the W-shape molecules (linked through a helical form) lead to the formation of an acicular or quasimolecular crystal [63]. On the other hand, the layered structure of orpiment is isostructural to that of $\alpha$-$As_2Se_3$ [64] and belongs to the same space group that the claudetite mineral ($\beta$-$As_2O_3$); however, claudetite is not isostructural to orpiment, because the layered structure of the former is due to the arrangement of $AsO_3$ molecules in a buckled structure [65] as in black phosphorous [66]. The layered structure of orpiment also bears some resemblance to the zigzag layered structure recently described for $Sb_2S_3$, $Sb_2Se_3$, and $Bi_2S_3$ [67-69], crystallizing in the orthorhombic *Pnma* structure, and to that of $\alpha$-$As_2Te_3$, crystallizing in the monoclinic *C2/m* phase **[70]**. In these former group-15 $A_2X_3$ sesquichalcogenides, the layers are clearly composed by spiral chains held together through long and weak *A-X* bonds, whereas in $As_2S_3$ the chains are connected through short and strong *A-X* bonds. Another difference found between these systems, is regarding the cation coordination in the different structures at room pressure. In α-$As_2S_3$, the two inequivalent As atoms are threefold-coordinated; in $Sb_2S_3$, $Bi_2S_3$ and $Sb_2Se_3$, there is an average fourfold coordination because one cation has threefold coordination and the other has fivefold coordination; and in α-$As_2Te_3$, As atoms have a coordination between five and six. This means that the larger cation or the larger anion favors an increase of cation coordination probably due to the smaller cation LEP effect **[71]**.

The experimental XRD pattern of our sample at room conditions (see **Figure 2(a)**) shows that the sample corresponds to the α phase without the presence of additional phases or impurities. The pattern was fitted to a monoclinic structure (*P2₁/c* space group), with small residuals and a correlation factor $R_{wp}$ of 10%, yielding the following lattice parameters: $a$ = 4.2626(5) Å, $b$ = 9.6056(7) Å, $c$ = 12.1836(12) Å and $\beta$ = 110.054(7) º with $V_0$ = 468.61(5) Å³. In **Table 1** we can observe that these values are in good agreement with previously reported experimental values **[10,11,13]**, with our theoretical calculated lattice parameters and with other previously published theoretical data **[33,72]**. The theoretical atomic parameters of the five inequivalent atomic sites (three S and two As atoms) of α-$As_2S_3$ are also given in **Table S1** in the Supporting Information (SI) section for comparison with experimental and theoretical values.

As regards to the vibrational properties of α-$As_2S_3$, group theory predicts sixty vibrational modes at the BZ center with the following mechanical representation **[73]**:

$$\Gamma_{60} = 15A_g(R) + 13B_u(IR) + 15B_g(R) + 14A_u(IR) + 1A_u + 2B_u \qquad (1)$$

where g (gerade) modes are Raman-active (R) and u (ungerade) modes are infrared-active (IR), except one $A_u$ and two $B_u$ modes, which are acoustic modes. Therefore, $\alpha$-$As_2S_3$ has 30 Raman-active modes and 27 IR-active modes.

**Figure 2(b)** shows the unpolarized RS spectrum of orpiment at room conditions together with the theoretically-predicted Raman-active mode frequencies at room pressure. The values of the experimental and theoretical frequencies of the Raman-active modes of $\alpha$-$As_2S_3$ at room pressure are summarized in **Table 2**. The RS spectrum of orpiment at room pressure clearly shows 19 out of the 30 theoretically-predicted Raman-active modes and a quite good agreement between our experimental and theoretical frequencies (including vdW interactions). The spectrum is similar to those previously reported, with a doublet between 60 and 70 $cm^{-1}$, a phonon gap between 200 and 290 $cm^{-1}$ and a triplet around 360 $cm^{-1}$**[14-29,33]**. Unfortunately, we have not been able to observe modes below 40 $cm^{-1}$ due to the limit imposed by the edge filter of our spectrometer.

The vibrational spectrum of orpiment at room pressure is separated between a low-frequency region (below 200 $cm^{-1}$) and a high-frequency region (above 260 $cm^{-1}$) with a phonon gap between these regions. Some Raman features are observed inside the phonon gap of $\alpha$-$As_2S_3$, which are assumed to be second-order modes, so they will not be further commented in this work. Among the measured RS modes below 250 $cm^{-1}$ and above 280 $cm^{-1}$, the more intense modes are those of the high-frequency region between 350 and 360 $cm^{-1}$. Due to the large number of Raman-active modes located within a small frequency region of the RS spectrum at room pressure and the broadening of experimental peaks, it is almost impossible to identify the modes by solely using the RS spectrum at room pressure and its comparison with lattice-dynamics *ab initio* calculations. Therefore, a tentative assignment of experimental peaks will be performed in a forthcoming section taking into account the HP dependence of the experimental and theoretical Raman-active mode frequencies.

The atomic vibrations of several characteristic Raman- and IR-active modes of α-$As_2S_3$ have been visualized with the aid of the Jmol Interface for Crystallographic and Electronic Properties (J-ICE) interface [74] (see **Figures S1 to S14** in the SI) and discussed in the SI. An interesting feature is that many vibrations evidence that atoms of the same chain (As1, S1, As2, and S2) vibrate in-phase or out-of-phase, whereas S3 atoms interconnecting the chains vibrate in a quite different fashion than the other four inequivalent atoms. This observation clearly evidences the chain-like nature of the layers of orpiment in good agreement with a previous vibrational study [24].

As a layered material, some of the most important vibrational modes of α-$As_2S_3$ are the rigid layer modes. Our calculations confirm that the two lowest-frequency modes located at 27 and 38 $cm^{-1}$, and attributed to the $A_g^1$ and $A_g^2$ modes, correspond to the rigid shear layer modes of orpiment (**Figure S1** in the SI). Moreover, our calculations show that the compressional or longitudinal rigid layer mode is the $B_g^1$ mode theoretically predicted at 62 $cm^{-1}$ (**Figure S2** in the SI). This frequency value is in good agreement with previous estimations of the compressional mode [22]. At room pressure, we have observed this mode is within a region where it is overlapped with inter-chain vibrational modes (see comments in the SI), as previously suggested [24].

As regards other vibrational modes of orpiment, those of the high-frequency region mainly correspond to As-S stretching modes located between 350 and 400 $cm^{-1}$ and to a mixture of As-S stretching and bending modes between 260 and 350 $cm^{-1}$. On the other hand, modes located below 200 $cm^{-1}$ correspond to pure As-S bending modes, being those below 100 $cm^{-1}$ mainly related to rotations and translations of the spiral chains, which can be seen as rigid units. This distribution of modes agrees well with previous works, that assigned the As-S bending modes and As-S stretching modes near 200 and 400 $cm^{-1}$, respectively [18,75]. In fact, the modes close to 380 and 350 $cm^{-1}$ can be assigned to the antisymmetric and symmetric As-S stretching vibrations inside the spiral chains, respectively, in good agreement with previous estimations **[28]**.

### 4.2.- Structural characterization of α-As$_2$S$_3$ at HP

**Figure 3** shows powder HP-XRD patterns of orpiment at selected pressures up to 26.6 GPa. XRD patterns were analyzed by means of Le Bail fits to the monoclinic *P2$_1$/c* structure, which is stable up to the maximum achievable pressure. Le Bail analysis of this material allowed the possibility of obtaining the structural parameters of α-As$_2$S$_3$ at different pressures (see **Figure 4**). A progressive increase of the Bragg peaks width was observed and explained in the framework of a progressive loss of the hydrostatic conditions.

A monotonous decrease of the unit-cell volume of orpiment up to 26.6 GPa is observed in **Figure 4(a)**, where experimental volume data at several pressures have been compared with DFT calculations, both including and without including vdW interactions. At low pressures, the pressure dependence of the experimental volume agrees quite well with the theoretical simulation when vdW interactions are included. However, experimental data above 10 GPa show a smaller agreement with these calculations and much better agreement with calculations that do not include vdW interactions. The above experimental result indicates that orpiment is more incompressible above 10 GPa. This fact is likely due to the loss of quasi-hydrostatic conditions of the pressure-transmitting medium above this pressure.

The fit of pressure vs. experimental volume up to 10 GPa through a third-order Birch-Murnaghan Equation of State (BM-EoS) **[76]**, according to the trend showed by its *F-f* plot (**Figure S15 in SI**) yields a zero-pressure volume, bulk modulus and pressure derivative of the bulk modulus of: $V_0$=468.61(5) Å$^3$, $B_0$=13.3(5) GPa and $B_0$'=8.9(5), respectively. The volume at zero pressure, $V_0$, was fixed to the value measured outside the DAC and the order of the BM-EoS was determined by the monotonously positive slope of the Eulerian strain vs normalized pressure plot of the theoretically simulated data. This plot agrees with highly compressible non-covalent compounds with large cation LEP activity, as it has been observed for structurally-related sesquichalcogenides, such as $Sb_2S_3$, $Sb_2Se_3$, $Bi_2S_3$ and $\alpha$-$As_2Te_3$ and characterized by bulk modulus pressure derivatives larger than 4 [69,70].

Our experimental bulk modulus and pressure derivative under hydrostatic conditions for orpiment can be compared to our theoretical data with and without vdW interactions. As expected in the description of **Figure 4(a)**, the theoretical bulk modulus is closer to the experimental one when calculations include vdW interactions ($B_0$= 12.7(5) GPa) than when calculations do not include them ($B_0$= 16.1(12) GPa). Nevertheless, the theoretical pressure derivative of the bulk modulus at zero pressure is considerably smaller than the experimental value ($B_0$'= 8.9(5)) in the case of simulations with vdW interactions ($B_0$'=7.0(14)) than in simulations without vdW interactions ($B_0$'= 7.9(6)). A comparison of these results with similar compounds, such as the claudetite polymorph of $As_2O_3$ or $\alpha$-$As_2Te_3$, indicates that orpiment presents a similar value of the bulk modulus to the former ($B_0$= 15.5(4) GPa) [77], but a much smaller value to the bulk modulus than the latter ($B_0$= 24(3) GPa) [70]. This result is consistent with the higher LEP activity of As atom in oxides and sulfides than in selenides and tellurides [71].

The pressure dependence of the experimental and theoretical lattice parameters of $\alpha$-$As_2S_3$ up to 26.6 GPa also shows a monotonous and smooth decrease (see **Figure 4(b)**). The good agreement of the theoretical behavior of volume and lattice parameters with our experimental data at room temperature allows us to conclude that no first-order phase transition occurs throughout the whole range of studied pressures. Furthermore, the pressure dependence of our lattice parameters do not evidence anomalies at the region close to 10 GPa, which could be indicative of any structural transition as already suggested [30]. Therefore, our results are in good agreement with recently published theoretical data [33].

The analysis of the axial compressibility in monoclinic structures requires a more complex analysis than for more symmetric phases because the $\beta$-angle is not 90º, so the directions of maximum and minimum compressibility of the compound are usually not along any of the three crystallographic axes. Therefore, we have analyzed the compressibility of the material by calculating and diagonalizing the experimental and theoretical isothermal compressibility tensor $\beta_{ij}$ at different pressures (details are given in the SI). This tensor is a symmetric second rank tensor that relates the state of strain of a crystal to the pressure change that induced it [78] and it has been obtained with the finite Eulerian approximation as implemented in the Win_Strain package [79].

**Figures 5(a)** and **5(b)** describe the pressure dependence of the tensor elements (numerical data are provided in **Tables S1 and S2** in the SI) corresponding to the compressibilities along the different axis and the direction of maximum compressibility with respect to the $c$-axis ($\psi$) or to the $a$-axis ($\theta$), respectively. According to these figures, the $a$-axis (the spiral axis) is more compressible than the $b$-axis (the axis perpendicular to the layers) up to 23 GPa. This is a surprising result for layered materials, where the axis perpendicular to the layers is usually the most compressible one; however, it is coherent with the spiral chain-like nature of the layers in orpiment and the small connectivity of the spiral layers along the $a$-axis.

The compressibility trend among the different crystallographic directions remains constant ($\beta_{11} > \beta_{22} > \beta_{33}$) up to 23 GPa, being all axial compressibilities positive, thus indicating a compression of the structure along the three main crystallographic axes up to 23 GPa. This result is somewhat in disagreement with previous estimates by Besson *et al.* [26], who found positive and negative values for the compressibilities along the $a$- and $c$-axis of the layer plane, respectively, and suggested a compression of the layers in the direction perpendicular to the spiral chains and an expansion of the spiral chains along the helicoidal axis, similar to what occurs in trigonal Se and Te **[80,81]**.

The reliability of our theoretical calculations allowed us to obtain the theoretical compressibility tensor up to 32 GPa. A sudden regularization of the three axial compressibilities to roughly the same value is attained around 23 GPa, with a quick change of the direction of maximum compressibility around this pressure. At 24 GPa, the direction of maximum compressibility is found to be along the *b*-axis ([010] direction); reason why we do not display any value on **Figure 5**. However, above 27 GPa, the direction of maximum compressibility is again found to be within the *ac*-plane, but now it is close to the *c*-axis. In particular, $\Psi = 15(4)^\circ$ and $\theta = 92(4)^\circ$ at 32 GPa. The drastic change of the direction of the maximum compressibility observed above 23 GPa can be considered as a more significant feature of the variation of properties associated to the low-pressure structure.

For completeness, the experimental and theoretical pressure dependence of the axial ratios in $\alpha$-As$_2$S$_3$ is shown in **Figure S16** in the SI. The monotonous trend of all the axial ratios is well reproduced by our theoretical calculations including vdW interactions. A closer look into the slopes reveals a clear change of tendency above 20 GPa, where all three ratios seem to become insensitive to pressure up to 50 GPa. A similar behavior is observed in the pressure dependence of the $\beta$-angle (inset of **Figure 4(b)**). In some previous works, the change in the axial ratios of several group-15 sesquichalcogenides has been considered as a proof for the occurrence of a pressure-induced electronic topological transition (ETT) **[67,82,83]**, since a minimum of the *c/a* ratio has previously been observed to be coincident with the occurrence of a pressure-induced ETT in $\alpha$-Bi$_2$Se$_3$, $\alpha$-Sb$_2$Te$_3$, and $\alpha$-Bi$_2$Te$_3$ **[84]**. However, in recent works **[69,85]**, this fact has been put into question because the minimum of the *c/a* ratio can be simply originated by a change of the ratio of inter-layer/intra-layer forces. Therefore, the change of the minimum of the *c/a* ratio does not necessarily warrant a change in the electronic density of states near the Fermi level leading to an ETT; however, it can be indicative of an IPT, as recently suggested for orpiment **[33]**.

To prove that the change of the axial ratios is indicative of a pressure-induced IPT, we have analyzed the pressure dependence of the internal atomic parameters of the five inequivalent atomic sites in α-$As_2S_3$ (see **Figure S17** in the SI). The most noticeable results are those observed for the three coordinates of the two inequivalent As atoms, which show a clear trend towards certain fixed coordinates above 18 (25) GPa, as obtained for calculations without (with) vdW interactions. This result occurs at similar pressures where the change of the direction of the maximum compressibility was observed, which clearly indicates a modification of the HP behavior of the low-pressure phase.

A remarkable result, derived from the pressure dependence of the atomic coordinates, is the grouping and regularization of many theoretical As-S interatomic distances around 18 (25) GPa, without (with) vdW interactions (see **Figure 6**). As a consequence, there is an increase of As1 coordination from threefold at 0 GPa to more than fivefold (5+2) above 20 GPa (see **Figure 1e**) because, on one hand, all As1-S1 and As1-S3 intra-layer distances become similar, thus giving a fivefold coordination and, on the other hand, two additional As1-S2 inter-layer distances become smaller than 3 Å. Similarly, there is an increase of As2 coordination from threefold at 0 GPa to fivefold above 20 GPa that stems from the equalization of the As2-S2 and As2-S3 intra-layer distances. In summary, we can conclude that above 20 GPa the As1 (As2) polyhedral units of orpiment pass from a threefold coordination towards a sevenfold (fivefold) coordination without a change in the space group.

A more detailed analysis of the As polyhedral units (see **Figures S18 to S20** in the SI) shows how both As1 and As2 polyhedral units remain in a threefold effective coordination below 10 GPa. However, above this critical pressure, there is a progressive increase of the effective coordination up to 22 GPa, where the As2 polyhedron remains in fivefold effective coordination, whereas that of the As1 polyhedral unit increases up to sixfold effective coordination (**Figure S20** in the SI). We must note that in distorted polyhedral units (values very different from 0 in **Figure S19** in the SI), the effective coordination number does not describe properly the coordination of the polyhedral unit because some of the interatomic distances are longer than others, thus leading to an effective coordination dominated by the next neighbors (**Figure S20** in the SI) but with large polyhedral volumes (**Figure S18** in the SI). Above 20 GPa, the effective coordination of As1 is slightly larger than five despite there is a higher real coordination of (5+2) due to the strong distortion of the polyhedral unit around As1, whereas for As2 the effective coordination (five) coincides with the real coordination due to the proximity of the interatomic distances.

The above results between 0 and 20 GPa are consistent with: i) the regularization of the axial compressibilities, ii) the trend found towards the fixed As coordinates, and iii) the similar $x$ and $z$ coordinates showed by the Wyckoff sites of As1, S2, and S3 atoms, above 18 (25) GPa in our calculations without (with) vdW interactions. All these results indicate the presence of an IPT above 20 GPa, in agreement with a recent paper **[33]** and also with by HP-XAS results that reported a change in the As environment above 10 GPa and an increase of As coordination above 30 GPa **[30]**.

In summary, HP-XRD measurements and DFT calculations carried out up to 32 GPa show that the effect of pressure upon the structure of α-$As_2S_3$ does not result in a first-order phase transition, but triggers a change of the trend of atomic coordinates, of axial ratios, and an increase of coordination of As atoms above 20 GPa; as well as a drastic change of the direction of the maximum compressibility above 23 GPa. All these features are in good agreement with published data **[30,33]** and could be attributed to an pressure-induced IPT in orpiment above 20 GPa.

### 4.3.- Vibrational characterization of α-As₂S₃ at HP

In order to study the effect of pressure on the vibrational properties of orpiment and better understand the HP behavior of α-As$_2$S$_3$, we have carried out HP-RS measurements on orpiment samples up to 14.5 GPa. Experimental results are compared to lattice-dynamics calculations of α-As$_2$S$_3$. On **Figure 7** we show the room-temperature RS spectra of orpiment at selected pressures under hydrostatic conditions. In this figure, it is clear to observe the absence of any first-order PT in orpiment up to 14.5 GPa, in good agreement with our HP-XRD measurements and previously published results [30,33]. However, RS spectra show a much larger number of Raman modes between 2.0 and 4.1 GPa, despite the disappearance of the Raman modes in the phonon gap between 210 and 250 cm$^{-1}$. As already commented, these modes in the phonon gap, together with a mode observed above 1 GPa near 120 cm$^{-1}$, are assumed to be second-order modes and will not be further discussed. No major changes of the RS spectra are observed between 4.1 and 14.5 GPa. In this context, it must be noted that the bandgap of orpiment (2.7 eV at room pressure) decreases under pressure at a rate of -0.14 eV/GPa [26,31]. This means that the bandgap equals the HeNe laser energy (1.96 eV) at 5.3 GPa, so some resonance effects could be observed above ca. 4 GPa. However, no damage in the sample or redshift of frequency modes after long exposure time to the laser, which indicates the lack of radiation damage in our samples.

As already mentioned, it is very difficult to assign the features that show up in the room-pressure RS spectrum of α-As$_2$S$_3$. However, it is possible to perform a tentative peak assignment by studying the pressure dependence of the Raman-active modes in combination with lattice-dynamics calculations on the basis of the correlation of the frequencies and their pressure coefficients (see **Table 2** and **Figure 8(a)**). Our observed frequencies and pressure coefficients are in good agreement with previous HP-RS studies [23,26,27,33]. Notably, the shear or transverse rigid modes A$_g^1$ and A$_g^2$ at 26 and 37 cm$^{-1}$ and the first B$_g$ mode (B$_g^2$ mode) at 69 cm$^{-1}$ show the largest relative increase of frequency under compression (0.24, 0.13 and 0.15 GPa$^{-1}$, respectively). Moreover, the $B_g^2$ mode shows the experimental and theoretical largest pressure coefficient and mode Grüneisen parameter of all Raman-active modes (around 10.5 cm$^{-1}$/GPa and 1.97, respectively).

While the above results regarding the shear rigid layer $A_g$ modes are rather common in layer materials, the result regarding the $B_g^2$ mode is quite surprising because the largest pressure coefficient is indeed expected for the compressional or longitudinal layer mode $B_g^1$ located at 62 cm$^{-1}$ (see **Figure S2** in the SI). It is well known that the compressional layer mode usually shows higher frequencies and pressure coefficients than those of the shear rigid layer modes in layered materials with vdW interactions between the layers. This common trend is due to the extraordinary increase of the stretching force constant between neighboring layers caused by the strong decrease of the interlayer distance, as discussed in a recent paper **[86]**. However, we have found that the $B_g^1$ mode in orpiment (not observed experimentally) shows a rather small theoretical pressure coefficient (see **Table 2**). This anomalous pressure dependence of the compressional mode of orpiment can be understood by considering a frequency anticrossing (occurring already at room pressure) between the $B_g^1$ and $B_g^2$ modes. In fact, the large pressure coefficient of the "bare" $B_g^1$ mode (see dashed line in **Figure 8(b)**) is so large, when compared to other Raman modes, that this mode undergoes anticrossings with up to four $B_g$ modes (up to $B_g^5$) in the pressure range between 0 and 15 GPa. This anticrossing allows us to explain the small pressure coefficient of the $B_g^1$ mode and the large pressure coefficient of the $B_g^2$ mode of orpiment. Moreover, this puzzling behavior may explain why the compressional mode was not previously assigned, despite the frequency of the compressional mode was well identified in a previous work **[22]** as well as the pressure coefficient of the $B_g^2$ mode **[23]**.

As observed in **Figure 8(a)**, experimental and theoretical frequencies of Raman-active modes do not show a simple monotonic behavior with increasing pressure. In fact, many Raman modes exhibit a complex behavior under pressure with crossings and anticrossings of modes with different and equal symmetries, respectively. This complex behavior stems from the presence of 30 Raman-active modes in a small frequency region between 30 and 400 $cm^{-1}$, similarly to what has been found in monoclinic $\alpha$-$As_2Te_3$ **[70]**. Despite the complex behavior observed in our theoretical Raman-active modes, several experimental modes have shown a behavior consistent with the theoretical modes. In particular, we have found an experimental reduction of the phonon gap with pressure, an anticrossing of Raman modes close to 150 $cm^{-1}$ around 2 GPa, and the splitting of several Raman modes located near 350 $cm^{-1}$, which are in good agreement with previous results **[26,27,33]**. Such behaviors are supported by our lattice-dynamics calculations for the pressure dependence of the $A_g^9$ and $A_g^{10}$ modes, the $A_g^5$ and $A_g^6$ modes and the $A_g^{12}$, $A_g^{13}$ and $A_g^{14}$ modes, respectively (see **Figure 8(a)**).

A striking feature of orpiment is the large number of Raman modes with negative pressure coefficients at room pressure. This is confirmed by both experimental and theoretical data. In particular, modes located at 155, 291, 325, 356 and 400 $cm^{-1}$ show softening under pressure in rather good agreement with previous measurements **[23,33]**. As noted by Besson et al. **[26]**, negative pressure coefficients of internal modes have been observed in Raman measurements of other chain-like structures, such as trigonal S, Se and Te **[87-89]**, but never for ring molecules, like those found in orthorhombic S **[23]**. In trigonal S, Se and Te, the soft Raman phonon is the $A_1$ mode (the breathing mode of the chains at the *ab*-plane **[90]**) and it can be related to the expansion of the *c*-axis, caused by the increase of the Se-Se intra-chain distance, at the expense of the contraction of the *a*-axis, caused by the decrease of the Se-Se inter-chain distance. In orpiment, the situation is quite different because all three *a*-, *b*-, and *c*-axes suffer a contraction under compression so the explanation for the negative pressure coefficients is not related to the elongation of any axis.

**Figure 6** shows that As1-S3, As2-S2, and As2-S3 (As1-S1) intra-layer distances increase above 0 (10) GPa, thus providing an explanation for the softening of several phonons in different pressure ranges. Notably, changes of the pressure coefficient of some experimental and theoretical Raman-active modes have been observed around 4 GPa in α-$As_2S_3$, which could be attributed to the strong changes of the interatomic distances around these pressures. The changes can be clearly observed in the experimental $B_g^1$ and $A_g^{10}$ modes and in the theoretical $A_g^2$, $B_g^1$, $A_g^8$, $A_g^9$, $A_g^{10}$ and $A_g^{15}$ modes (see **Figure 8(a)**). They can also be observed in the theoretical $B_u^1$, $A_u^2$, $A_u^3$, $A_u^8$, $B_u^7$, $B_u^8$ and $A_u^9$ modes (see **Figure S21** in the SI), which do not evidence anticrossings between 0 and 6 GPa.

**Figures 8(b)** and **S21** show a notable softening of the theoretical vibrational modes of orpiment in the region between 18 and 26 GPa, where the pressure-induced IPT occurs according to our calculations. In order to probe the possibility of a second-order IPT occurring in α-$As_2S_3$, we have studied the phonon dispersion curves calculated at different pressure values, ranging up to 30 GPa (see **Figures S22 and S23** in the SI), because a second-order IPT is related to the presence of a soft phonon mode according to Landau theory. Since we do not observe any soft mode up to 30 GPa, we can conclude that the pressure-induced IPT occurring in α-$As_2S_3$ above 20 GPa must be of order higher than 2; i.e. an IPT of electronic origin, such as the pressure-induced ETTs observed in other group-15 sesquichalcogenides [84].

### 4.4.- Electronic band structure calculations of α-$As_2S_3$ at HP

To complete the picture of the behavior of orpiment at HP and in order to assess whether a pressure-induced ETT could be observed in orpiment, we have performed *ab initio* electronic band-structure calculations of $\alpha$-As$_2$S$_3$ at different pressures to identify possible changes of the band extrema that could be related to a pressure-induced ETT (see **Figure 9**). Our calculations show that orpiment is an indirect bandgap (1.7 eV) semiconductor at 0 GPa, with the valence band maximum (VBM) and conduction band minimum (CBM) being located along the high-symmetry segment of Y-$\Gamma$ and $\Gamma$-Z directions, respectively. The value of the bandgap is clearly underestimated by our DFT calculations, since orpiment is known to have a bandgap around 2.7 eV at room pressure **[26,31]**, but the structure of the electronic bands and the bandgap evolution with pressure are correctly described by this method. The bandgap energy value in our calculations is similar to the values of recent *ab initio* calculations, however differing in the precise locations of the VBM and CBM **[33,72]**.

There are considerable changes in the VBM and CBM of $\alpha$-As$_2$S$_3$ at HP. Both valence and conduction bands show a low dispersion across the BZ at 0 GPa, but a high dispersion above 20 GPa. The flat bands at 0 GPa reflect the 2D character of orpiment, while the highly-dispersed bands above 20 GPa reflect the 3D character of orpiment above this pressure. It can be observed that above 20 GPa, the VBM shifts to the $\Gamma$-point and along the $\Gamma$-B direction, whereas the CBM moves to the B-point. Additionally, a variation of the VBM occurs between 20 and 30 GPa, with the VBM at 30 GPa shifted along the high-symmetry $\Gamma$-Y direction. Since we observe the crossing of an extremum through the Fermi level above 20 GPa, thus modifying of the topology of the Fermi-surface, we thus suggest a possible pressure-induced ETT above this pressure in $\alpha$-As$_2$S$_3$.

The calculated indirect bandgap exhibits a strong decrease at HP from 1.7 eV at 0 GPa to 0 eV at 26 GPa (see solid line in **Figure 10**). Therefore, our calculations provide evidence for a semiconducting-metallic transition in orpiment above 26 GPa. A recent experiment suggests that metallization in orpiment should occur above 42 GPa **[33]**. However, if metallization occurred above 42 GPa, the underestimation of the bandgap should be only 0.3 eV in our calculations (see long dashed line in **Figure 10**). On the other hand, if we take into account the value of the experimental bandgap at room pressure (2.7 eV) and shift the calculated bandgap to match the latter, the evolution of the bandgap (see short dashed line in **Figure 10**) matches remarkably well with the behavior of the optical bandgap (symbols in **Figure 10**) **[26,27]** and the extrapolation of the shifted calculated bandgap to 0 eV yields a closure of the bandgap to 0 eV at pressures that go beyond the studied range (above 50 GPa). In fact, the decrease of the electronic bandgap shows a theoretical pressure coefficient of -0.11 eV/GPa, which is in very good agreement with a previous experimental estimation of the optical bandgap pressure coefficient (-0.14 eV/GPa) **[26,27,31]**. Therefore, our theoretical estimation of the pressure dependence of the bandgap suggests that metallization should actually occur at much higher pressure values of 40 GPa. On the other hand, we found an energetically more favorable structural phase (**Figure S24**) than the *P2₁/n* above 40 GPa, which may indicate the existence of a phase transition towards a metallic phase of the $As_2S_3$. Thus, the metallization experimentally observed above 42 GPa **[33]**, could be explained by the non-hydrostatic conditions in the sample during electric measurements or by a phase transition towards another phase.

In summary, $\alpha$-$As_2S_3$ is a semiconductor with an indirect bandgap (2.7 eV at room pressure) that decreases at HP at a considerable rate until it becomes a metal well above 50 GPa. A clear change in the VBM is observed above 20 GPa that could lead to a pressure-induced ETT; i.e. an IPT of order $2^{1/2}$ according to Ehrenfest notation **[91]**.

### 4.5.- Metavalent bonding in $\alpha$-$As_2S_3$

In this section, we will show that the results obtained so far for the pressure dependence of the structural, vibrational, optical and electric properties of $\alpha$-$As_2S_3$ can be understood on the light of a special case of resonant bonding formalism **[36-41]**, recently termed as metavalent bonding in order to distinguish it from the resonant bonding occurring in benzene and graphite [40]. In particular, the HP behavior of orpiment between 0 and 20 GPa can be considered as the process of change from p-type covalent bonding towards metavalent bonding at HP. Moreover, we will show that the concept of metavalent bonding could be extended to understand the pressure behavior of other group-15 sesquichalcogenides and of trigonal Se and Te.

Metavalent bonding is a recently proposed new class of bonding formalism, mainly located between p-type covalent bonding and metallic bonding, which is characteristic of a new type of materials termed as "incipient metals" [40]. This kind of bonding occurs in materials where there is a deficiency of valence electrons in the unit cell to form a large number of bonds, such as in octahedrally-coordinated rocksalt-related structures as those found for GeTe, SnTe, PbSe, PbTe, Sb, Bi, $Sb_2Te_3$, $Bi_2Se_3$, $Bi_2Te_3$, $AgSbTe_2$, $AgBiSe_2$, $AgBiTe_2$, and $GeSb_2Te_4$ [37,40,41]. Under these circumstances, the few valence electrons available in the formula unit must be shared (resonate) between several bonds.

The main characteristics of this type of bonding are: i) a cation coordination much higher than that assumed with the 8-N rule; ii) high Born effective charges that are much larger than the valence of the atoms; iii) higher optical dielectric constants than typical covalent materials; iv) high mode Grüneisen parameters of phonons and lower wavenumbers of optical phonons than typical covalent materials, thus revealing a high lattice anharmonicity causing a very small thermal conductivity; and v) a moderately high electrical conductivity caused by a very small bandgap. All these features stem from the partial delocalization of electrons that are shared between several bonds. Due to these characteristics, materials featuring metavalent bonding have been named "incipient metals" because they show characteristics close to those of metals. In fact, they exhibit extraordinary properties that make them ideal candidates for phase-change materials, thermoelectric materials, and topological insulators.

Regarding orpiment, the equalization of the As-S intralayer distances above 18 (25) GPa according to our calculations without (with) considering vdW interactions is in agreement with the bonding variation nature of As-S bonds from covalent to metavalent **[39-41]**. To prove such fact, we have probed the bonding character of orpiment at different pressures; e.g. the electron localization function (ELF) along the seven next-neighbor As-S distances, the dielectric tensor, and the Born effective charges, as suggested in Ref. **40**.

The increase of the metavalent bonding in orpiment under compression can be traced by the increase of the theoretical dielectric constants and the Born effective charges (see **Figure 11**), as commented in Refs. **36** and **40**. A strong increase in the absolute value of most of these magnitudes is observed between 0 and 20 GPa **[92]**. The only Born effective charge component that does not show similar behavior within the range of 0 and 20 GPa is the $Z_{zz}$ component, which is related to the *c*-axis, the direction of smallest change in bonding character because it is the direction of smaller compressibility (see **Figure 5**). These changes are a clear indication of the decrease (increase) of the covalent (metavalent) bonding of the layers of orpiment. Besides, the strong decrease of the optical bandgap with pressure and the metallization of orpiment well above 50 GPa is consistent with the metavalent bonding observed for chalcogenides **[40,41]**. The rapid delocalization of the electron charge density over the plane of the layers caused by the steep increase of coordination of As atoms due to the strong compression of the chains mainly along the *a*-axis is the origin of the pronounced negative slope of the bandgap with increasing pressure.

**Figure 12** shows the ELF values along the seven next-neighbor As1-S and As2-S distances obtained from our theoretical results at 0 and 25 GPa. At 0 GPa, both the As1 and As2 atoms have high ELF values (around 0.8) close to the center of the As-S bonds for each of the three bonds of each As atom (As1-S1, As1-S2, As1-S3, As2-S1, As2-S2, and As2-S3); thus evidencing the threefold coordination of As1 and As2 atoms and the covalent bonding in orpiment at room pressure. However, at 25 GPa there are five intra-layer As2-S bonds with high ELF values (As2-S1 bond above 0.8 and two As2-S2 and two As2-S3 bonds above 0.6) showing fivefold coordination for As2 in orpiment. A similar situation occurs for the As1 atoms at 25 GPa, with five intra-layer As1-S bonds with a relatively high ELF (As1-S2 bond near 0.8, two As1-S1 above 0.7 and two As1-S3 above 0.5), but also with two inter-layer As1-S2 bonds with an intermediate value of the ELF (slightly above 0.2), which can be considered to support 5+2 coordination. Note that the typical metavalent bonding in the layer plane (*ac*-plane) of orpiment is clear for the As2 atoms, where all four intra-layer bonds (As2-S2, As2-S3, As2-S4, and As2-S5) have the same ELF value because they have the same As2-S length. On the other hand, the character of As1-S bonds is more complex because the ELF values and lengths of all the intra-layer bonds are not equal. Such behavior suggests that the intra-layer metavalent bonding of As1 atoms is sacrificed by a larger coordination number taking into account S atoms of adjacent layers. Note that these S atoms are not linked to As2 atoms of the monoclinic structure even up to 50 GPa, what evidences the different behavior of As1 and As2 atoms on increasing pressure. The larger coordination of As1 atoms at HP is indicative of a tendency of As1 atoms to metallic bonding, which is usually characterized by a much larger effective coordination number than covalent and metavalent bondings [40,41].

The decrease of the bond force (ELF value) of some As-S bonds in orpiment is likely related to the length increase of those bonds (see **Figure 6**), because the charge of valence electrons in As atoms must progressively be redistributed among several equidistant As-S bonds within the context of metavalent bonding [40,41]. Therefore, the charge at each of the short bonds at low pressure is redistributed between two equal bonds at HP leading to a smaller ELF for the initial short bonds at HP than room pressure. Such a feature is confirmed by the equal values of the ELF in As-S bonds of equal length at HP already commented. Therefore, the decrease of the ELF of the original covalent bonds at HP clearly shows the delocalization of electronic charge in these bonds to form metavalent bonds at HP.

This bond enlargement and charge redistribution typical of the formation of metavalent bonds is expected to cause a decrease of some of the optical phonon frequencies. In particular, a softening of the transverse optical (TO) and longitudinal optical (LO) modes has been predicted in rocksalt compounds with metavalent bonding as compared to the same compound in absence of metavalent bonding **[37,38]**. For a monoclinic compound, such as orpiment, the situation is more complex, because the TO and LO modes of the rocksalt structure are split into many components due to the decrease in symmetry. Therefore, we expect that many optical modes of orpiment will soften with increasing pressure due to the establishment of long-range interactions between neighboring atoms upon increasing the metavalent bonding at increasing pressure. In fact, **Table 2** and **Figure 8** show that one of the most characteristic modes that exhibit a negative pressure coefficient is the $A_g^{10}$ mode (see atomic vibrations in **Figure S11** in the SI), that is experimentally observed at 291 cm$^{-1}$ at room pressure. Other soft Raman- and IR-active modes can be observed in **Figures 8(b) and S21** in the SI, respectively. Moreover, these last figures show a clear decrease of the phonon bandgap (between $A_g^9$ and $A_g^{10}$ and between $B_u^7$ and $B_u^8$) from 0 to 20 GPa. This decrease of the phonon bandgap is also expected in the context of metavalent bonding **[37,38,40]**. Finally, we observed a considerable softening of several modes taking place between 18 and 26 GPa and clear positive slopes of most vibrational modes above 26 GPa; i.e. once the bonding transition to metavalent bonding takes over. These features are in agreement with the development of metavalent bonding since low-frequency values of optical vibrational modes and positive slopes of all Raman-active and IR-active modes have been observed in $Bi_2Se_3$, $Sb_2Te_3$ and $Bi_2Te_3$, which are metavalent compounds at room pressure **[84]**.

A closer look into the HP behavior of other chalcogenides evidences that the softening of optical modes at HP has been observed in a number of chalcogenides, thus suggesting that metavalent bonding is more common than expected at HP in these compounds. Several modes of the low-pressure trigonal phase of Se and Te exhibit a pronounced softening at HP. However, no softening has been found in the HP phases of these two elements [93,94], as it happens in $Bi_2Se_3$, $Bi_2Te_3$ and $Sb_2Te_3$ [84]. This result is consistent with the formation of metavalent bonding in both materials at HP [40]. Therefore, we expect a similar increase of dielectric constants and Born effective charges at HP in the low-pressure trigonal phase of Se and Te, as it has been shown in orpiment and other compounds [36,40,41].

Finally, we want to emphasize that the analysis of the different group-15 sesquichalcogenides (**Table S3** in the SI) reveals that their behavior at HP can also be understood in the framework of metavalent bonding. In particular, some of these materials possess metavalent bonding at room pressure and some others tend to a metavalent bonding at HP. For instance, a similar situation to that of $\alpha$-$As_2S_3$ is expected to occur for isostructural $\alpha$-$As_2Se_3$ and for $\beta$-$As_2Se_3$ at HP since both compounds feature a threefold coordination at room pressure. Unfortunately, there are almost no HP studies of these compounds. On the other hand, $\alpha$-$Sb_2S_3$, $\alpha$-$Sb_2Se_3$ and $\alpha$-$Bi_2S_3$, which crystallize in the orthorhombic *Pnma* phase, feature an average fourfold cation coordination, because one cation has threefold coordination and the other has fivefold coordination. Therefore, they show an intermediate cation coordination between the threefold coordination of $\alpha$-$As_2S_3$ and the sixfold coordination of $\alpha$-$Sb_2Te_3$. They exhibit softening of some high-frequency modes [67-69]. In particular, softening of two high-frequency modes has been observed in $\alpha$-$Sb_2S_3$ up to 12 GPa; pressure at which Sb1 and Sb2 can be considered to have almost a sevenfold coordination due to a pressure-induced metavalent bonding of both Sb atoms [69].

An intermediate case between covalent and metavalent bonding is also that of α-As$_2$Te$_3$, which shows an average cation coordination of 5.5, because one cation has 3+2 coordination and the other has 3+3 coordination. Therefore, it shows an intermediate coordination between the fivefold coordination of α-Sb$_2$Se$_3$ and the sixfold coordination of α-Sb$_2$Te$_3$. Consequently, α-As$_2$Te$_3$ shows almost no soft phonons. Only a few of high-frequency phonons show a soft behavior with almost negligible pressure coefficient between 0 and 4 GPa and a notable hardening above this pressure [70]. An inspection of the evolution of the As-Te distances in this pressure range shows that the As1 (As2) coordination changes from almost five (six) at room pressure to real six (six) above 4 GPa. Therefore, the changes observed in the pressure coefficients of zone-center vibrational modes in α-As$_2$Te$_3$ are indicative of a change of bonding type from an almost metavalent bonding inside the layers to a full metavalent bonding that takes into account also interactions between the layers. Finally, β-As$_2$Te$_3$ with $R\overline{3}m$ symmetry and sixfold coordination is expected to show no soft modes, as α-Bi$_2$Se$_3$, α-Bi$_2$Te$_3$ and α-Sb$_2$Te$_3$ [84].

It must be noted that a notable decrease of the bandgap with pressure has also been obtained in DFT calculations for α-Sb$_2$S$_3$, α-Sb$_2$Se$_3$ and α-Bi$_2$S$_3$ [69], and at a smaller rate also for α-As$_2$Te$_3$ [84] and Bi$_2$Te$_3$ [95]. In fact, we have observed a metallization around 26 GPa in our DFT calculations for α-As$_2$S$_3$, similar to the metallization found around 4 GPa for α-As$_2$Te$_3$ and above 12 GPa for α-Sb$_2$S$_3$. Therefore, it is expected that a similar increase of dielectric tensors and the Born effective charges at HP occurs in these group-15 sesquichalcogenides, thus supporting the occurrence of metavalent bonding in these compounds at HP.

**CONCLUSIONS**

The monoclinic structure of orpiment ($\alpha$-As$_2$S$_3$) at room conditions is composed of layers formed by AsS spiral chains, where the As atoms have threefold coordination. We have studied the structural, vibrational, and electronic properties of orpiment at room temperature and HP and have found that orpiment undergoes a strong compression up to 20 GPa, followed by an IPT of electronic origin, that leads to a coordination of As which is higher than five above that pressure. The most striking feature of the pressure-induced IPT is that several As-S bond distances become equal above 20 GPa.

The lattice dynamics of orpiment has been studied and all the rigid layer modes and the main inter-chain and intra-chain modes have been fully described. Our measurements and calculations evidence the softening of many vibrational modes and the decrease of the phonon gap (it closes above 20 GPa) at HP. Moreover, our calculations confirm the metallization of orpiment above 26 GPa due to the strong decrease of the optical bandgap, thus supporting metallization in laboratory under hydrostatic conditions well above 50 GPa when taking into consideration the underestimation of the energy bandgap in DFT calculations. All these changes are related to the formation of metavalent bonding occurring in orpiment above 20 GPa due to the delocalization of electronic clouds, mainly in the *ac*-plane of the layers, without the need for a first-order structural phase transition as in Se and Te but similar to GeSe **[39]**.

It has been shown that metavalent bonding already occurs in other group-15 sesquichalcogenides at room pressure. Full metavalent bonding occurs in topological insulators and good thermoelectric materials as $\alpha$-Bi$_2$Se$_3$, $\alpha$-Bi$_2$Te$_3$ and $\alpha$-Sb$_2$Te$_3$ with tetradymite layered structure that feature a sixfold cation coordination. In these compounds, metallic bonding is formed at HP after a first-order phase transition to a structure with a cation coordination larger than the original sixfold one. On the other hand, a partial metavalent bonding occurs in $\alpha$-Bi$_2$S$_3$, $\alpha$-Sb$_2$S$_3$ and $\alpha$-Sb$_2$Se$_3$, with an orthorhombic *Pnma* structure, and in $\alpha$-As$_2$Te$_3$. These compounds show an intermediate coordination between three and six. In all these compounds, pressure increases cation coordination leading to a considerable bandgap reduction and partial phonon softening consistent with metavalent bonding. Finally, we predict that a similar situation than for $As_2S_3$ is expected for $As_2Se_3$ polymorphs and also in the low-pressure trigonal Se and Te at HP. These compounds featuring threefold coordination at room pressure are expected to exhibit metavalent bonding at HP.

In summary, our present results on $\alpha$-$As_2S_3$ confirm that pressure is able to tune the metavalent bonding in group-15 sesquichalcogenides with strong LEP activity, such as occurs for orpiment, thus turning common semiconductors into "incipient metals" with promising phase-change, thermoelectric and topological insulating properties at extreme conditions. Since $Sb_2Te_3$ and $Bi_2Te_3$ are topological insulators and two of the best known thermoelectric materials at room conditions due to metavalent bonding, this work paves the way to design new group-15 sesquichalcogenides and related compounds with thermoelectric or topological insulating properties both at room pressure and at extreme conditions.

**Figure Captions**

**Figure 1.** α-$As_2S_3$ structure at room pressure. Big yellow and small purple spheres correspond to S and As atoms, respectively. **(a)** 3D structure. All solid lines correspond to bonds of similar length (~2.3 Å) and dashed lines correspond to longer As-S distances (> 3.0 Å). **(b)** Structure of the layer in the *ac* plane showing the heart-shaped rings. The upper part shows the $AsS_3$ pyramids due to the short bonds that result in the threefold coordination of As and twofold coordination of S at room pressure. Inside the cell edges we show the long intralayer bonds that would contribute to fivefold coordination of As above 25 GPa. **(c)** Structure viewed as AsS spiral chains (in As1-S1-As2-S2 sequence) joined by inter-chain S3 atoms. Only short bonds are displayed. Solid (dashed) lines correspond to intra (inter)-chain bonds. **(d)** Projection of the α-$As_2S_3$ structure onto the *bc* plane. **(e)** 3D structure of $As_2S_3$ at 30 GPa, showing polyhedral units with fivefold coordination of As1 and As2 atoms (up and middle) and with 5+2 coordination for As1 atoms (down).

**Figure 2.** Structural and vibrational characterization of α-$As_2S_3$ at room conditions: **(a)** Powder XRD pattern (circles). Le Bail analysis (black solid line) and residuals (red solid line) are also plotted. **(b)** RS spectrum. Bottom marks indicate the theoretical Raman-active mode frequencies from calculations including vdW interactions.

**Figure 3.** Powder HP-XRD patterns of α-$As_2S_3$ at selected pressures up to 26.6 GPa. Patterns are shifted vertically for comparison.

**Figure 4. (a)** Experimental (symbols) and theoretical (solid lines) pressure dependence of the unit-cell volume in α-$As_2S_3$. Black (blue) solid lines represent data from calculations including (do not including) vdW interactions. Dashed lines correspond to experimental data fit to a 3[rd] order BM-EoS. **(b)** Experimental (symbols) and theoretical (solid lines) pressure dependence of the lattice parameters and monoclinic β angle (inset) in α-$As_2S_3$. All calculations include vdW interactions.

**Figure 5. (a)** $β_{xx}$ coefficients of the compressibility tensor that indicates the compressibility along the different crystallographic axes. **(b)** Angle of maximum compressibility ψ relative to the *c*-axis (from *c* to *a*) or equivalently θ relative to the *a*-axis (from *a* to *c*). Solid lines represent data from calculations and symbols data from our experiments.

**Figure 6.** Evolution under pressure of the theoretical As-S interatomic distances of α-As$_2$S$_3$. **(a)** As1 and its next-neighbor S atoms. **(b)** As2 and its next-neighbor S atoms. Dashed (solid) lines represent data from theoretical calculations that include (do not include) vdW interactions.

**Figure 7.** Room-temperature RS spectra of orpiment at selected pressures up to 15 GPa.

**Figure 8. (a)** Experimental (symbols) and theoretical (lines) pressure dependence of the Raman-active mode frequencies of As$_2$S$_3$ up to 16 GPa. Black (red) color represents A$_g$ (B$_g$) Raman-active modes. Theoretical curves correspond to calculations with vdW interaction. **(b)** Pressure dependence of the theoretical Raman-active mode frequencies of As$_2$S$_3$ up to 36 GPa. Red dotted line represents the tentative pressure dependence of the "bare" B$_g^1$ mode (the compressional rigid layer mode) if no anticrossing would occur with other modes of the same symmetry. Note that a change in the pressure coefficient of the "bare" B$_g^1$ mode observed near 4 GPa and that a considerable softening of some vibrational modes is observed between 18 and 26 GPa.

**Figure 9.** Calculated band structure of As$_2$S$_3$ at different pressures: **(a)** 0 GPa, **(b)** 5 GPa, **(c)** 10 GPa, **(d)** 15 GPa, **(e)** 20 GPa, and **(f)** 30 GPa.

**Figure 10.** Pressure dependence of the energy bandgap of orpiment. Solid, long dashed and short dashed lines correspond to results of theoretical calculations including vdW interaction, theoretical calculations shifted by 0.3 eV (to match the metallization observed above 42 GPa in Ref. 33) and theoretical calculations shifted by 1.0 eV (to match the optical bandgap at 0 GPa as in Refs. 26 and 27), respectively. Symbols correspond to experimental data from Refs. 26 and 27.

**Figure 11.** Pressure dependence of the dielectric functions **(a)** along the main crystallographic axes and Born effective charges **(b,c,d)** of the different As1, As2, S1, S2 and S3 atoms along the main crystallographic axes.

**Figure 12.** Pressure dependence of the ELF along the As1-S (top) and As2-S (bottom) bonds at 0 GPa **(a)** and 25 GPa **(b)**.

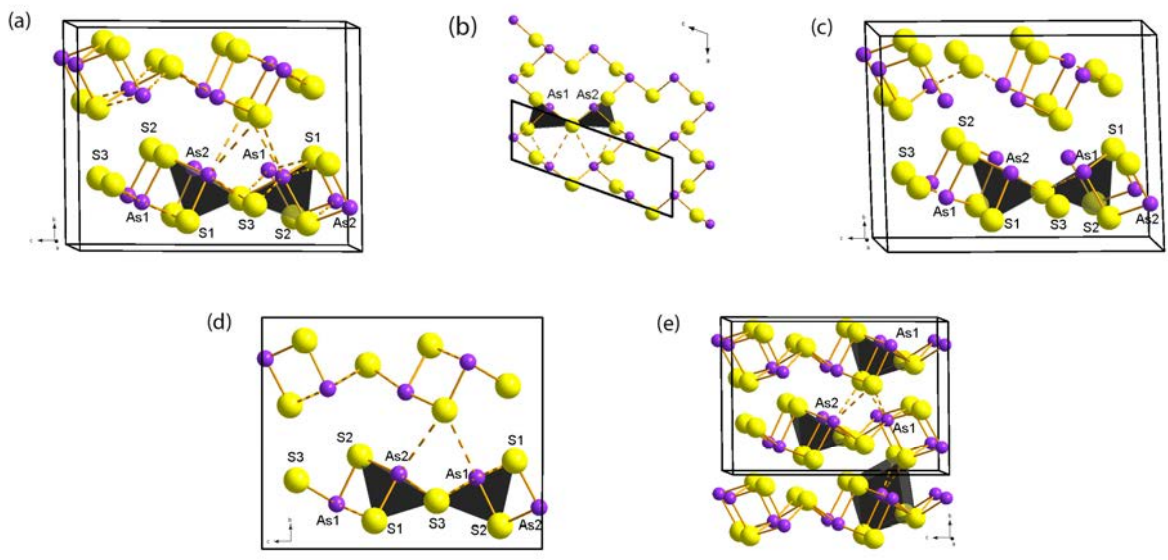

Figure 1

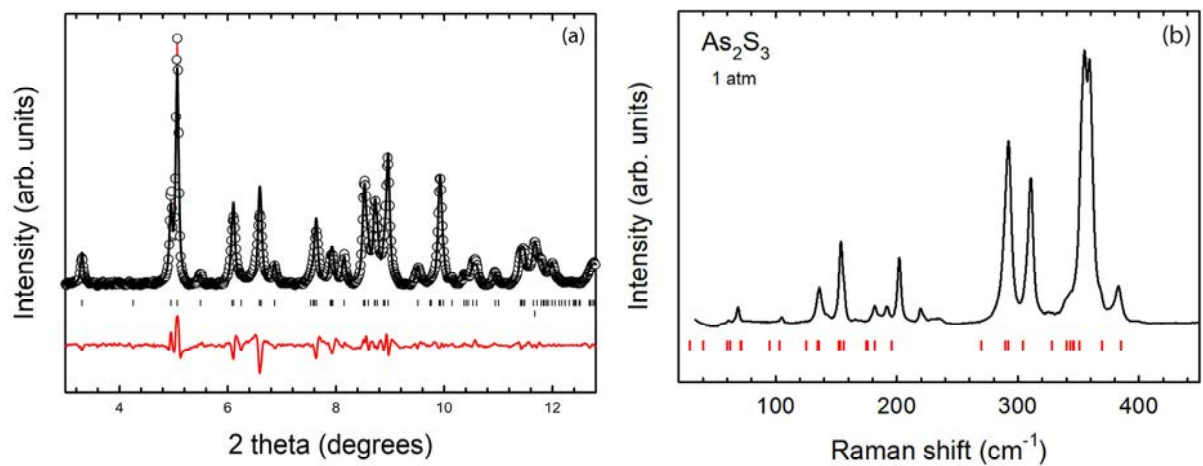

Figure 2

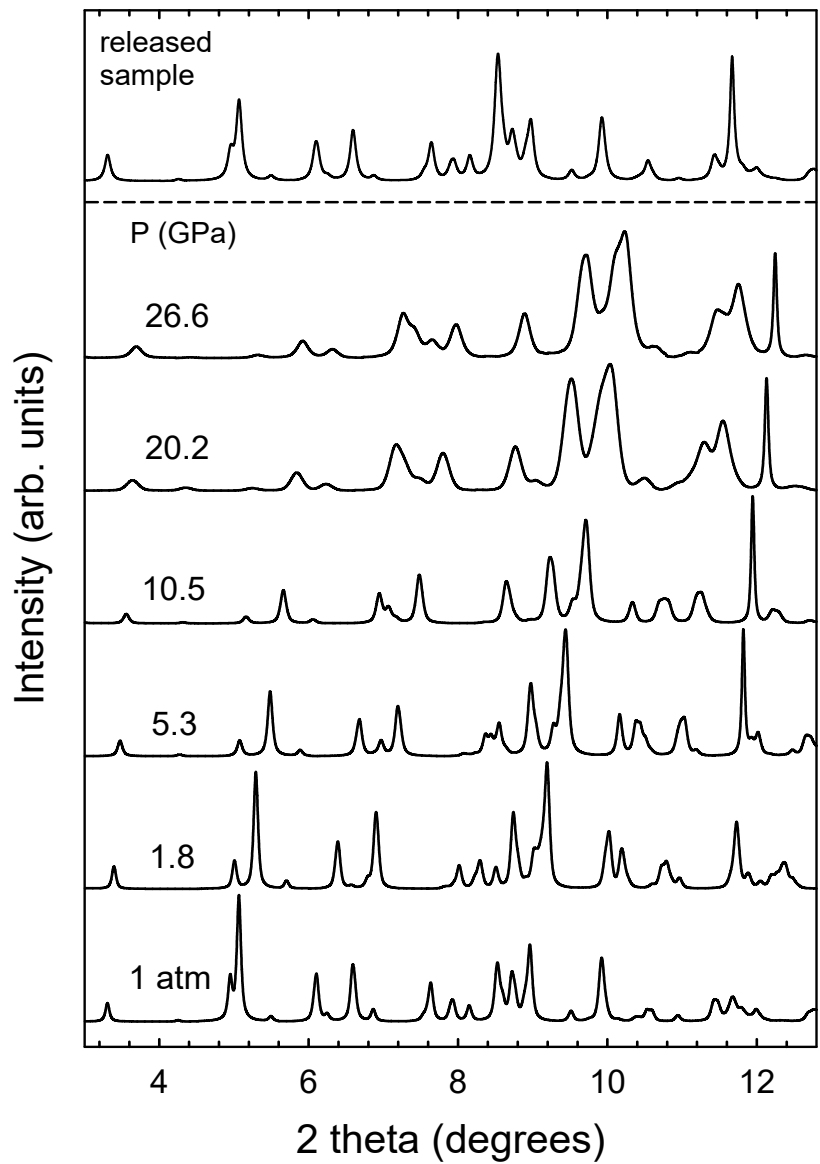

Figure 3

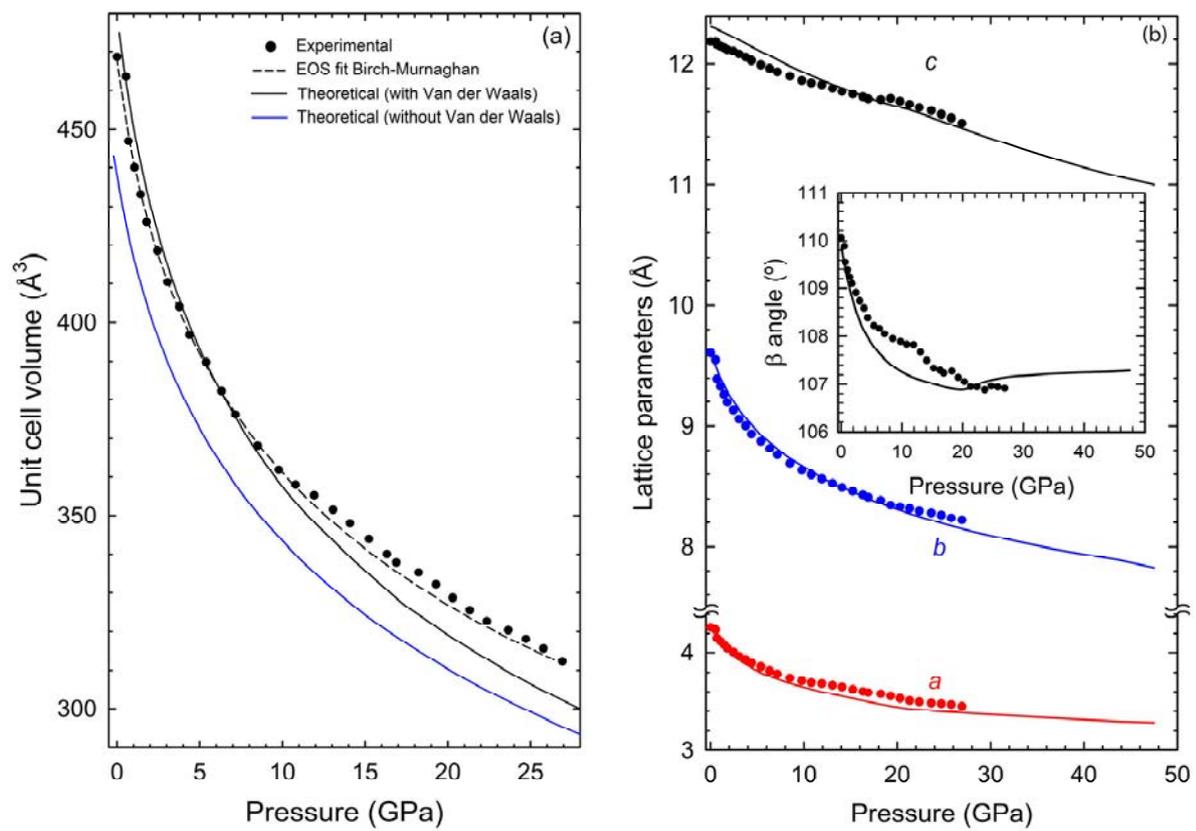

Figure 4

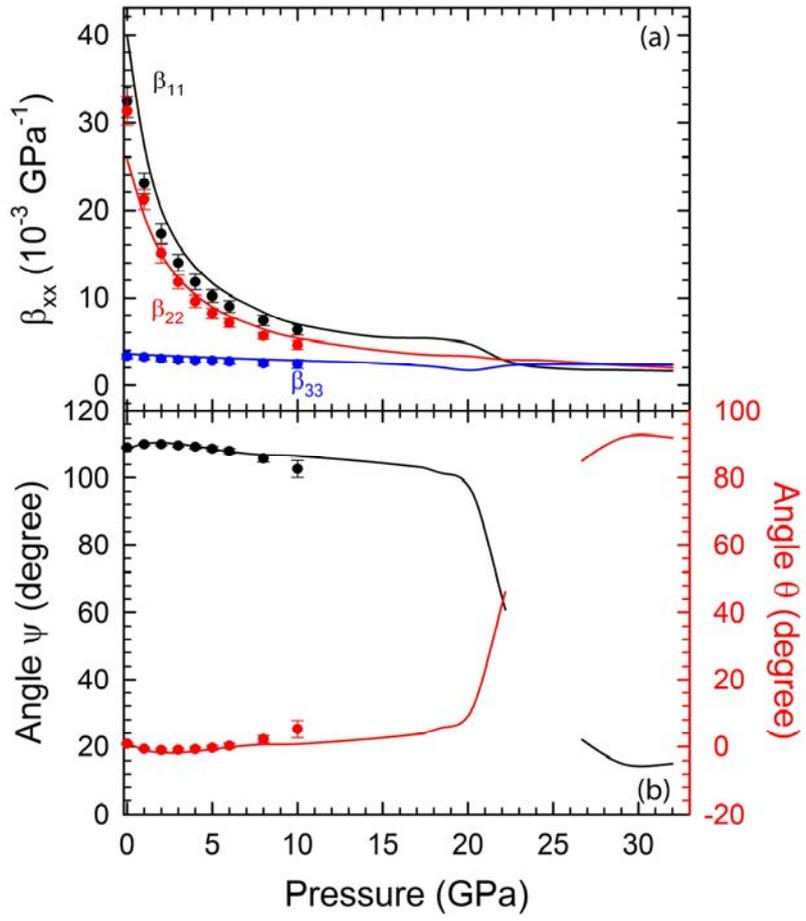

Figure 5

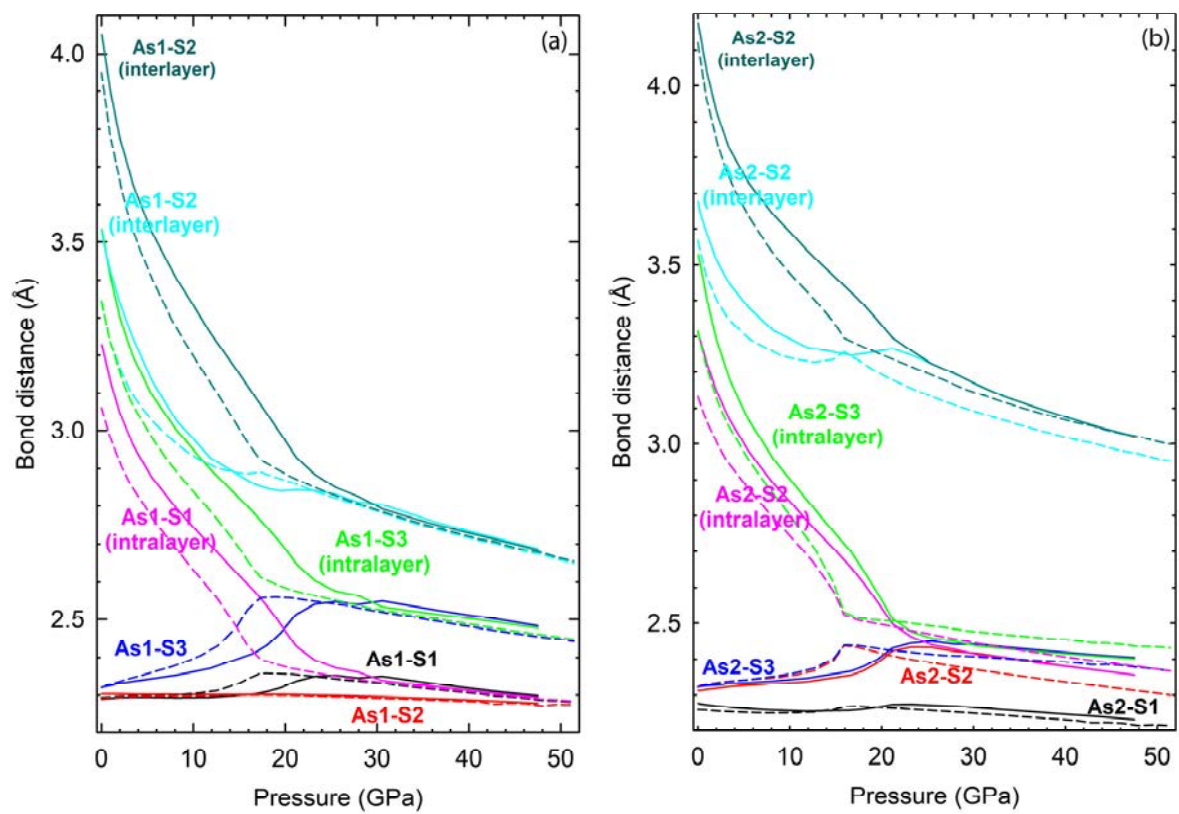

Figure 6

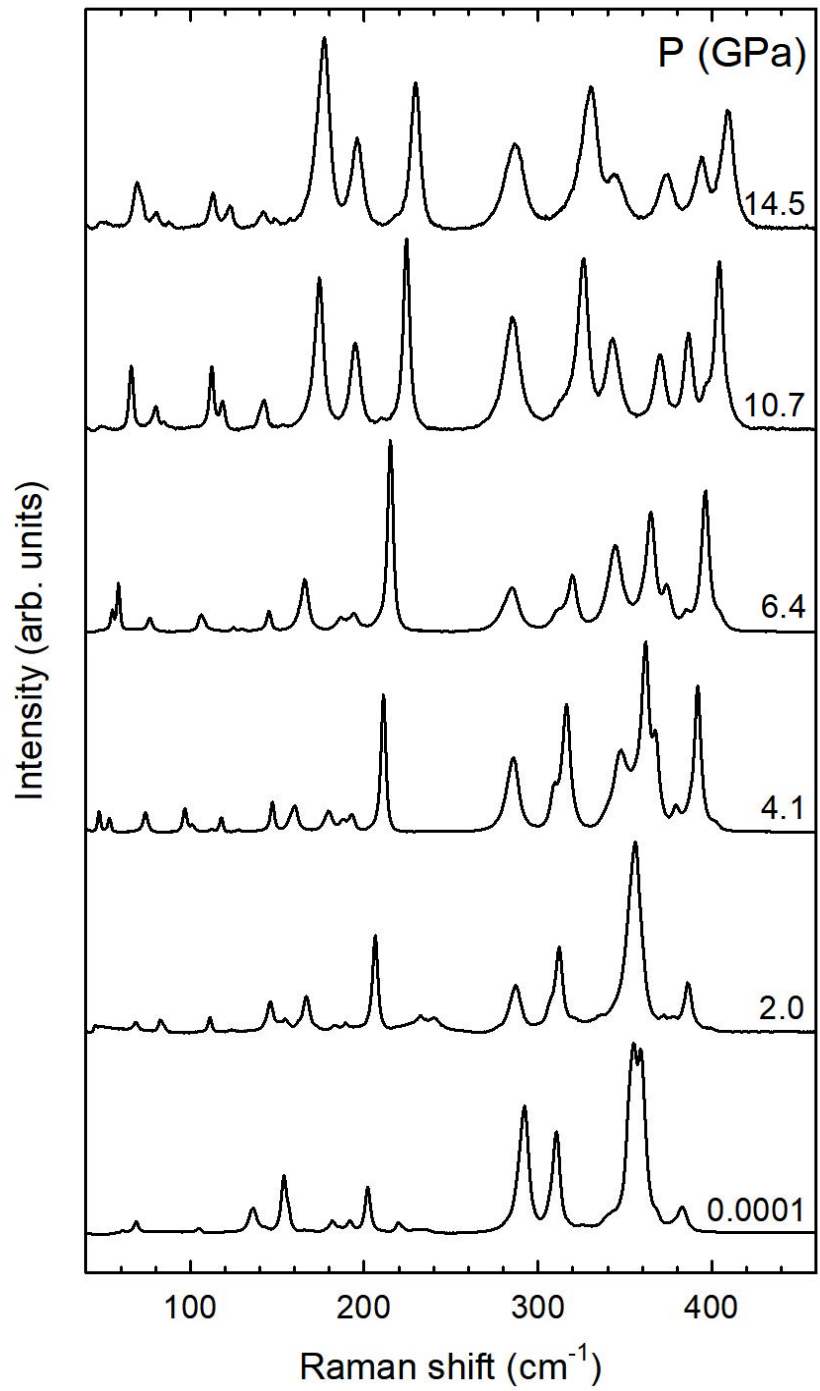

P (GPa)

14.5

10.7

6.4

4.1

2.0

0.0001

Intensity (arb. units)

Raman shift (cm$^{-1}$)

Figure 7

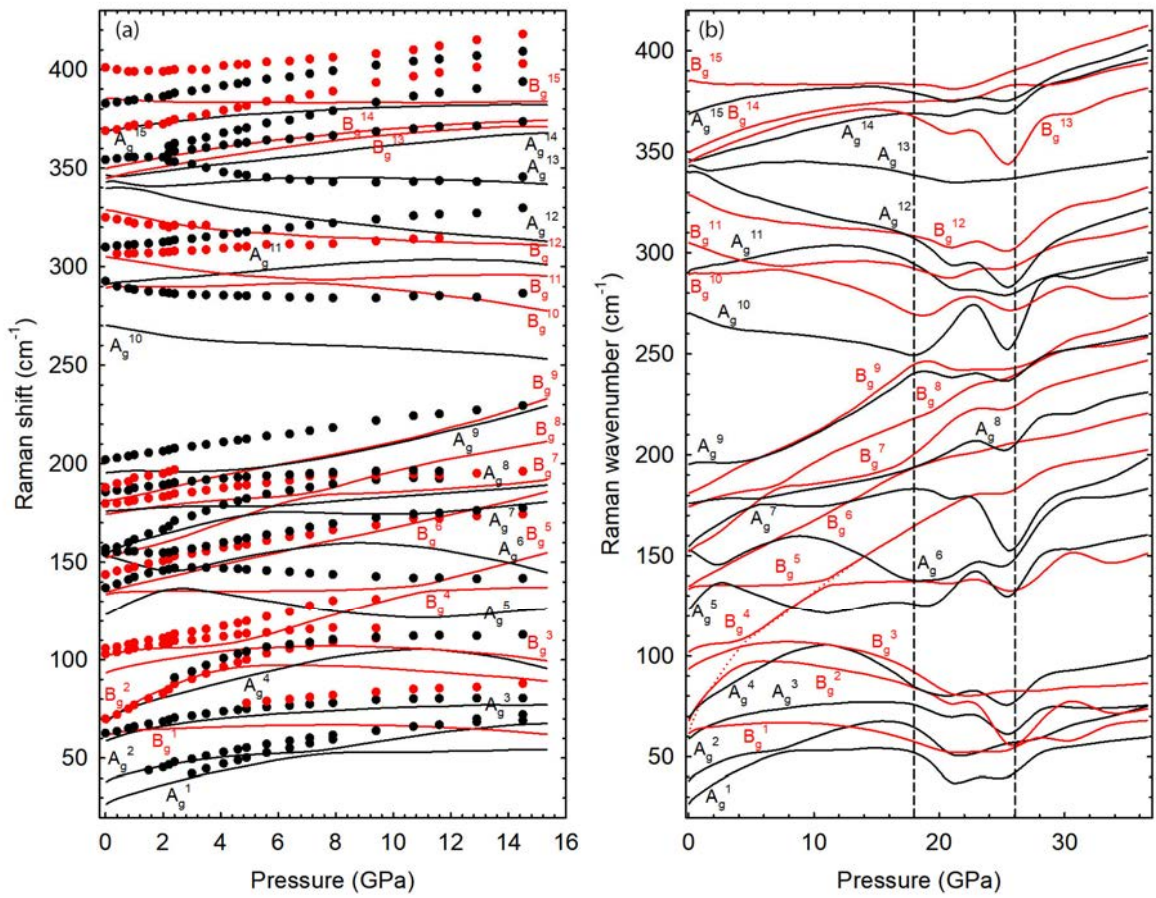

Figure 8

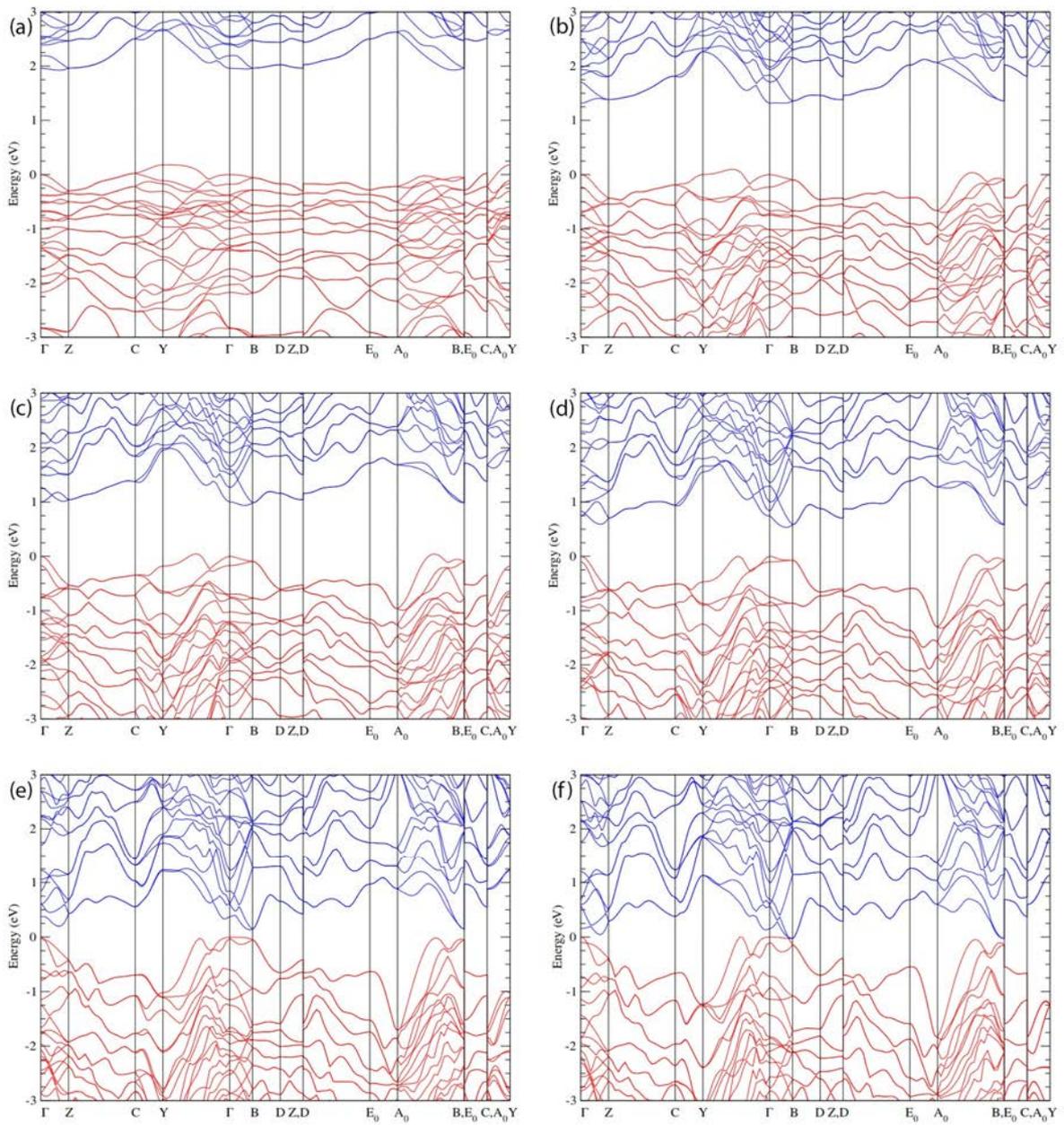

Figure 9

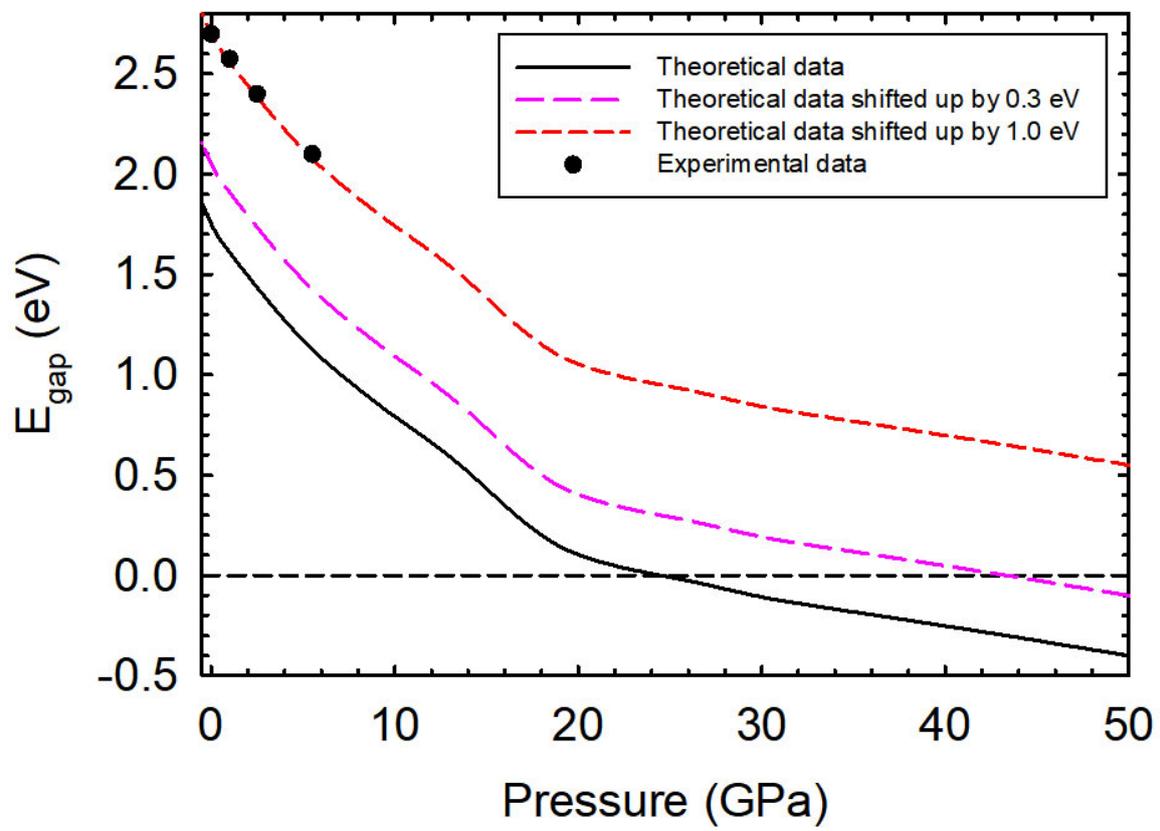

Figure 10

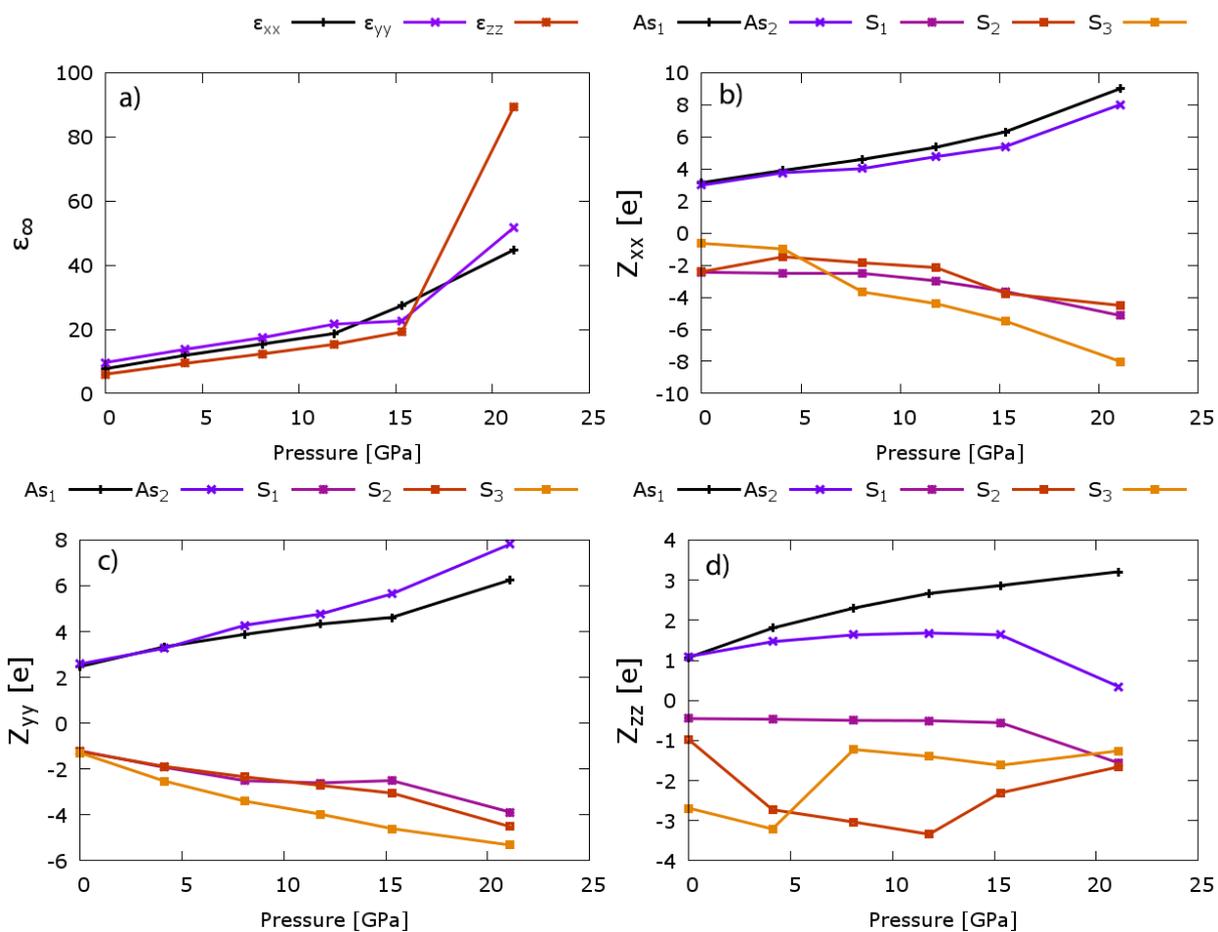

Figure 11

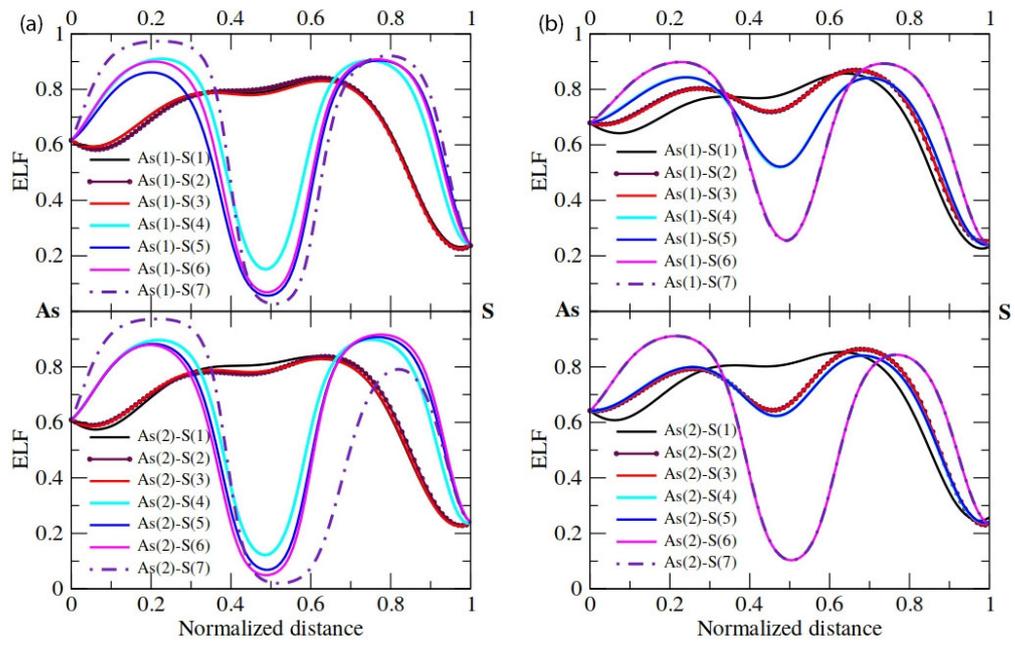

Figure 12

**TABLES**

**Table 1.** Experimental and theoretical (with vdW) lattice parameters corresponding to the $P2_1/c$ phase of α-$As_2S_3$ at ambient conditions. Experimental and theoretical values from Refs. **10, 11, 33** and **72** are given for comparison. All data have been expressed in the standard $P2_1/c$ setting for comparison.

| Lattice parameters | Experimental (This work) | Theoretical (This work) | Experimental (Ref. 11) | Experimental (Ref. 10) | Theoretical (Ref. 33) | Theoretical (Ref. 72) |
|---|---|---|---|---|---|---|
| $a$ (Å) | 4.2626(5) | 4.2608 | 4.256(2) | 4.22(5) | 4.22 | 4.22 |
| $b$ (Å) | 9.6056(7) | 9.6289 | 9.577(5) | 9.57(2) | 9.65 | 9.57 |
| $c$ (Å) | 12.1836(12) | 12.3084 | 12.191 (5) | 12.18(4) | 12.27 | 12.1839 |
| $\beta$ (º) | 110.054(7) | 109.84 | 109.75(8) | 109.8(5) | 109.59 | 109.85 |
| $V_0$ (Å$^3$) | 468.61(5) | 475.00 | 467.68 | 462.8 | 471.2 | 462.80 |

**Table 2.** Theoretical (with vdW interactions) and experimental Raman-active mode frequencies and their respective pressure coefficients for α-As$_2$S$_3$ at room temperature, as fitted with equation $\omega(P) = \omega_0 + \alpha \cdot P$. Experimental values from Refs. **23, 25 and 33** have been added for comparison.

| Mode | Theoretical | | Experimental | | | |
|---|---|---|---|---|---|---|
| | $\omega_0 \ (cm^{-1})$[a] | $\alpha \left(\frac{cm^{-1}}{GPa}\right)$[a] | $\omega_0 \ (cm^{-1})$[a] | $\alpha \left(\frac{cm^{-1}}{GPa}\right)$[a] | $\omega_0 \ (cm^{-1})$ | $\alpha \left(\frac{cm^{-1}}{GPa}\right)$ |
| $A_g^1$ | 27 (1) | 5.1 (3) | 26 (1) | 6.32 (37) | 25[b], 26[c] | 9.0[b] |
| $A_g^2$ | 38 (1) | 4.5 (2) | 37 (1) | 4.9 (3) | 36[b], 37[c] | 7.2[b] |
| $A_g^3$ | 59 (1) | 4.1 (2) | 62 (1) | 3.7 (1) | | |
| $B_g^1$ | 62 (1) | 1.6 (2) | | | 62[b,c] | 5.0[b] |
| $B_g^2$ | 68 (1) | 10.3 (4) | 69 (3) | 10.5 (8) | 69[b], 70[c] | 11.0[b] |
| $A_g^4$ | 70 (1) | 5.9 (3) | 70 (3) | 8.1 (3) | | |
| $B_g^3$ | 94 (2) | 3.1 (1) | 103 (2) | 2.9 (1) | | |
| $B_g^4$ | 103 (2) | 2.1 (5) | 106 (2) | 2.8 (1) | 107[b], 106[c] | 4.9[b] |
| $A_g^5$ | 124 (2) | 7.7 (7) | 138 (5) | 5.1 (4) | 136[b], 137[c], 135[d] | 7.3[b], -0.42[d] |
| $B_g^5$ | 134 (2) | 0.7 (2) | 120 (1) | 2.9 (5) | | |
| $B_g^6$ | 135 (2) | 3.6 (1) | 144 (1) | 3.9 (2) | 145[c] | |
| $B_g^7$ | 152 (3) | 3.7 (2) | | | | |
| $A_g^6$ | 153 (3) | -4.6 (7) | 155 (2) | -0.3 (3) | 158[c] | |
| $A_g^7$ | 155 (3) | 5.5 (3) | 156 (2) | 6.9 (4) | 154[b,c], 153[d] | 8.5[b], 0.24[d] |
| $B_g^8$ | 174 (3) | 2.5 (2) | 179 (2) | 2.2 (1) | 180[c], 177[d] | -2.22[d] |
| $A_g^8$ | 176 (3) | 1.2 (1) | 185 (2) | 2.3 (1) | | |
| $B_g^9$ | 181 (3) | 3.3 (2) | 188 (2) | 4.6 (9) | 188[c] | |
| $A_g^9$ | 196 (3) | 1.3 (1) | 202 (1) | 2.5 (2) | 202[b], 204[c], 201[d] | 3.2[b] |
| $A_g^{10}$ | 270 (4) | -3.1 (2) | 291 (1) | -2.6 (3) | 292 [b], 293 [c], 290 [d] | -3.8 [b], -0.89 [d] |
| $B_g^{10}$ | 290 (4) | 0.3 (1) | 306 (2) | 0.5 (1) | 307 [c], 308 [d] | -0.2 [d] |
| $A_g^{11}$ | 291 (4) | 1.3 (1) | 310 (2) | 1.4 (1) | 310 [b], 312 [c] | 1.0 [b] |
| $B_g^{11}$ | 305 (4) | -2.7 (1) | 325 (2) | -3.5 (4) | 326 [b,c] | -4.6 [b] |
| $B_g^{12}$ | 329 (5) | -3.7 (1) | | | 343 [c] | |
| $A_g^{12}$ | 341 (5) | 1.0 (5) | 356 (4) | -0.7 (6)[*] | 357 [b], 360 [c] | 0.96 [b] |
| $A_g^{13}$ | 343 (5) | 3.4 (3) | 354 (4) | 1.8 (4)[*] | 356 [c] | |
| $B_g^{13}$ | 345 (5) | 3.1 (1) | | | | |
| $A_g^{14}$ | 346 (5) | -0.5 (1) | 353 (4) | 4.1 (6)[*] | 354 [c], 353 [d] | -0.19 [d] |
| $B_g^{14}$ | 350 (5) | 3.1 (4) | 369 (2) | 2.2 (3) | 370 [c] | |
| $A_g^{15}$ | 369 (5) | 2.4 (1) | 383 (2) | 2.2 (2) | 383 [b], 384 [c], 380 [d] | 1.7 [b], 0.47 [d] |
| $B_g^{15}$ | 385 (5) | -0.9 (2) | 400 (2) | -0.7 (3) | 401 [c] | |

[a] This work. [b] Ref. 23. [c] Ref. 25. [d] Ref. 33. * The pressure coefficient of these modes has been measured above 2 GPa.

## ASSOCIATED CONTENT

**Supporting Information**. Supporting Information provides an analysis of the structural and vibrational properties of $As_2S_3$ at ambient and high pressure. Details of the structural evolution of compressed $As_2S_3$ show the evolution of the axial ratios, the theoretical Wyckoff free coordinates and the polyhedral distortion under pressure. On the other hand, an extensive description of the experimental and theoretical isothermal compressibility tensor is given. Details of atomic vibrations of many zone-center phonons are provided and the pressure dependence of the theoretical Raman- and IR-active modes are given. Finally, a table with the known group-15 sesquichalcogenides and their particular cation coordination and type of bonding is provided. This material is available free of charge via the Internet at http://pubs.acs.org.

## AUTHOR INFORMATION

### Corresponding Author

*(J.A. Sans) E-mail: juasant2@upv.es and (F.J. Manjón) e-mail: fjmanjon@upv.es. Telephone: (+34) 963877000 ext.(75287)

### Author Contributions

The manuscript was written through contributions of all authors. All authors have given approval to the final version of the manuscript.

## ACKNOWLEDGMENT

Authors thank the financial support from Spanish Ministerio de Economia y Competitividad (MINECO) through MAT2016-75586-C4-2/3-P and FIS2017-83295-P and from Generalitat Valenciana under project PROMETEO/2018/123-EFIMAT. ELDS acknowledges the European

Union Horizon 2020 research and innovation programme under Marie Sklodowska-Curie for grant agreement No. 785789-COMEX. JAS also acknowledges Ramón y Cajal program for funding support through RYC-2015-17482. AM, SR and ELDS thank interesting discussions with J. Contreras-García who taught them how to analyze the ELF. Finally, authors thank ALBA Light Source for beam allocation at beamline MSPD (Experiment No. 2013110699) and acknowledge computing time provided by MALTA-Cluster and Red Española de Supercomputación (RES) through computer resources at MareNostrum with technical support provided by the Barcelona Supercomputing Center (QCM-2018-3-0032).

**REFERENCES**

(1) Smith, J. D.; Bailar, J. C.; Emeléus H. J.; Nyholm, R. The Chemistry of Arsenic, Antimony and Bismuth, Pergamon Texts in Inorganic Chemistry (1973).

(2) Pliny the Elder, Naturalis Historia, Chapter 22, ed. by J. Bostock, M.D., F.R.S. H.T. Riley, Esq., B.A. London (Taylor and Francis, London, 1855).

(3) Fitzhugh, E. W. Orpiment and Realgar, in Artists' Pigments, A Handbook of Their History and Characteristics, Vol 3 (Oxford University Press, 1997), p. 47−80.

(4) Spurrell, F.C.J. Notes on Egyptian Colors, *Archaeological J.* **1895**, 52, 222−239.

(5) Burgio, L.; Clark, R. J. H. Comparative pigment analysis of six modern Egyptian papyri and an authentic one of the 13th century BC by Raman microscopy and other techniques, *J. Raman Spectrosc.* **2000**, 31, 395−401.

(6) Waxman, S.; Anderson, K.C. History of the development of arsenic derivatives in cancer therapy. *Oncologist* **2001**, 6 (Suppl 2), 3−10.

(7) Ding, W.; Tong, Y.; Zhang, X.; Pan, M.; Chen, S. Study of Arsenic Sulfide in Solid Tumor Cells Reveals Regulation of Nuclear Factors of Activated T-cells by PML and p53, *Sci. Rep.* **2016**, 6, 19793.

(8) Heo, J.; Chung, W. J. Rare-earth-doped chalcogenide glass for lasers and amplifiers, in *Chalcogenide Glasses: Preparation, Properties and Applications* (Woodhead Publishing, 2014), pp. 347−380.

(9) Hewak, D. W.; Zheludev, N. I.; MacDonald, K. F. Controlling light on the nanoscale with chalcogenide thin films, in *Chalcogenide Glasses: Preparation, Properties and Applications* (Woodhead Publishing, 2014), pp. 471−508.

(10) Morimoto, N. The crystal structure of orpiment ($As_2S_3$) refined. *Miner. J.* **1954**, 1, 160−169.

(11) Mullen, D. J. E.; Nowacki, W. Refinement of the crystal structures of realgar, AsS and orpiment, $As_2S_3$. *Z. Kristall.* **1972**, 136, 48−65.

(12) Kampf, A. R.; Downs, R. T.; Housley, R. M.; Jenkins, R. A.; Hyršl, J. Anorpiment, $As_2S_3$, the triclinic dimorph of orpiment. *Mineral Mag.* **2011**, 75, 2857−2867.

(13) Gibbs, G. V.; Wallace, A. F.; Zallen, R.; Downs, R. T.; Ross, N. L.; Cox, D. F.; Rosso, K. M. Bond Paths and van der Waals Interactions in Orpiment, $As_2S_3$. *J. Phys. Chem. A* **2010**, 114, 6550−6557.

(14) Cheng, H.F.; Zhou, Y.; Frost, R. L. Structure comparison of Orpiment and Realgar by Raman spectroscopy, *Spectros. Lett.* **2017**, 50, 23−29.

(15)  Porto, S. P. S.; Wood, D. L. Ruby Optical Maser as a Raman source. *J. Opt. Soc. Am.* **1962**, 52, 251−252.

(16)  Weber, A.; Porto, S. P. S. He-Ne Laser as a Light Source for High Resolution Raman Spectroscopy. *J. Opt. Soc. Am.* **1965**, 55, 1033−1034.

(17)  Ward, A. T. Raman spectroscopy of sulfur, sulfur-selenium, and sulfur-arsenic mixtures. *J. Phys. Chem.* **1968**, 72, 4133−4139.

(18)  Forneris, R. The infrared and Raman spectra of realgar and orpiment. *Am. Miner.* **1969**, 54, 1062−1074.

(19)  Scheuermann, W.; Ritter, G. J. Raman spectra of cinnabar (HgS), realgar ($As_4S_4$) and orpiment ($As_2S_3$). *Z. Naturforsch.* **1969**, A 24, 408−411.

(20)  Mathieu, J. M.; Poulet, H. Spectres de vibration de l'orpiment $As_2S_3$. *Bull. Soc. Fr. Mineral. Cristall.* **1970**, 93, 532−535.

(21)  Zallen, R.; Slade, M.; Ward, A. T. Lattice vibrations and Interlayer Interactions in Crystalline $As_2S_3$ and $As_2Se_3$. *Phys. Rev. B* **1971**, 3, 4257−4273.

(22)  Zallen R.; Slade, M. Rigid-layer modes in chalcogenide crystals. *Phys. Rev. B* **1974**, 9, 1627−1637.

(23)  Zallen, R. Pressure-Raman effects and vibrational scaling laws in molecular crystals: $S_8$ and $As_2S_3$. *Phys. Rev. B* **1974**, 9, 4485−4496.

(24)  DeFonzo, A.P.; Tauc, J. Network dynamics of 3:2 coordinated compounds, *Phys. Rev. B* **1978**, 18, 6957−6972.

(25)  Razzetti, C.; Lottici, P. P. Polarization analysis of the Raman spectrum of $As_2S_3$ crystals, *Solid State Commun.* **1979**, 29, 361−364.

(26)  Besson, J. M.; Cernogora, J.; Zallen, R. Effect of pressure on optical properties of crystalline $As_2S_3$. *Phys. Rev. B* **1980**, 22, 3866−3876.

(27)  Besson, J. M.; Cernogora, J.; Slade, M. L.; Weinstein, B. A.; Zallen, R. Pressure effects on the absorption edge, refractive index, and Raman spectra of crystalline and amorphous $As_2S_3$. *Physica B* **1981**, 105, 319−323.

(28)  Frost, R. L.; Martens, W. N.; Kloprogge, J. T. Raman spectroscopic study of cinnabar (HgS), realgar ($As_4S_4$), and orpiment ($As_2S_3$) at 298 and 77K. *N. Jb. Miner. Mh. Jg.* **2002**, 10, 469−480.

(29)  Mamedov, S.; Drichko, N. Characterization of 2D $As_2S_3$ crystal by Raman spectroscopy. *MRS Advances* **2018**, 3, 385−390.

(30) Itié, J. P.; Polian, A.; Grimsditch, M.; Susman, S. X-ray absorption spectroscopy Investigation of Amorphous and Crystalline As$_2$S$_3$. *Jpn. J. Appl. Phys.* **1992**, 32, Suppl. 2, 719−721.

(31) Zallen, R. Effect of Pressure on Optical Properties of Crystalline As$_2$S$_3$. *High Press. Res.* **2004**, 24, 117−118.

(32) Bolotina, N. B.; Brazhkin, V. V.; Dyuzheva, T. I.; Katayama, Y.; Kulikova, L. F.; Lityagina, L. V.; Nikolaev, N. A. High-Pressure Polymorphism of As$_2$S$_3$ and New AsS$_2$ Modification with Layered Structure. *JETP Lett.* **2013**, 98, 539−543.

(33) Liu, K. X.; Dai, L. D.; Li, H. P.; Hu, H. Y.; Yang, L. F.; Pu, C.; Hong, M. L.; Liu, P.F. Phase Transition and Metallization of Orpiment by Raman Spectroscopy, Electrical Conductivity and Theoretical Calculation under High Pressure. *Materials* **2019**, 12, 784.

(34) Kravchenko, E. A.; Timofeeva, A. V.; Virogradova, G. Z. Crystals Modifications of Arsenic and Antimony Sulphides at High Pressure and Temperature. *J. Mol. Str.* **1980**, 58, 253−262.

(35) Radescu, S.; Mujica, A.; Rodríguez-Hernández, P.; Muñoz, A.; Ibáñez, J.; Sans, J. A.; Cuenca-Gotor, V. P.; Manjón, F. J. Study of the orpiment and anorpiment phases of As$_2$S$_3$ under pressure. *J. Phys.: Conf. Series* **2017**, 950, 042018.

(36) Shportko, K.; Kremers, S.; Woda, M.; Lencer, D.; Robertson, J.; Wuttig, M. Resonant bonding in crystalline phase-change materials. *Nat. Mat.* **2008**, 7, 653−658.

(37) Lee, S.; Esfarjani, K.; Luo, T. F.; Zhou, J. W.; Tian, Z. T.; Chen, G. Resonant bonding leads to low lattice thermal conductivity. *Nat. Comm.* **2014**, 5, 3525.

(38) Li, C. W.; Hong, J.; May, A. F.; Bansal, D.; Chi, S.; Hong, T.; Ehlers, G.; Delaire, O. Orbitally driven giant phonon anharmonicity in SnSe. *Nat. Phys.* 2015, 11, 1063–1069.

(39) Xu, M.; Jakobs, S.; Mazzarello, R.; Cho, J.-Y.; Yang, Z.; Hollermann, H.; Shang, D. S.; Miao, X. S.; Yu, Z. H.; Wang, L.; Wuttig, M. Impact of Pressure on the Resonant Bonding in Chalcogenides. *J. Phys. Chem. C* **2017**, 121, 25447−25454.

(40) Wuttig, M.; Deringer, W. L.; Gonze, X.; Bichara, C.; Raty, J.-Y. Incipient Metals: Functional Materials with a Unique Bonding Mechanism. *Adv. Mat.* **2018**, 30, 1803777.

(41) Raty, J.-Y.; Schumacher, M.; Golub, P.; Deringer, V. L.; Gatti, C.; Wuttig, M. A Quantum-Mechanical Map for Bonding and Properties in Solids. *Adv. Mater.* **2018**, 31, 1806280.

(42) Fauth, F.; Peral, I.; Popescu, C.; Knapp, M. The New Material Science Powder Diffraction Beamline at ALBA Synchrotron. *Powder Diffr.* **2013**, 28, S360−S370.

(43) Hammersley, A. P.; Svensson, S. O.; Hanfland, M.; Fitch, A. N.; Hausermann, D. Two-Dimensional Detector Software: From Real¨Detector to Idealised Image or Two-Theta Scan. *High Pressure Res.* **1996**, 14, 235−248.

(44) Larson, A. C.; von Dreele, R.B. General Structure Analysis System (GSAS). *LANL Report* **2004**, 86, 748.

(45) Toby, B. H. EXPGUI, A Graphical User Interface for GSAS. *J. Appl. Crystallogr.* **2001**, 34, 210−213.

(46) Momma, K.; Izumi, F. VESTA 3 for three-dimensional visualization of crystal, volumetric and morphology data. *J. Appl. Crystallogr.* **2011**, 44, 1272−1276.

(47) Dewaele, A.; Loubeyre, P.; Mezouar, M. Equations of State of Six Metals above 94 GPa. *Phys. Rev. B* **2004**, 70, 094112.

(48) Mao, M. K.; Xu, J.; Bell, P. M. Calibration of the Ruby Pressure Gauge to 800 kbar under Quasi-Hydrostatic Conditions. *J. Geophys. Res.* **1986**, 91, 4673−4676.

(49) Piermarini, G.J.; Block, S.; Barnett, J.D. Hydrostatic limits in liquids and solids to 100 kbar. *J. Appl. Phys.* **1973**, 44, 5377−5382.

(50) Klotz, S.; Chervin, J.-C.; Munsch, P.; Le Marchand, G. Hydrostatic limits of 11 pressure transmitting media. *J. Phys. D: Appl. Phys.* **2009**, 42, 075413.

(51) Hohenberg, P.; Kohn, W. Inhomogeneous Electron Gas. *Phys. Rev.* **1964**, 136, B864−B871.

(52) Kresse, G.; Hafner, J. Ab Initio Molecular Dynamics for Liquid Metals. *Phys. Rev. B* **1993**, 47, 558−561.

(53) Kresse, G.; Furthmüller, J. Efficiency of ab-initio total energy calculations for metals and semiconductors using a plane-wave basis set. *Comput. Mat. Sci.* **1996**, 6, 15-50.

(54) Blöchl, P. E. Projector Augmented-Wave Method. *Phys. Rev. B* **1994**, 50, 17953−17979.

(55) Perdew, J. P.; Burke, K.; Ernzerhof, M. Generalized Gradient Approximation Made Simple. *Phys. Rev. Lett.* **1996**, 77, 3865−3868.

(56) Perdew, J. P.; Ruzsinszky, A.; Csonka, G. I.; Vydrov, O. A.; Scuseria, G. E.; Constantin, L. A.; Zhou, X.; Burke, K. Restoring the Density-Gradient Expansion for Exchange in Solids and Surfaces. *Phys. Rev. Lett.* **2008**, 100 136406.

(57) Monkhorst, H. J.; Pack, J. D. Special points for Brillouin-zone integrations. *Phys. Rev. B* **1976**, 13, 5188.

(58) Grimme, S. Semiempirical GGA-type density functional constructed with a long-range dispersion correction. *J. Comp. Chem.* **2006**, 27, 1787-1799.

(59) Contreras-García, J.; Martin Pendás, A.; Silvi, B.; Recio, J.M. Useful applications of the electron localization function in high-pressure crystal chemistry. *J. Phys. Chem. Solids* **2008**, 69, 2204−2207.

(60) Contreras-García, J.; Silvi, B.; Martín Pendás, A.; Recio, J.M. Computation of local and global properties of the electron localization function topology in crystals. *J. Chem. Theory Comput.* **2009**, 5, 164–173.

(61) Parlinski, K.; Li, Z. Q.; Kawazoe, Y. First-Principles Determination of the Soft Mode in Cubic $ZrO_2$. *Phys. Rev. Lett.* **1997**, 78, 4063−4066.

(62) Alfè, D. PHON: A program to calculate phonons using the small displacement method. *Comp. Phys. Commun.* **2009**, 180, 2622−2633.

(63) Svensson, C. The crystal structure of orthorhombic antimony trioxide, $Sb_2O_3$. *Acta. Cryst. B* **1974**, 30, 458−461.

(64) Stergiou A. C.; Rentzeperis, P. J. The crystal structure of arsenic selenide, $As_2Se_3$. *Z. Kristall.* **1985**, 173, 185−191.

(65) Pertlik, F. Verfeinerung der Kristallstruktur des Minerals Claudetit, $As_2O_3$ ("Claudetit I"). *Monatshefte für Chemie* **1978**, 109, 277–282.

(66) Brown, A.; Rundqvist, S. Refinement of the crystal structure of black phosphorus. *Acta Cryst.* **1965**, 19, 684–685.

(67) Efthimiopoulos, I.; Zhang, J. M.; Kucway, M.; Park, C. Y.; Ewing, R. C.; Wang, Y. $Sb_2Se_3$ Under Pressure. *Sci. Rep.* **2013**, 3, 2665−2672.

(68) Efthimiopoulos, I.; Kemichick, J.; Zhou, X.; Khare, S. V.; Ikuta, D.; Wang, Y. High-Pressure Studies of $Bi_2S_3$. *J. Phys. Chem. A* **2014**, 118, 1713−1720.

(69) Ibañez, J.; Sans, J. A.; Popescu, C.; López-Vidrier, J.; Elvira-Betanzos, J. J.; Cuenca-Gotor, V. P.; Gomis, O.; Manjón, F. J.; Rodríguez-Hernández, P.; Muñoz, A. Structural, Vibrational, and Electronic Study of $Sb_2S_3$ at High Pressure. *J. Phys. Chem. C* **2016**, 120, 10547−10558.

(70) Cuenca-Gotor, V. P.; Sans, J. A.; Ibáñez, J.; Popescu, C.; Gomis, O.; Vilaplana, R.; Manjón, F. J.; Leonardo, A.; Sagasta, E.; Suárez-Alcubilla, A.; Gurtubay, I. G.; Mollar, M.; Bergara, A. Structural, Vibrational, and Electronic Study of α-$As_2Te_3$ under Compression. *J. Phys. Chem. C* **2016**, 120, 19340−19352.

(71) Walsh, A.; Payne, D. J.; Egdell, R. G.; Watson, G.W. Stereochemistry of post-transition metal oxides: revision of the classical lone pair model. *Chem. Soc. Rev.* **2011**, 40, 4455−4463.

(72) Srivastava, P.; Singh Mund, H.; Sharma, Y. Investigation of electronic properties of crystalline arsenic chalcogenides: Theory and experiment. *Physica B* **2011**, 406, 3083−3088.

(73) Kroumova, E.; Aroyo, M.I.; Perez-Mato, J.M.; Kirov, A.; Capillas, C.; Ivantchev, S.; Wondratschek, H. Bilbao Crystallographic Server: Useful Databases and Tools for Phase-Transition Studies. *Phase Transitions* **2003**, 76, 155−170.

(74) Canepa, P.; Hanson, R. M.; Ugliengo, P.; Alfredsson, M. J-ICE: a new Jmol interface for handling and visualizing crystallographic and electronic properties. *J. Appl. Cryst.* **2011**, 44, 225−229.

(75) Siebert, H. Struktur der Sauerstoffsäuren, Z. anorg. Allg. Chemie **1954**, 275, 225−240.

(76) Birch, F. The Effect of Pressure Upon the Elastic Parameters of Isotropic Solids, According to Murnaghan's Theory of Finite Strain. *J. Appl. Phys.* **1938**, 9, 279−288.

(77) Guńka, P. A.; Dranka, M.; Hanfland, M.; Dziubek, K. F.; Katrusiak, A.; Zachara, J. Cascade of High-Pressure Transitions of Claudetite II and the First Polar Phase of Arsenic(III) Oxide. *Cryst. Growth Des.* **2015**, 15, 3950−3954.

(78) Haussühl, S. Physical Properties of Crystals. An Introduction (Wiley-VCH, 2007).

(79) Angel, R. J. http://www.rossangel.com/text_strain.htm

(80) Minomura S.; Aoki K.; Koshizuka, N.; Tsushima T. The Effect of Pressure on The Raman Spectra in Trigonal Se and Te. in High-Pressure Science and Technology (Springer, 1979), p. 435.

(81) Bandyophadhay, A. K.; Singh, D. B. Pressure induced phase transformations and band structure of different high pressure phases in tellurium. *Pramana* **1999**, 52, 303−319.

(82) Sorb, Y.A.; Rajaji, V.; Malavi, P. S.; Subbarao, U.; Halappa, P.; Peter, S. C.; Karmakar, S.; and Narayana, C. Pressure-induced electronic topological transition in $Sb_2S_3$. *J. Phys.: Condens. Mat.* **2016**, 28, 015602.

(83) Efthimiopoulos, I.; Buchan, C.; Wang, Y. Structural properties of $Sb_2S_3$ under pressure: evidence of an electronic topological transition. *Sci. Rep.* **2016**, 6, 24246.

(84) Manjón, F.J.; Vilaplana, R.; Gomis, O.; Perez-Gonzalez, E.; Santamaria-Perez, D.; Marin-Borras, V.; Segura, A.; Gonzalez, J.; Rodriguez-Hernandez, P.; Muñoz, A. High-pressure studies of topological insulators $Bi_2Se_3$, $Bi_2Te_3$, and $Sb_2Te_3$. *Phys.Status Solidi B* **2013**, 250, 669-676.

(85) Sans, J.A.; Manjón, F.J.; Pereira, A.L.J.; Vilaplana, R.; Gomis, O.; Segura, A.; Muñoz, A.; Rodríguez-Hernández, P.; Popescu, C.; Drasar, C.; Ruleova, P. Structural, vibrational, and electrical study of compressed BiTeBr. *Phys. Rev. B* **2016**, 93, 024110.

(86) Pereira, A. L. J.; Santamaría-Pérez, D.; Ruiz-Fuertes, J.; Manjón, F. J.; Cuenca-Gotor, V. P.; Vilaplana, R.; Gomis, O.; Popescu, C.; Muñoz, A.; Rodríguez-Hernández, P.; Segura, A.; Gracia, L.; Beltrán, A.; Ruleova, P.; Drasar, C.; Sans, J. A. Experimental and Theoretical Study of Bi$_2$O$_2$Se Under Compression. *J. Phys. Chem. C* **2018**, 122, 8853−8867.

(87) Degtyareva, O.; Hernández, E. R.; Serrano, J.; Somayazulu, M.; Mao, H.-K.; Gregoryanz, E.; Hemley, R. J. Vibrational dynamics and stability of the high-pressure chain and ring phses in S and Se. *J. Chem. Phys.* **2007**, 126, 084503.

(88) Richter, W.; Renucci, J. B.; Cardona, M. Hydrostatic Pressure Dependence of First-Order Raman Frequencies in Se and Te. *Phys. Status Solidi B* **1973**, 56, 223−229.

(89) Aoki, K.; Shimomura, O.; Minomura, S.; Koshizuka, N.; Tsushima, T. Raman Scattering of Trigonal Se and Te at Very High Pressure. *J. Phys. Soc. Jpn.* **1980**, 48, 906−911.

(90) Lucogsky, G. Comparison of the Long Wave Optical Phonons in Se and Te. *Phys. Stat. Sol. B* **1972**, 49, 633−641.

(91) Lifshitz, I. M. Anomalies of Electron Characteristics of a Metal in the High Pressure Region. *Sov. Phys. JETP* **1960**, 11, 1130.

(92) No value is obtained at higher pressure since the metallization of the monoclinic structure above 25 GPa in our calculations yields unreasonably high values of these parameters.

(93) Properzi, L.; Polian, A.; Munsch, P.; Di Cicco, A. Investigation of the phase diagram of selenium by means of Raman spectroscopy. *High Pressure Research* **2013**, 33, 35−39.

(94) Marini, C.; Chermisi, D.; Lavagnini, M.; Di Castro, D.; Petrillo, C.; Degiorgi, L.; Scandolo, S.; Postorino, P. High-pressure phases of crystalline tellurium: A combined Raman and ab initio study. *Physical Review B* **2012**, 86, 064103.

(95) Vilaplana, R.; Gomis, O.; Manjón, F. J.; Segura, A.; Pérez-González, E.; Rodríguez-Hernández, P.; Muñoz, A.; González, J.; Marín-Borrás, V.; Muñoz-Sanjosé, V.; Drasar, C.; Kucek, V. High-pressure vibrational and optical study of Bi$_2$Te$_3$. *Phys. Rev. B* **2011**, 84, 104112.

**Table of Contents Graphic and Synopsis**

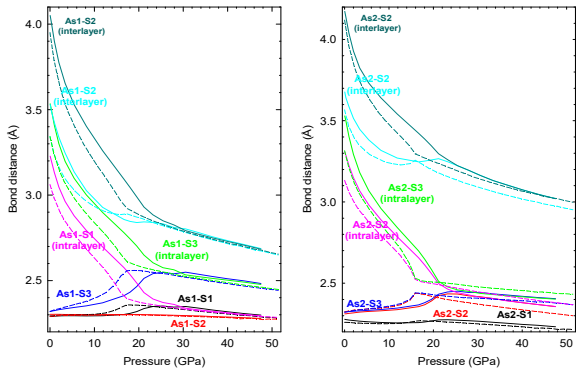

This manuscript reports a joint experimental and theoretical study of the structural, vibrational and electronic properties of orpiment ($\alpha$-$As_2S_3$) under compression. A strong change in the As-S distances and consequently in the coordination of As atoms is observed between room pressure and 25 GPa. This change can be ascribed to the existence of an isostructural phase transition above 20 GPa. The changes observed are consistent with the development of metavalent or resonant bonding in orpiment above 20 GPa.

# Supplementary Information of

# Orpiment under compression: metavalent bonding at high pressure

V.P. Cuenca-Gotor,[1] J.A. Sans,[1,*] O. Gomis,[2] A. Mujica,[3] S. Radescu,[3] A. Muñoz,[3] P. Rodríguez-Hernández,[3] E. Lora da Silva,[1] C. Popescu,[4] J. Ibañez,[5] R. Vilaplana,[2] and F.J. Manjón[1,*]

[1] Instituto de Diseño para la Fabricación y Producción Automatizada, Universitat Politècnica de València, 46022 Valencia (Spain)

[2] Centro de Tecnologías Físicas, Universitat Politècnica de València, 46022 Valencia (Spain)

[3] Departamento de Física, Instituto de Materiales y Nanotecnología, MALTA Consolider Team, Universidad de La Laguna, 38207 San Cristóbal de La Laguna (Spain)

[4] ALBA-CELLS, 08290 Cerdanyola, Barcelona (Spain)

[5] Institute of Earth Sciences Jaume Almera, CSIC, 08028 Barcelona (Spain)

**Structural properties of orpiment at room conditions**

**Table S1.** Theoretical (the, with vdW) and experimental (exp) atomic parameters of the $P2_1/c$ structure of α-$As_2S_3$ at room conditions.

| Atoms | Site | x | Y | z |
|---|---|---|---|---|
| As1 | 4e | 0.10723 (the)[a]<br>0.09805 (exp)[b] | 0.30752 (the)[a]<br>0.30829 (exp)[b] | 0.23420 (the)[a]<br>0.23531 (exp)[b] |
| As2 | 4e | 0.38070 (the)[a]<br>0.37395 (exp)[b] | 0.17785 (the)[a]<br>0.17878 (exp)[b] | 0.01176 (the)[a]<br>0.01323 (exp)[b] |
| S1 | 4e | 0.61732 (the)[a]<br>0.60660 (exp)[b] | 0.37901 (the)[a]<br>0.37872 (exp)[b] | 0.09901 (the)[a]<br>0.09849 (exp)[b] |
| S2 | 4e | 0.1719 (the)[a]<br>0.16272 (exp)[b] | 0.0999 (the)[a]<br>0.10277 (exp)[b] | 0.15133 (the)[a]<br>0.15262 (exp)[b] |
| S3 | 4e | 0.05703 (the)[a]<br>0.06338 (exp)[b] | 0.70532 (the)[a]<br>0.70646 (exp)[b] | 0.12304 (the)[a]<br>0.12234 (exp)[b] |

[a] This work. [b] Data from Ref. 1 converted with VESTA from the original $P2_1/n$ setting to the standard $P2_1/c$ setting with coordinate standardization.

**Vibrational properties of orpiment at room conditions**

Group theory predicts sixty zone-center vibrational modes at the BZ center for $\alpha$-As$_2$S$_3$ with mechanical representation **[2]**:

$$\Gamma_{60} = 15 \ (A_g(R) + B_u \ (IR)) + 15 \ (B_g \ (R) + A_u \ (IR)) \tag{1}$$

where g (gerade) modes are Raman-active (R) and u (ungerade) modes are infrared-active (IR). Note that all modes are paired up, $(A_g, B_u)$ and $(B_g, A_u)$, where the first is R and the second is IR. Therefore, $\alpha$-As$_2$S$_3$ has 30 Raman-active modes and 27 IR-active modes because one $A_u$ and two $B_u$ modes are acoustic modes. $A_g$ and $B_u$ modes correspond mainly to vibrations of atoms in the *ac*-plane, while $B_g$ and $A_u$ modes correspond mainly to atomic vibrations along the *b*-axis.

It is well-known that in layered materials with layers piled along the *c*-axis, which usually crystallize either in a hexagonal or tetragonal space groups, the lowest-frequency E (doubly degenerated) and A (or B) modes at the $\Gamma$-point can be classified as inter-layer modes (out-of-phase vibrations of atoms corresponding to adjacent layers), also known as rigid layer modes, or intra-layer modes (out-of-phase vibrations of atoms inside the layers) **[3]**. Inter-layer E and A (or B) modes are usually related to shear or transversal vibrations between adjacent layers along the *a*- or *b*-direction of the *ab*-layer plane and to longitudinal vibrations of one layer against the adjacent ones (along the *c*-axis), respectively.

Both E and A (or B) inter-layer modes arise from the transverse acoustic (TA) and longitudinal acoustic (LA) modes, respectively, due to the band folding of the Brillouin zone boundaries into the $\Gamma$-point due to the decrease of symmetry from cubic to hexagonal or tetragonal. Similarly, E and A (or B) intra-layer modes originate from the transverse optical (TO) and longitudinal optical (LO) modes at $\Gamma$-point and from additional modes due to the folding of the BZ boundaries into the $\Gamma$-point **[3]**.

The number of inter-layer and intra-layer modes in layered materials depends on the complexity of the unit cell. In the simplest case, there should be two inter-layer modes and four intra-layer modes. For $\alpha$-As$_2$S$_3$, which crystallizes on the monoclinic phase, there are only modes with A and B symmetry, therefore there are three inter-layer rigid modes. Since the acoustic modes are one $A_u$ and two $B_u$ modes, corresponding to inter-layer rigid modes where both layers of the unit cell have in-phase translations, the out-of-phase translations of the two layers corresponding to the three optical inter-layer rigid modes must be Raman-active and must be two $A_g$ and one $B_g$ modes. We have visualized the atomic motions of the normal vibrations of $\alpha$-As$_2$S$_3$ with the use of J-ICE program **[4]**. Two of these are the two shear rigid modes with lowest-frequency (27 and 389 cm$^{-1}$) represented in **Figure S1** and that are attributed to the $A_g^1$ and $A_g^2$ modes as indicated in previous papers (see **Ref. 1** and references therein). On the other hand, the compressional or longitudinal rigid mode is the $B_g^1$ mode located around 62 cm$^{-1}$ (see **Figure S2**). As commented in previous papers, this compressional rigid mode has a similar frequency to other optical modes; i.e. there is a mix between this inter-layer mode and several inter-chain modes (see **Figure S3**) as previously suggested **[2]**.

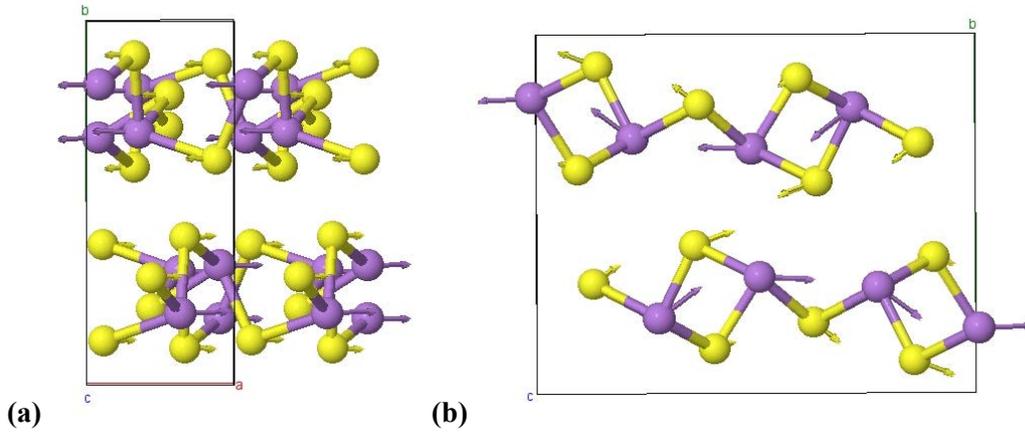

**(a)**          **(b)**

**Figure S1.** Atomic vibrations of the $A_g^1$ **(a)** and $A_g^2$ **(b)** inter-layer modes located around 27 and 38 cm$^{-1}$ in the *ab-* (*bc*)-plane. These are the shear rigid layer modes in $\alpha$-As$_2$S$_3$. As: Purple spheres, S: yellow spheres.

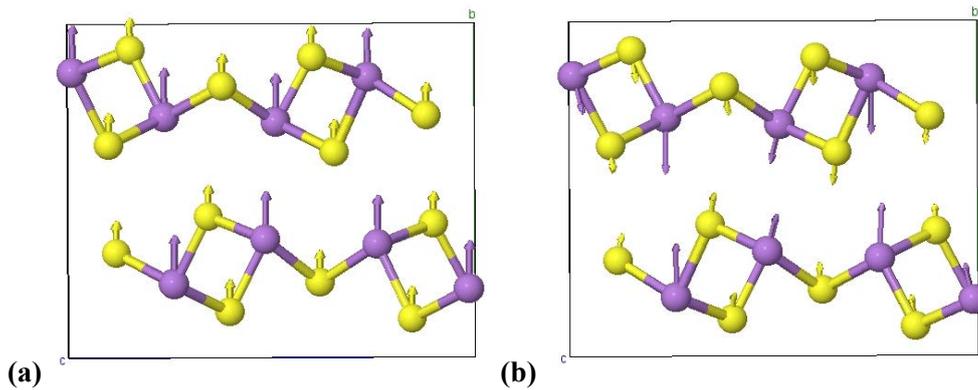

**(a)**          **(b)**

**Figure S2.** Atomic vibrations of the acoustic $A_u$ inter-layer mode **(a)** and the related $B_g^1$ inter-layer mode located around 62 cm$^{-1}$ **(b)** in the *bc*-plane. The $B_g^1$ mode is the longitudinal or compressional rigid layer mode in $\alpha$-As$_2$S$_3$. As: Purple spheres, S: yellow spheres.

Intra-layer modes can be divided into high-frequency intra-chain modes and low-frequency inter-chain modes **[2]**. The mixture of inter-layer and inter-chain modes found is in agreement with the analysis of Defonzo and Tauc **[2]**. In fact, low-frequency inter-chain modes mixed with the compressional rigid mode are those related to the rotation of the AsS spiral chains around the *a*-axis (see **Figures S3** and **S4**), and there are also modes related to the translation of the spiral chains inside each layer (see **Figures S5** to **S8**).

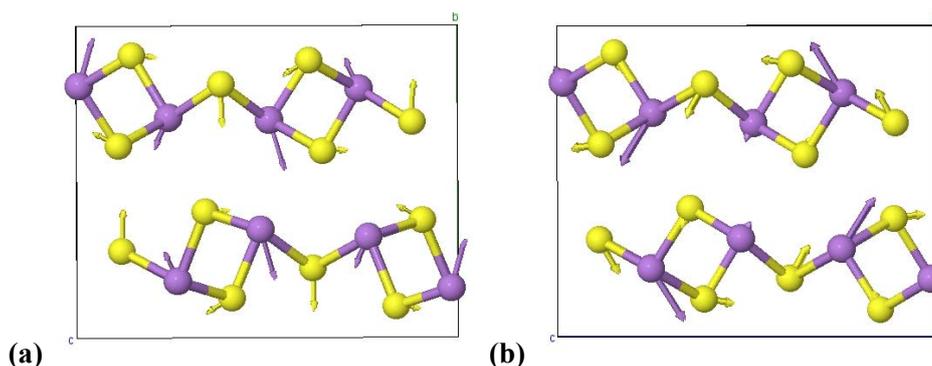

**(a)**                                    **(b)**

**Figure S3.** Atomic vibrations of the B$_u^1$ **(a)** and A$_g^3$ **(b)** modes located around 50 and 59 cm$^{-1}$, respectively, in the *bc*-plane. These are inter-chain modes of α-As$_2$S$_3$ corresponding to rotational modes of the spiral AsS chains around the *a*-axis (perpendicular to the page). Atoms of one chain rotate around the *a*-axis in the opposite sense (clockwise vs. counter-clockwise) than the neighbour chains inside the same layer. Rotations of the neighbour chains in the adjacent layers are in the same direction in the B$_u$ mode and in the opposite direction in the A$_g$ mode. These are two of the "inter-chain rolling" modes defined in Ref. **[2]**. As: Purple spheres, S: yellow spheres.

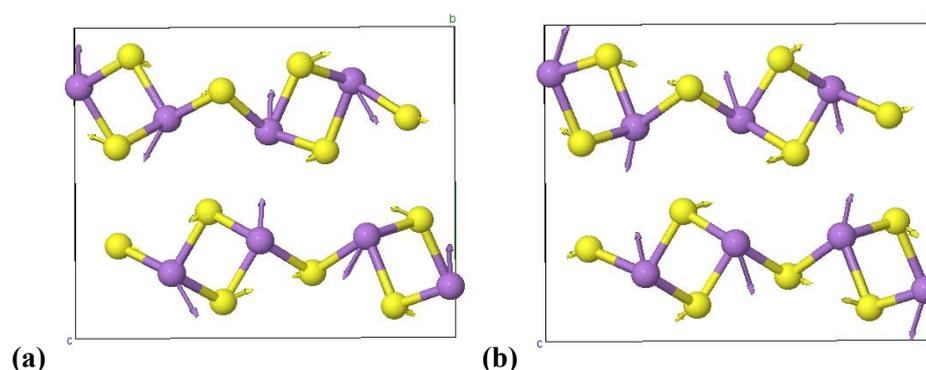

**(a)**                                    **(b)**

**Figure S4.** Atomic vibrations of the A$_u^2$ mode **(a)** and the B$_g^4$ mode **(b)** located around 93 and 103 cm$^{-1}$, respectively, in the *bc*-plane. These are inter-chain modes corresponding to rotational modes of the AsS spiral chains around the *a*-axis. Atoms of one chain rotate around the *a*-axis in the same direction as the neighbour chains inside the same layer. Rotations of the chains in the adjacent layers are in opposite direction (clockwise vs. counterclockwise) in the A$_u$ mode and in the same direction in the B$_g$ mode. These are other two "inter-chain rolling" modes defined in Ref. **[2]**. As: Purple spheres, S: yellow spheres.

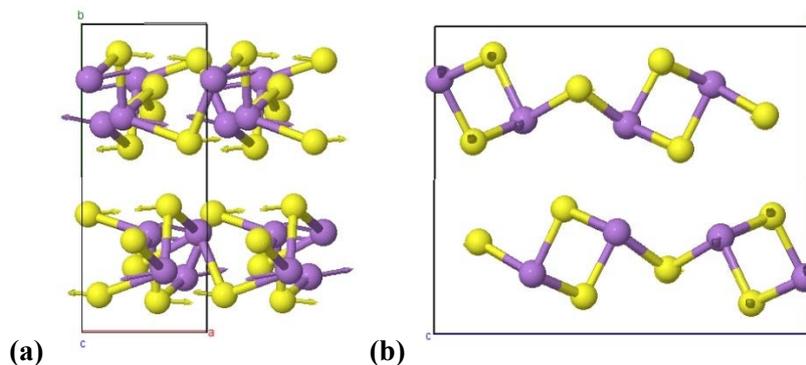

**(a)**                                    **(b)**

**Figure S5.** Atomic vibrations of the mode A$_g^4$ located around 70 cm$^{-1}$ in the *ba*- **(a)** and *bc*-planes **(b)**. This is an inter-chain mode corresponding to translations of the spiral chains along the *a*-axis. Atoms of one spiral chain translate along the *a*-axis in oposite phase with respect to those of the neighbour chains inside the same layer. Translations in the neighbour layers are opposite to those of the first layer. The S atoms linking the chains (S3 atoms) are almost at rest. As: Purple spheres, S: yellow spheres.

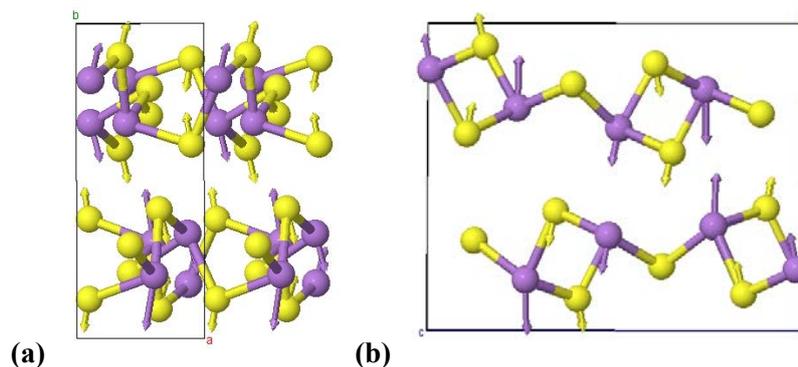

**(a)**                    **(b)**

**Figure S6.** Atomic vibrations of the $A_u^1$ mode located around 63 cm$^{-1}$ in the *ba-* **(a)** and *bc*-planes **(b)**. This is an inter-chain mode corresponding to translations of the spiral chains mainly along the *b* axis, together with a minor rotation around the *a*-axis. Atoms of one chain translate in opposite phase with respect to those of the neighbour chains both in the same layer and in neighbour layers. The S atoms (S3) linking the chains are almost at rest. As: Purple spheres, S: yellow spheres.

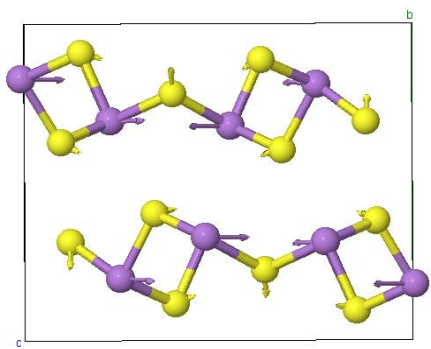

**Figure S7.** Atomic vibrations of the $B_u^2$ mode located around 94 cm$^{-1}$ in the *bc* plane. This is an inter-chain mode corresponding to translations of the spiral chains along the *c*-axis. Atoms of one chain translate in opposite phase with respect to atoms of the neighbour chains in the same layer. This is the "inter-chain beating" mode of orpiment **[2]** and it is the pair of the $A_g^4$ mode around 70 cm$^{-1}$. As: Purple spheres, S: yellow spheres.

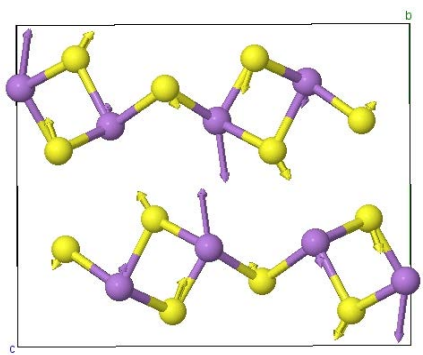

**Figure S8.** Atomic vibrations of the $B_g^2$ mode located around 69 cm$^{-1}$ in the *bc*-plane. This is an inter-chain mode corresponding to a mixture of translations of the spiral chains mainly along the *b*-axis together with a small rotation around the *a*-axis. Atoms of one chain translate in opposite phase with respect to those of the neighbour chains in the same layer. The S atoms linking the chains are almost at rest. This is the pair of the $A_u^1$ mode around 63 cm$^{-1}$. As: Purple spheres, S: yellow spheres.

Several modes located around 150 cm$^{-1}$ are related to vibrations of the S3 atoms (outside the chains), which vibrate in a completely different manner to the S1 and S2 atoms (inside the chains). These modes evidence the chain-like nature of the layers (see **Figure S9**). Finally, it can be stressed that there are typical bending modes in the intermediate-frequency region (**Figure S10**), modes involving partial bending and stretching (**Figure S11**) and typical stretching modes in the high-frequency region (**Figures S12, S13, and S14**).

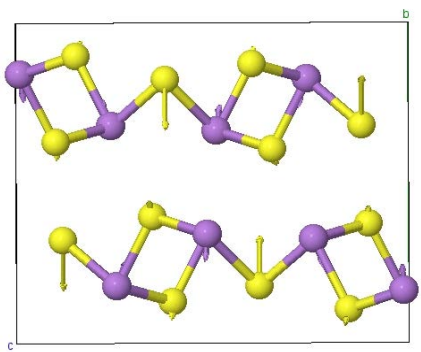

**Figure S9.** Atomic vibrations of the $A_g^6$ mode located at 153 cm$^{-1}$ in the *bc* plane. This mode corresponds to an intra-layer mode mainly characterized by vibrations of the S atoms between the chains (S3 atoms).

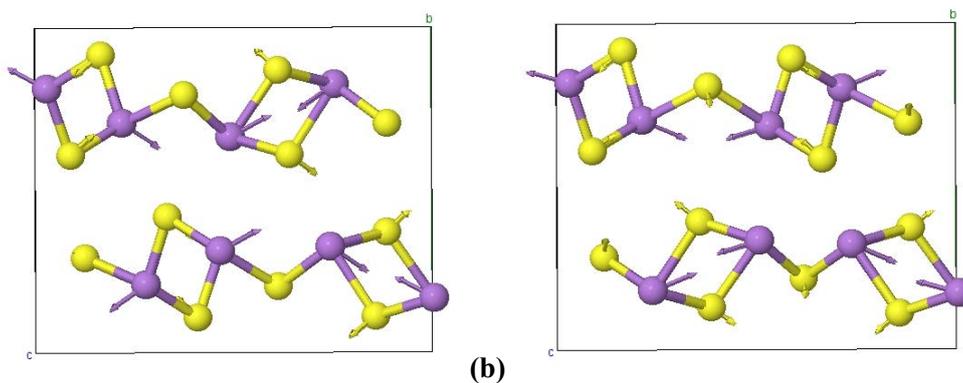

**(a)**　　　　　　　　　　　　**(b)**

**Figure S10.** Atomic vibrations of the $A_u^7$ mode (a) and $B_u^7$ mode (b) located at 167 and 196 cm$^{-1}$, respectively, in the *bc* plane. They correspond to intra-chain modes where atoms of a semi-rod vibrate in opposite phase against the other semi-rod. As: Purple spheres, S: yellow spheres.

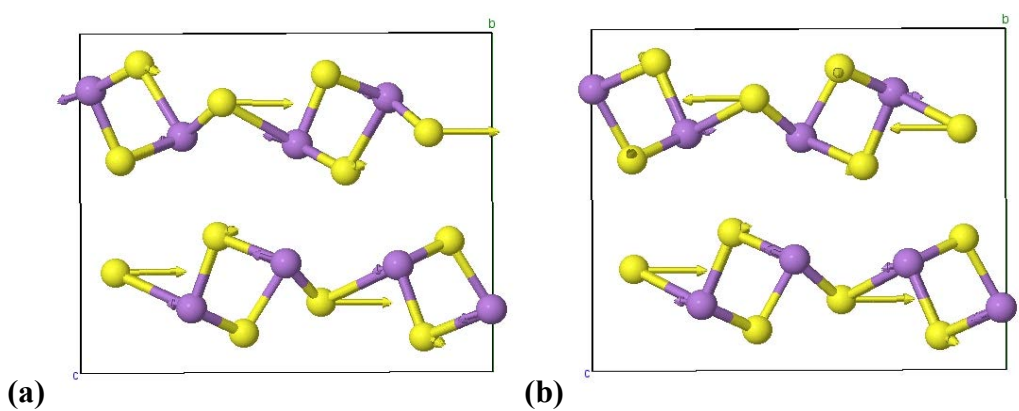

**(a)**　　　　　　　　　　　　**(b)**

**Figure S11.** Atomic vibrations of the $B_u^8$ mode (a) and the $A_g^{10}$ mode (b) located around 269 and 270 cm$^{-1}$, respectively, in the *bc* plane. These high-frequency modes correspond to intra-layer modes mainly characterized by bending As-S vibrations. In particular, these two modes are related to vibrations of S3 atoms in the plane of the layers.

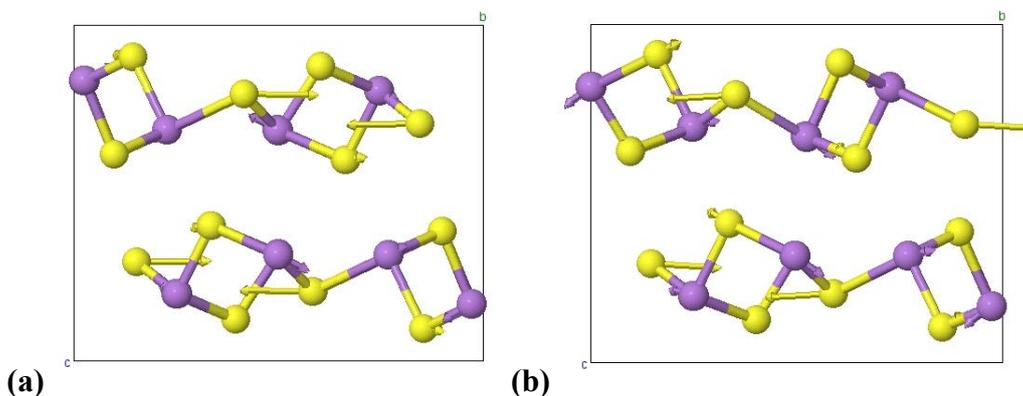

**(a)** **(b)**

**Figure S12.** Atomic vibrations of the $B_g^{11}$ mode (a) and the $A_u^{10}$ mode (b) located around 305 and 307 cm$^{-1}$, respectively, in the *bc* plane. These high-frequency modes correspond to intra-layer modes mainly characterized by bending As-S vibrations. In particular, these two modes are related to vibrations of S3 atoms in the plane of the layers.

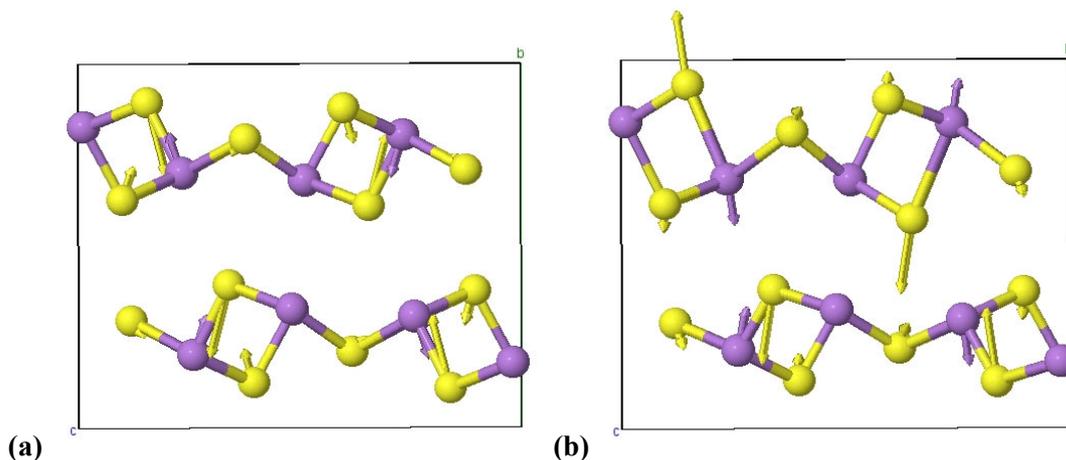

**(a)** **(b)**

**Figure S13.** Atomic vibrations of the $A_g^{14}$ mode (a) and the $B_u^{12}$ mode (b) located around 346 and 350 cm$^{-1}$, respectively, in the *bc* plane. These high-frequency modes correspond to intra-chain modes mainly characterized by symmetric stretching As-S vibrations. All vibrational modes around 340-350 cm$^{-1}$ are characterized by similar symmetric stretching intra-chain As-S vibrations.

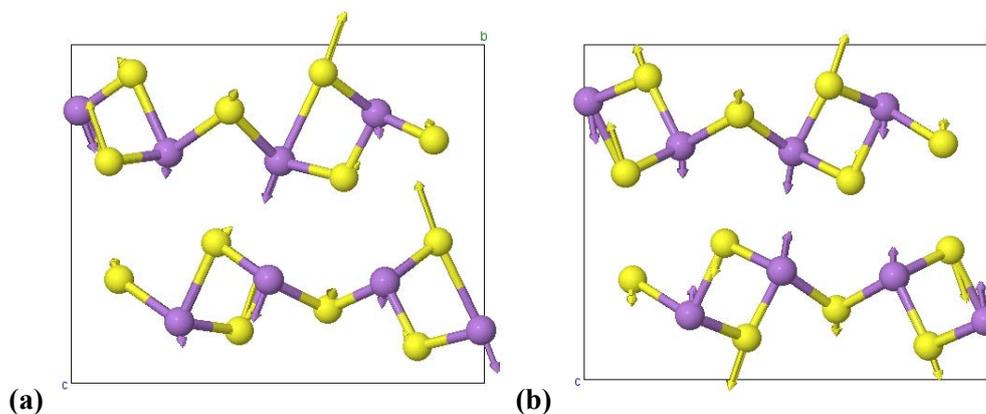

**(a)** **(b)**

**Figure S14.** Atomic vibrations of the $A_u^{14}$ mode (a) and the $B_g^{15}$ mode (b) located around 374 and 385 cm$^{-1}$, respectively, in the *bc* plane. These high-frequency modes correspond to intra-chain

modes mainly characterized by stretching As-S vibrations. All four vibrational modes located above 370 cm$^{-1}$ are characterized by similar asymmetric stretching intra-chain As-S vibrations.

## Structural properties of orpiment under high pressure

<u>Calculation of the experimental and theoretical compressibility tensor at different pressures</u>

The isothermal compressibility tensor, $\beta_{ij}$, is a symmetric second rank tensor that relates the state of strain of a crystal to the change in pressure that induced it [5]. The tensor coefficients for a monoclinic crystal with *b* as the unique crystallographic axis are:

$$\beta_{ij} = \begin{pmatrix} \beta_{11} & 0 & \beta_{13} \\ 0 & \beta_{22} & 0 \\ \beta_{13} & 0 & \beta_{33} \end{pmatrix}$$

We have obtained the isothermal compressibility tensor coefficients for monoclinic As$_2$S$_3$ at several pressures using the Institute of Radio Engineers (IRE) convention for the orthonormal basis for the tensor: $e_3||c, e_2||b*, e_1||e_2 \times e_3$. The tensor has been obtained with the finite Eulerian approximation as implemented in the Win_Strain package [6].

The change of the $\beta$ monoclinic angle (always perpendicular to the *b*-axis) with pressure implies that, in this monoclinic compound, the direction of the *a*-axis changes with pressure assuming both *b*- and *c*-axis constant. Furthermore, the variation of this monoclinic angle from 90º indicates that the direction of maximum compressibility is not exactly that of the *a*-axis. Therefore, in order to evaluate the direction of maximum compressibility as a function of pressure, we have calculated and diagonalized the experimental and theoretical isothermal compressibility tensor, $\beta_{ij}$, at different pressures.

The experimental and theoretical elements of this tensor at different pressures are reported in Tables 1 and 2, up to 10.0 and 32.0 GPa, respectively, where the directions of the maximum, intermediate and minimum compressibility and the values of the compressibility along those directions are given by the eigenvectors ($ev_i$, i=1-3) and eigenvalues ($\lambda_i$, i=1-3), respectively.

First of all, we have to note that there is a reasonable good agreement between the experimental and calculated axial compressibilities ($\beta_{ii}$ coefficients) at room pressure because $\beta_{11} \geq \beta_{22} > \beta_{33}$ in both cases. This result shows that the compressibility along the *a*-axis (layer plane) is greater or similar than that of the *b*-axis (axis perpendicular to the layer) and much larger than the *c*-axis. This is an unexpected result for a layered crystal whose layers extend in the *ac* plane and it is also a clear indication of the acicular (quasi-molecular) character of the layers.

Diagonalization of the $\beta_{ij}$ tensor at room pressure yields for our experiments the maximum, intermediate and minimum compressibilities $36.4(1.8) \cdot 10^{-3}$, $31.3(1.6) \cdot 10^{-3}$ and $-0.7(3) \cdot 10^{-3}$ GPa$^{-1}$, respectively; whereas for the case of our calculations the obtained values for the compressibilities are $44.7(2.4) \cdot 10^{-3}$, $25.8(1.3) \cdot 10^{-3}$ and $-1.14(15) \cdot 10^{-3}$ GPa$^{-1}$. These experimental (theoretical) results indicate that around 48% (58%) of the total compression at room pressure is being accommodated along the direction of maximum compressibility. The direction of the maximum compressibility at zero pressure, given by the eigenvector $ev_1$, occurs at the (0 1 0) plane and it is defined by angle $\Psi$ (see **Tables 1 and 2**) relative to the $c$-axis (from $c$ to $a$) or equivalently by an angle $\theta$ relative to the $a$-axis (from $a$ to $c$). The direction of intermediate compressibility at zero pressure, given by eigenvector $ev_2$, is along the $b$-axis. Finally, the direction of minimum compressibility at zero pressure, given by eigenvector $ev_3$, is at the (0 1 0) plane and perpendicular to the direction of maximum compressibility within the same plane. In particular, the experimental direction of maximum compressibility at room pressure is at $\theta =0.8(3)^{\circ}$ from the $a$-axis, whereas the calculated one is at $1.12(20)^{\circ}$. This means that the direction of maximum (minimum) compressibility at room pressure is close to the $a$-axis ($c$-axis).

As regards the behaviour of the experimental (theoretical) compressibility tensor under pressure, we start with the study between 0 and 10 GPa where both the theoretical and experimental $\beta_{ij}$ values have been obtained. The most notable feature is that $\beta_{11} \geq \beta_{22} > \beta_{33}$ is maintained as pressure increases. Therefore, our experiment (*ab initio* calculations) shows that the $a$-axis has compressibility greater or similar (greater) than that of the $b$-axis. In this sense, a greater or similar compressibility in a direction along the layer with respect to that of the inter-layer (along the $b$-axis) is maintained with compression. On the other hand, the eigenvalue $\lambda_3$ is negative below 0.9 (1.3) GPa in the case of the experimental (theoretical) $\beta_{ij}$ tensor. This means that the material slightly expands under compression (positive strain) along the direction of minimum compressibility, given by the eigenvector $ev_3$. On the other hand, the direction of major compression is close to the $a$-axis under compression.

Above 10 GPa, the theoretical compressibility tensor has been obtained up to 32.0 GPa. In this case, the direction of maximum compressibility changes rapidly for pressures greater than 20 GPa. For instance, $\theta =46(13)^{\circ}$ at 22.2 GPa and the direction of maximum compressibility is along the $b$-axis ([010] direction) at 24.3 GPa. However, the direction of maximum compressibility is again at the $ac$ plane above 26.7 GPa, but it is close to the $c$-axis instead. In particular, $\Psi = 15(4)$ and $\theta =92(4)^{\circ}$ at 32.0 GPa.

**Table 1.** Experimental isothermal compressibility tensor coefficients, $\beta_{ij}$, for As$_2$S$_3$ and their respective eigenvalues, $\lambda_i$, and eigenvectors, ev$_i$, at different pressure values. The results are obtained by employing the finite Eulerian method. The eigenvalues are given in decreasing value along a column.

| P(GPa) | 0.0 | 1.0 | 2.0 | 3.0 | 4.0 | 5.0 | 6.0 | 8.0 | 10.0 |
|---|---|---|---|---|---|---|---|---|---|
| $\beta_{11}$ ($10^{-3}$ GPa$^{-1}$) | 32.4(1.8) | 23.0(1.2) | 17.3(1.2) | 14.0(1.0) | 11.8(9) | 10.2(7) | 9.0(7) | 7.4(6) | 6.3(6) |
| $\beta_{22}$ ($10^{-3}$ GPa$^{-1}$) | 31.3(1.6) | 21.2(1.1) | 15.1(1.1) | 11.8(0.8) | 9.6(7) | 8.2(6) | 7.1(5) | 5.6(4) | 4.6(5) |
| $\beta_{33}$ ($10^{-3}$ GPa$^{-1}$) | 3.3(3) | 3.2(3) | 3.0(3) | 2.9(3) | 2.8(3) | 2.8(3) | 2.7(3) | 2.5(3) | 2.4(5) |
| $\beta_{13}$ ($10^{-3}$ GPa$^{-1}$) | -11.5(5) | -8.4(4) | -6.0(4) | -4.5(3) | -3.5(3) | -2.84(23) | -2.31(20) | -1.51(15) | -0.94(22) |
| $\lambda_1$ ($10^{-3}$ GPa$^{-1}$) | 36.4(1.8) | 26.1(1.3) | 19.5(1.4) | 15.6(1.1) | 13.0(9) | 11.2(8) | 9.8(7) | 7.8(6) | 6.6(6) |
| ev$_1$ ($\lambda_1$) | (0.94,0,-0.33) | (0.94,0,-0.34) | (0.94,0,-0.34) | (0.94,0,-0.34) | (0.94,0,-0.33) | (0.95,0,-0.32) | (0.95,0,-0.31) | (0.96,0,-0.27) | (0.98,0,-0.22) |
| $\lambda_2$ ($10^{-3}$ GPa$^{-1}$) | 31.3(1.6) | 21.2(1.1) | 15.1(1.1) | 11.8(8) | 9.6(7) | 8.2(6) | 7.1(5) | 5.6(4) | 4.6(5) |
| ev$_2$ ($\lambda_2$) | (0,1,0) | (0,1,0) | (0,1,0) | (0,1,0) | (0,1,0) | (0,1,0) | (0,1,0) | (0,1,0) | (0,1,0) |
| $\lambda_3$ ($10^{-3}$ GPa$^{-1}$) | -0.7(3) | 0.09(24) | 0.84(25) | 1.3(3) | 1.6(3) | 1.8(3) | 1.9(3) | 2.1(3) | 2.2(5) |
| ev$_3$ ($\lambda_3$) | (0.33,0,0.94) | (0.34,0,0.94) | (0.34,0,0.94) | (0.34,0,0.94) | (0.33,0,0.94) | (0.32,0,0.95) | (0.31,0,0.95) | (0.27,0,0.96) | (0.22,0,0.98) |
| $\Psi$, $\theta$ (°)[a] | 109.1(3), 0.8(3) | 110.1(3), -0.7(3) | 110.1(4), -1.1(4) | 109.7(5), -1.0(5) | 109.3(6), -0.8(6) | 108.7(7), -0.4(7) | 108.0(8), 0.2(8) | 105.8(1.1), 2.2(1.1) | 102.6(2.6), 5.2(2.6) |

[a] The direction of maximum compressibility is located at the (0 1 0) plane at the given angles $\Psi$ to the $c$-axis (from $c$ to $a$) and $\theta$ to the $a$-axis (from $a$ to $c$).

**Table 2.** Theoretical isothermal compressibility tensor coefficients, $\beta_{ij}$, for $As_2S_3$ and their respective eigenvalues, $\lambda_i$, and eigenvectors, $ev_i$, at different pressures. The results are obtained by employing the finite Eulerian method. The eigenvalues are given in decreasing value along a column.

| P(GPa) | 0.0 | 1.0 | 2.0 | 3.0 | 4.0 | 5.0 | 6.0 | 8.0 | 10.0 |
|---|---|---|---|---|---|---|---|---|---|
| $\beta_{11}$ ($10^{-3}$ GPa$^{-1}$) | 40.0(2.2) | 27.4(1.4) | 20.1(1.4) | 16.1(1.2) | 13.5(1.0) | 11.7(8) | 10.3(8) | 8.3(6) | 7.0(6) |
| $\beta_{22}$ ($10^{-3}$ GPa$^{-1}$) | 25.8(1.4) | 19.4(1.0) | 15.0(1.1) | 12.2(0.9) | 10.3(7) | 9.0(6) | 7.9(6) | 6.4(5) | 5.4(4) |
| $\beta_{33}$ ($10^{-3}$ GPa$^{-1}$) | 3.58(24) | 3.49(21) | 3.4(3) | 3.3(0.3) | 3.2(3) | 3.1(3) | 3.07(24) | 2.93(24) | 2.8(3) |
| $\beta_{13}$ ($10^{-3}$ GPa$^{-1}$) | -13.9(7) | -10.2(5) | -7.3(5) | -5.4(0.4) | -4.2(3) | -3.25(24) | -2.60(20) | -1.80(15) | -1.36(17) |
| $\lambda_1$ ($10^{-3}$ GPa$^{-1}$) | 44.7(2.4) | 31.1(1.6) | 22.8(1.6) | 18.1(1.3) | 15.0(1.1) | 12.8(9) | 11.2(8) | 8.9(7) | 7.4(7) |
| $ev_1$ ($\lambda_1$) | (0.95,0,-0.32) | (0.94,0,-0.35) | (0.94,0,-0.35) | (0.94,0,-0.35) | (0.94,0,-0.33) | (0.95,0,-0.32) | (0.95,0,-0.31) | (0.96,0,-0.29) | (0.96,0,-0.28) |
| $\lambda_2$ ($10^{-3}$ GPa$^{-1}$) | 25.8(1.3) | 19.4(1.0) | 15.0(1.1) | 12.2(0.9) | 10.3(7) | 9.0(6) | 7.9(6) | 6.3(5) | 5.4(4) |
| $ev_2$ ($\lambda_2$) | (0,1,0) | (0,1,0) | (0,1,0) | (0,1,0) | (0,1,0) | (0,1,0) | (0,1,0) | (0,1,0) | (0,1,0) |
| $\lambda_3$ ($10^{-3}$ GPa$^{-1}$) | -1.14(15) | -0.27(0.13) | 0.67(13) | 1.31(16) | 1.76(18) | 2.05(19) | 2.23(20) | 2.39(21) | 2.4(3) |
| $ev_3$ ($\lambda_3$) | (0.32,0,0.95) | (0.35,0,0.94) | (0.35,0,0.94) | (0.35,0,0.94) | (0.33,0,0.94) | (0.32,0,0.95) | (0.31,0,0.95) | (0.29,0,0.96) | (0.28,0,0.96) |
| $\Psi$, $\theta$(°)[a] | 108.73(20), 1.12(20) | 110.25(20), -1.03(20) | 110.56(20), -1.84(20) | 110.2(3), -1.9(3) | 109.5(4), -1.4(4) | 108.6(4), -0.8(4) | 107.9(5), -0.2(5) | 106.8(7), 0.6(7) | 106.5(1.7), 0.7(1.7) |

[a] The direction of maximum compressibility is located at the *ac* plane; i.e. the (0 1 0) plane, at the given angle $\Psi$ to the *c*-axis (from *c* to *a*) and $\theta$ to the *a*-axis (from *a* to *c*).

| P(GPa) | 14.3 | 17.4 | 18.8 | 20.4 | 22.2 | 24.3[b] | 26.7 | 29.3 | 32.0 |
|---|---|---|---|---|---|---|---|---|---|
| $\beta_{11}$ ($10^{-3}$ GPa$^{-1}$) | 5.57(21) | 5.4(8) | 5.2(7) | 4.5(6) | 2.8(5) | 2.0(4) | 1.8(4) | 1.7(3) | 1.6(3) |
| $\beta_{22}$ ($10^{-3}$ GPa$^{-1}$) | 4.11(13) | 3.5(4) | 3.4(4) | 3.3(3) | 2.90(24) | 2.82(21) | 2.52(18) | 2.26(16) | 2.00(14) |
| $\beta_{33}$ ($10^{-3}$ GPa$^{-1}$) | 2.53(9) | 2.2(4) | 1.9(4) | 1.7(3) | 2.2(3) | 2.4(3) | 2.40(24) | 2.37(23) | 2.32(21) |
| $\beta_{13}$ ($10^{-3}$ GPa$^{-1}$) | -0.86(4) | -0.77(16) | -0.65(14) | -0.23(12) | 0.46(10) | 0.38(9) | 0.30(8) | 0.19(7) | 0.20(20) |
| $\lambda_1$ ($10^{-3}$ GPa$^{-1}$) | 5.80(22) | 5.5(8) | 5.3(8) | 4.6(6) | 3.0(5) | 2.82(21) | 2.52(18) | 2.42(23) | 2.37(21) |
| $ev_1$ ($\lambda_1$) | (0.97,0,-0.25) | (0.97,0,-0.23) | (0.98,0,-0.19) | (1.00,0,-0.08) | (0.87,0,0.49) | (0,1,0) | (0.38,0,0.93) | (0.26,0,0.97) | (0.26,0,0.97) |
| $\lambda_2$ ($10^{-3}$ GPa$^{-1}$) | 4.11(13) | 3.5(4) | 3.4(4) | 3.3(3) | 2.90(24) | 2.6(3) | 2.52(18) | 2.26(16) | 2.00(14) |
| $ev_2$ ($\lambda_2$) | (0,1,0) | (0,1,0) | (0,1,0) | (0,1,0) | (0,1,0) | (0.55,0,0.84) | (0,1,0) | (0,1,0) | (0,1,0) |
| $\lambda_3$ ($10^{-3}$ GPa$^{-1}$) | 2.30(9) | 2.0(4) | 1.8(4) | 1.7(3) | 2.0(3) | 1.8(3) | 1.7(3) | 1.7(3) | 1.6(3) |
| $ev_3$ ($\lambda_3$) | (0.25,0,0.97) | (0.23,0,0.97) | (0.19,0,0.98) | (0.08,0,1.00) | (0.49,0,-0.87) | (0.84,0,-0.55) | (0.93,0,-0.38) | (0.97,0,-0.26) | (0.97,0,-0.26) |
| $\Psi$, $\theta$(°)[a] | 104.8(6), 2.3(6) | 103.1(1.8), 3.8(1.8) | 101.1(1.6), 5.8(1.6) | 94.7(1.8), 12.2(1.8) | 61(13), 46(13) |  | 22.0(1), 85.1(1) | 15(5), 92(5) | 15(4), 92(4) |

[a] The direction of maximum compressibility is located at the *ac* plane; i.e. the (0 1 0) plane, at the given angle $\Psi$ to the *c*-axis (from *c* to *a*) and $\theta$ to the *a*-axis (from *a* to *c*).

[b] At 24.3 GPa, the direction of maximum compressibility is along *b*-axis; i.e. the [010] direction.

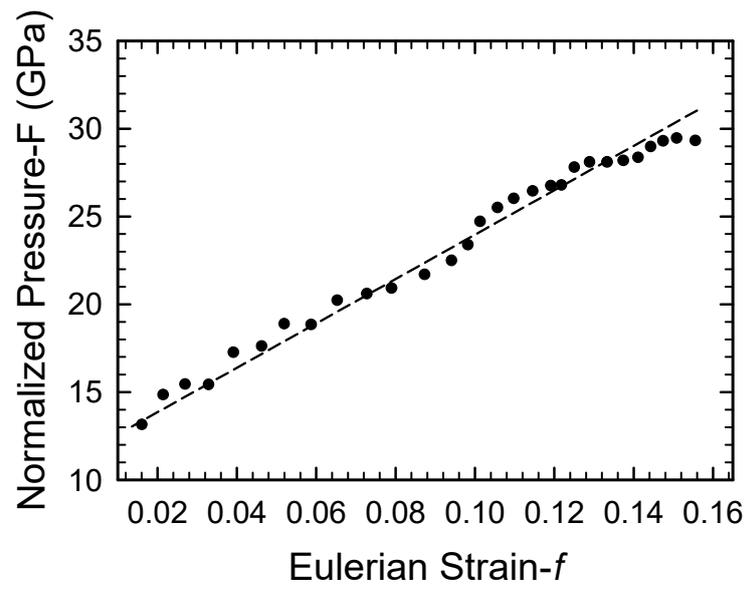

**Figure S15.** Normalized pressure vs Eulerian strain (F-f plot).

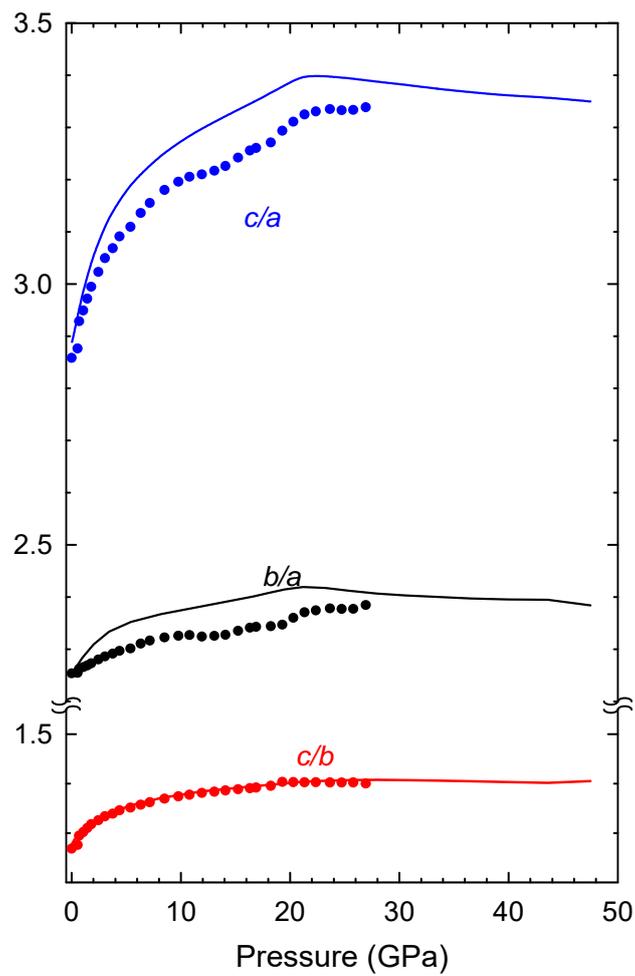

**Figure S16.** Pressure dependence of the experimental (symbols) and theoretical (lines) lattice parameter ratios as a function of pressure. Theoretical calculations include vdW dispersion corrections.

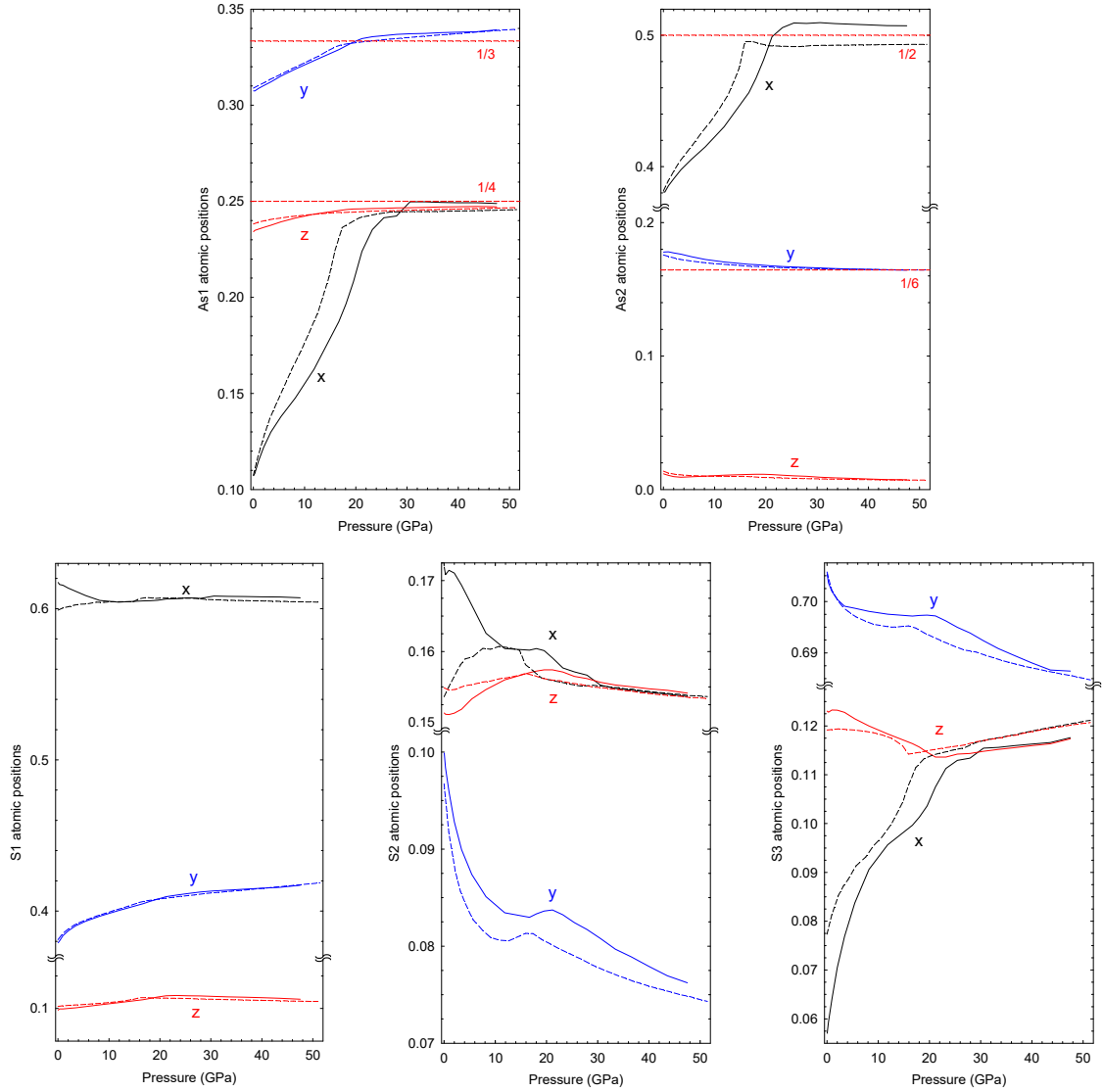

**Figure S17.** Pressure dependence of the theoretical free coordinates belonging to the different Wyckoff sites of the α-As₂S₃. Calculations with (without) vdW interactions are depicted as solid (dashed) lines. Compared to the As atomic coordinates, the S atomic coordinates show a very complex tendency with pressure. However, all of these show considerable changes above 18 and 25 GPa for calculations that do not consider and consider vdW interactions, respectively. Curiously, the erratic behavior of the S atomic coordinates is not reflected in the smooth HP dependence of the As-S interatomic distances (see Figure 6 in the main paper) that are mainly dominated by the strong changes of the x coordinate of As atoms.

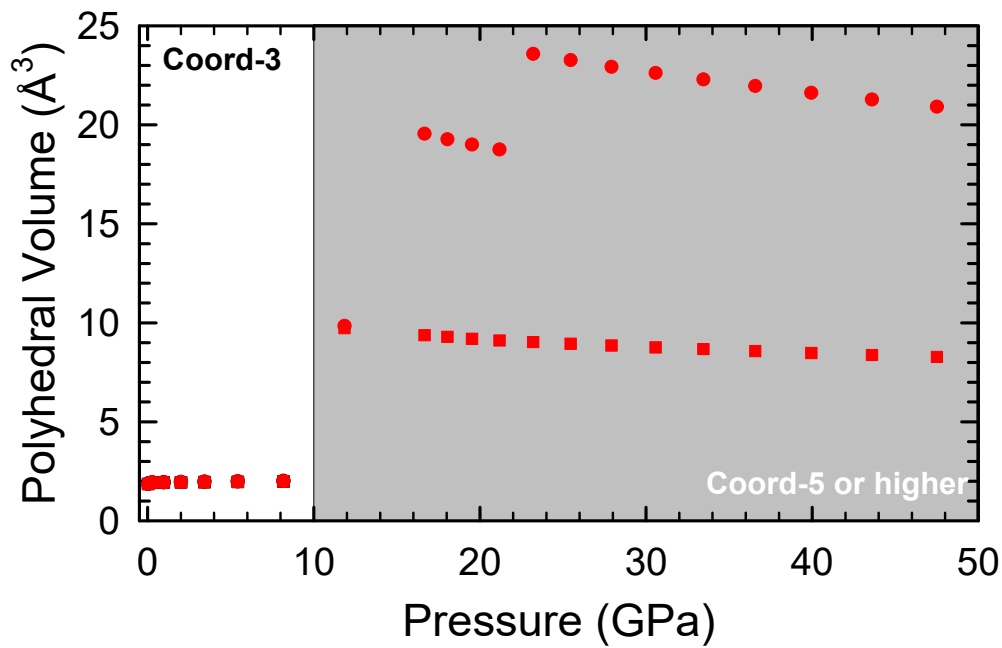

**Figure S18.** Pressure dependence of the theoretically predicted polyhedral unit volume around As1 (circles) and As2 (squares).

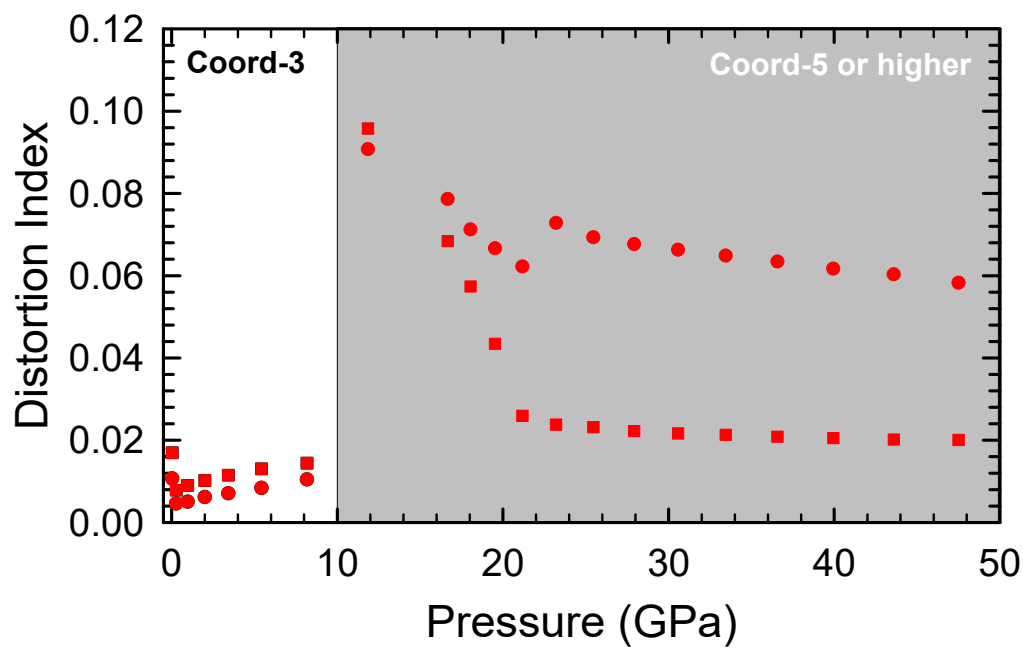

**Figure S19.** Pressure dependence of the theoretically predicted distortion index of the polyhedral units around As1 (circles) and As2 (squares).

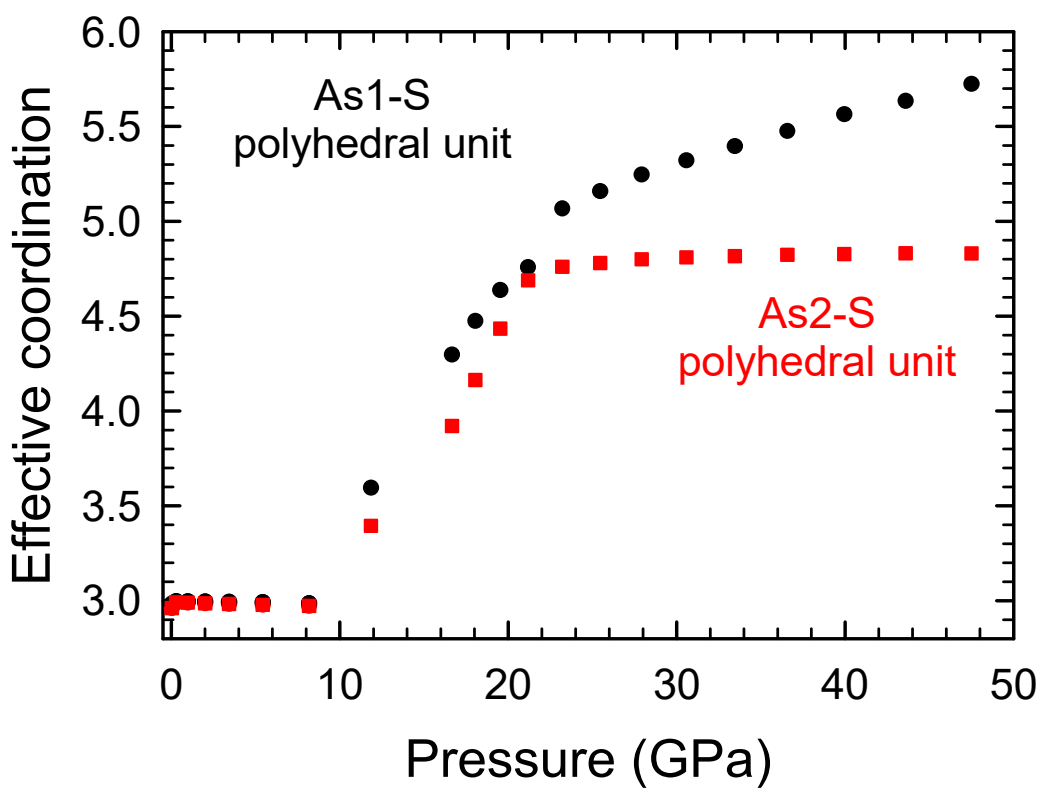

**Figure S20.** Pressure dependence of the theoretically predicted effective coordination number of polyhedral units around As1 (circles) and As2 (squares). Effective coordination number is calculated according to Ref. **[7]**.

## Vibrational properties of orpiment under compression

In layered compounds with typical van der Waals gap between the layers, the low-frequency inter-layer shear mode exhibits a much smaller pressure coefficient than other modes, whereas the low-frequency A (or B) mode displays the largest pressure coefficient. For example, the E and A modes with frequencies around 40 (60) cm$^{-1}$ and 116 (133) cm$^{-1}$ in InSe (GaSe) have pressure coefficients of 0.68 (0.85) cm$^{-1}$/GPa and 5.41 (5.78) cm$^{-1}$/GPa, respectively **[8,9]**. Similar behavior is found for layered topological insulators Bi$_2$Se$_3$, Bi$_2$Te$_3$ and Sb$_2$Te$_3$ **[10-12]**.

Usually, the small pressure coefficient of the low-frequency E mode in layered materials is ascribed to the weak bending force constant due to weak van der Waals forces between the neighboring layers. On the other hand, the large pressure coefficient of the low-frequency A mode is due to the extraordinary increase of the stretching force constant between neighboring layers due to the strong decrease of the inter-layer distance **[8,9]**.

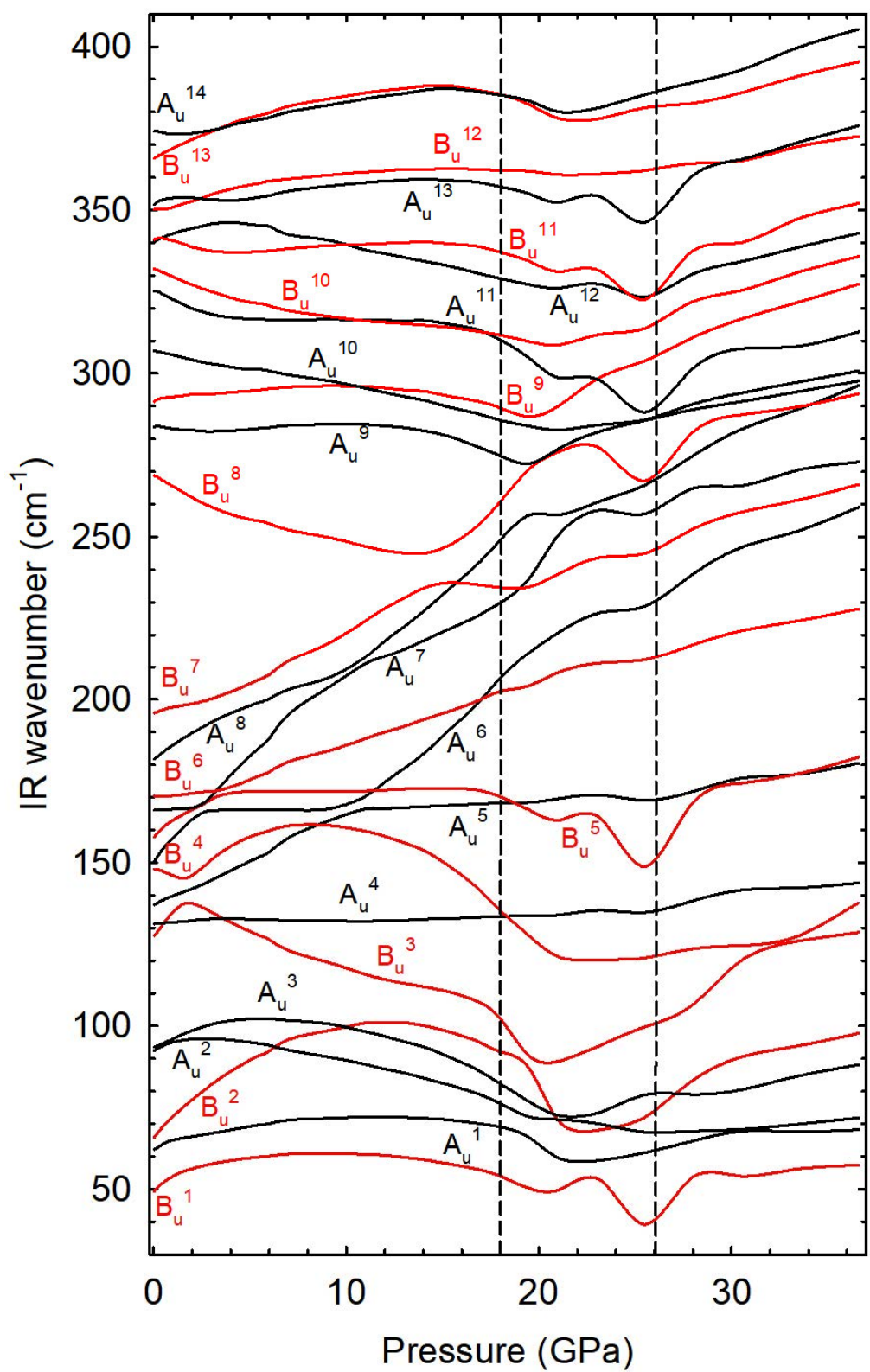

**Figure S21.** HP dependence of the theoretical wavenumbers of the IR-active modes in α-As₂S₃. Considerable softening of some vibrational modes is observed above 4 GPa and between 18 and 26 GPa.

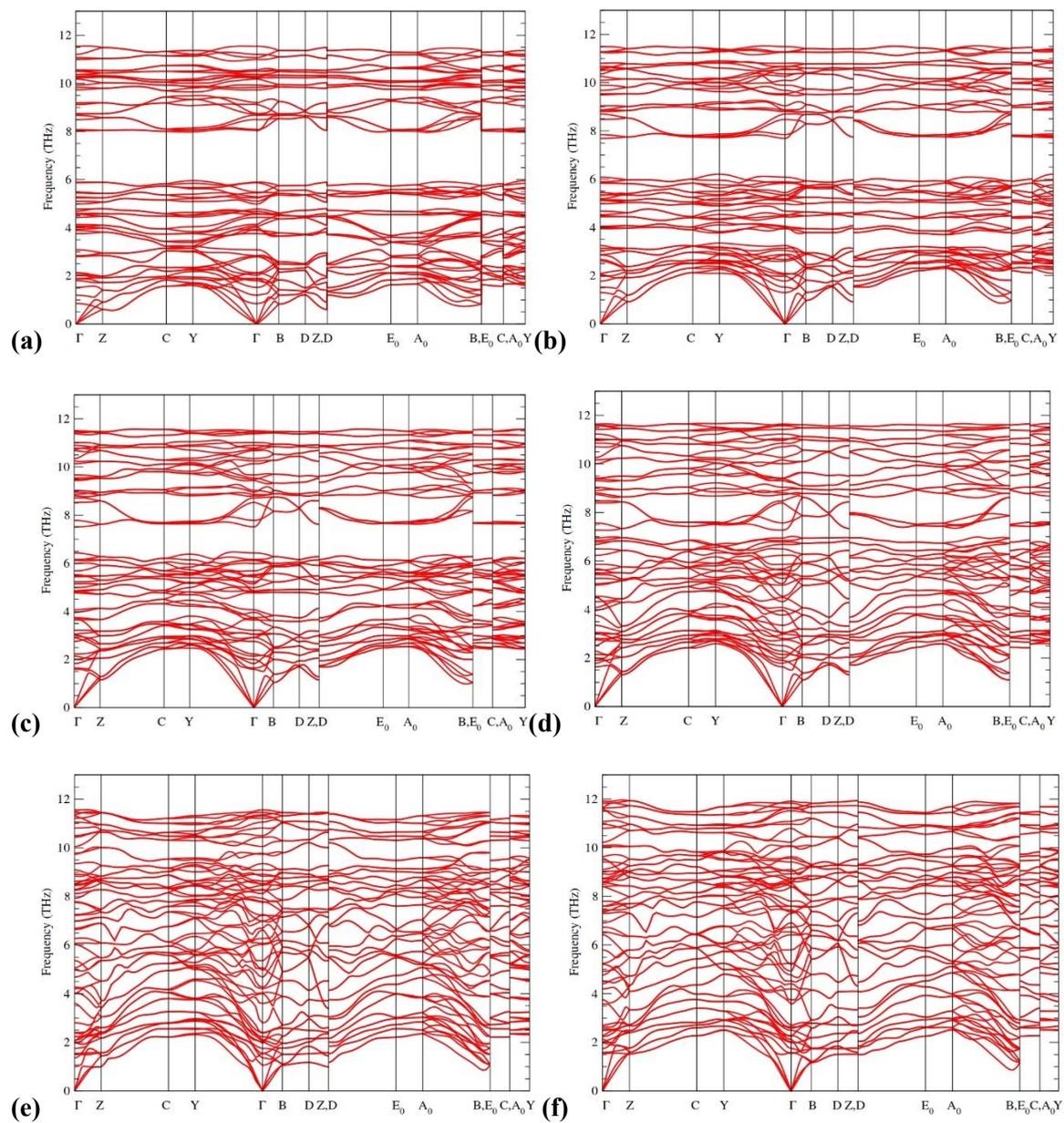

**Figure S22.** Calculated phonon dispersion curves of As$_2$S$_3$ at different pressures: **(a)** 0 GPa, **(b)** 5 GPa, **(c)** 10 GPa, **(d)** 15 GPa, **(e)** 21 GPa, and **(f)** 30 GPa.

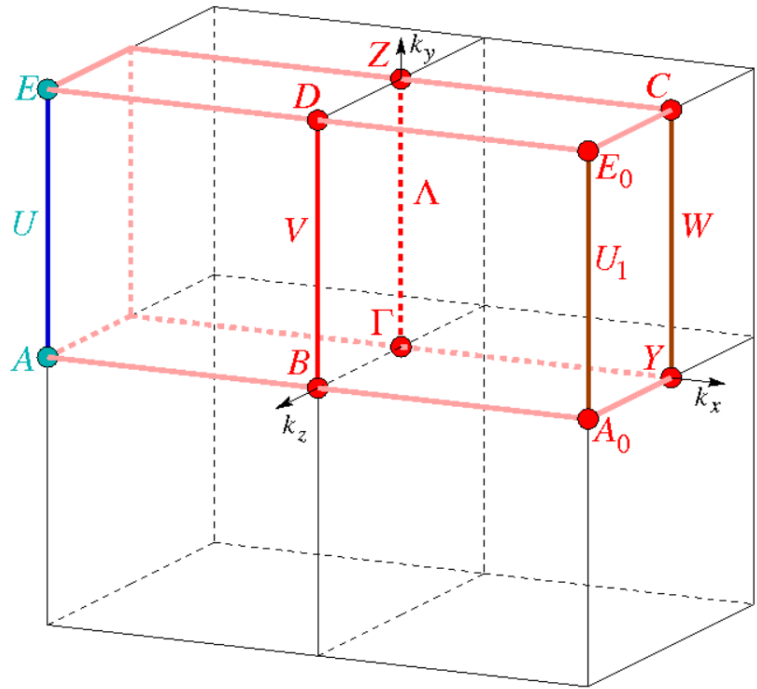

**Figure S23.** Representation of the points and paths used in the first Brillouin zone of monoclinic α-As$_2$S$_3$ in order to perform calculations of phonon dispersion curves and electronic band structures. Taken from the Bilbao Crystallographic Server database **[13]**.

**Table S2.** Theoretical (with vdW interactions) and experimental IR-active mode frequencies and their respective pressure coefficients for $\alpha$-$As_2S_3$ at room temperature, as fitted with equation $\omega(P) = \omega_0 + \alpha \cdot P$. Experimental values have been added for comparison.

| Mode (Sym) | Theoretical | | Experimental |
|---|---|---|---|
| | $\omega_0 \ (cm^{-1})$ [a] | $\alpha \left(\frac{cm^{-1}}{GPa}\right)$ [a] | $\omega_0 \ (cm^{-1})$ |
| $B_u^1$ | 50 (1) | 4.3 (3) | 52 [c] |
| $A_u^1$ | 63 (1) | 2.3 (3) | |
| $B_u^2$ | 66 (1) | 6.5 (2) | |
| $A_u^2$ | 93 (1) | 2.5 (2) | |
| $A_u^3$ | 94 (1) | 3.2 (1) | |
| $B_u^3$ | 128 (2) | 8 (1) | |
| $A_u^4$ | 131 (2) | 0.7 (1) | |
| $A_u^5$ | 137 (2) | 2.3 (1) | |
| $B_u^4$ | 148 (2) | -2.5 (9) | 140 [b], 139 [c] |
| $A_u^6$ | 150 (2) | 8.8 (3) | |
| $B_u^5$ | 158 (2) | 5.2 (3) | 159 [b], 160 [c] |
| $A_u^7$ | 167 (2) | -1.8 (4) | |
| $B_u^6$ | 170 (2) | -0.1 (2) | |
| $A_u^8$ | 182 (2) | 4.6 (1) | 181 [b], 183 [c] |
| $B_u^7$ | 196 (2) | 1.1 (2) | 198 [b], 202 [c] |
| $B_u^8$ | 269 (3) | -4.0 (1) | 279 [b] |
| $A_u^9$ | 284 (3) | -1.0 (1) | 299 [b] |
| $B_u^9$ | 292 (3) | 1.1 (2) | 311 [b], 305 [c] |
| $A_u^{10}$ | 307 (4) | -1.6 (1) | |
| $A_u^{11}$ | 326 (4) | -3.5 (2) | 345 [b], 348 [c] |
| $B_u^{10}$ | 332 (4) | -2.8 (1) | 354 [b,c] |
| $A_u^{12}$ | 340 (4) | 2.7 (1) | |
| $B_u^{11}$ | 342 (4) | -1.7 (4) | |
| $B_u^{12}$ | 350 (4) | 1.2 (3) | 361 [c] |
| $A_u^{13}$ | 352 (4) | 1.2 (4) | 375 [b] |
| $B_u^{13}$ | 366 (4) | 3.1 (1) | 383 [b], 381 [c] |
| $A_u^{14}$ | 374 (4) | -1.0 (1) | 393 [c] |

[a] This work. [b] Ref. 14. [c] Ref. 15.

**Table S3.** Known crystalline phases of group-15 $A_2B_3$ sesquichalcogenides at room pressure, their space groups, stability at room pressure, average cation coordination and bonding type.

| Compound | Space group | Stability at room pressure | Average cation coordination | Bonding-type |
|---|---|---|---|---|
| α-As$_2$S$_3$ | P2$_1$/c | Stable | 3 | p-type covalent |
| α-As$_2$Se$_3$ | P2$_1$/c | Stable | 3 | p-type covalent |
| β-As$_2$Se$_3$ | C2/m | Metastable | 3 | p-type covalent |
| γ-As$_2$Se$_3$ | R-3m | Metastable | 6 | metavalent |
| α-As$_2$Te$_3$ | C2/m | Stable | 5.5 | mixed* |
| β-As$_2$Te$_3$ | R-3m | Metastable | 6 | metavalent |
| α-Sb$_2$S$_3$ | Pnma | Stable | 4 | mixed* |
| α-Sb$_2$Se$_3$ | Pnma | Stable | 4 | mixed* |
| α-Sb$_2$Te$_3$ | R-3m | Stable | 6 | metavalent |
| α-Bi$_2$S$_3$ | Pnma | Stable | 4 | mixed* |
| α-Bi$_2$Se$_3$ | R-3m | Stable | 6 | metavalent |
| α-Bi$_2$Te$_3$ | R-3m | Stable | 6 | metavalent |

\* It is a mixture between p-type covalent and metavalent

### Possible high pressure phase

Above 42 GPa, Liu and co-workers [16] found a possible 1$^{st}$ order phase transition by a drastic decrease of the electrical resistivity above this pressure. This feature was assigned to a pressure-induced metallization but there was not identification of the high pressure phase. We do not want to finish the full picture of the behaviour of orpiment under compression without providing a hint of this possible high pressure phase. According to our *ab initio* theoretical simulations above 40 GPa, we found a new structure competitive with the low pressure phase (see **Figure S24**).

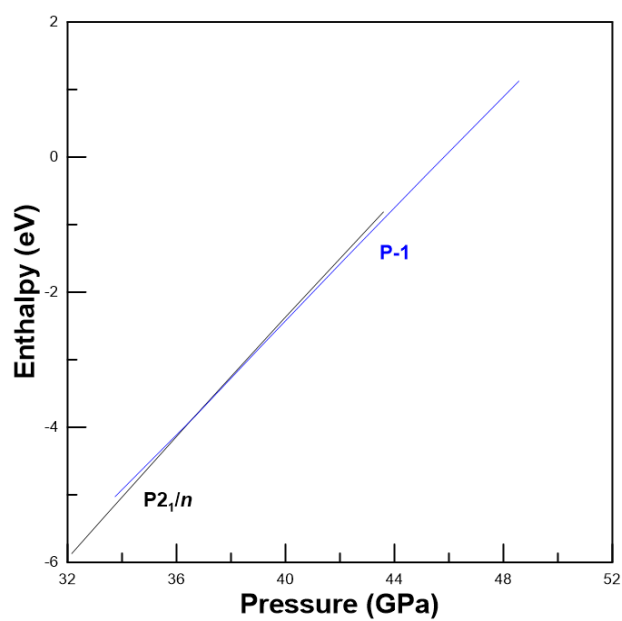

**Figure S24.** Enthalpy vs pressure around 40 GPa of low pressure phase (*P2$_1$/n*) and a proposed high pressure phase (*P-1*).

**References**

(1)     Mullen, D.J.E.; and Nowacki, W. Refinement of the crystal structures of realgar, AsS and orpiment, $As_2S_3$. *Z. Kristall.* **1972**, 136, 48.

(2)     Defonzo, A.P.; and Tauc, J. Network dynamics of 3:2 coordinated compounds. *Phys. Rev. B* **1978**, 18, 6957.

(3)     Zallen, R.; and Slade, M. Rigid-layer modes in chalcogenide crystals. *Phys. Rev. B* **1974**, 9, 1627.

(4)     Canepa, R.; Hanson, R. M.; Ugliengo, P.; Alfredsson, M. J-ICE: a new Jmol interface for handling and visualizing crystallographic and electronic properties. *J. Appl. Cryst.* **2011**, 44, 225-229.

(5)     Haussühl, S. Physical Properties of Crystals. An Introduction (Wiley-VCH, Weinheim, 2007).

(6)     Angel, R. J. http://www.rossangel.com/text_strain.htm

(7)     Hoppe, R. Effective coordination numbers (ECoN) and mean Active fictive ionic radii (MEFIR). *Z. Kristall.* **1979**, 150, 23.

(8)     Ulrich, C.; Mroginski, M.; Goñi, A.R.; Cantarero, A.; Schwarz, U.; Muñoz, V.; and Syassen, K. Vibrational Properties of InSe under Pressure: Experiment and Theory. *Phys. Stat. Sol. (b)* **1996**, 198, 121.

(9)     Kulibekov, A.M.; Olijnyk, H. P.; Jephcoat, A. P.; Salaeva, Z. Y.; Onari, S.; and Allakverdiev, K.R. Raman Scattering under Pressure and the Phase Transition in ε-GaSe. *Phys. Stat. Sol (b)* **2003**, 235, 517.

(10)    Vilaplana, R.; Gomis, O.; Manjón, F.J.; Segura, A.; Pérez-González, E.; Rodríguez-Hernández, P.; Muñoz, A.; González, J.; Marín-Borrás, V.; Muñoz-Sanjosé, V.; Drasar, C.; and Kucek, V. High-Pressure Vibrational and Optical Study of $Bi_2Te_3$. *Phys. Rev. B* **2011**, 84, 104112.

(11)    Gomis, O.; Vilaplana, R.; Manjón, F.J.; Rodríguez-Hernández, P.; Pérez-González, E.; Muñoz, A.; Kucek, V.; and Drasar, C. Lattice Dynamics of $Sb_2Te_3$ at High Pressures. *Phys. Rev. B* **2011**, 84, 174305.

(12)    Vilaplana, R.; Santamaría-Pérez, D.; Gomis, O.; Manjón, F.J.; González, J.; Segura, A.; Muñoz, A.; Rodríguez-Hernández, P.; Pérez-González, E.; Marín-Borrás, V.; Muñoz-Sanjosé, V.; Drasar, C.; Kucek, V. Structural and Vibrational Study of $Bi_2Se_3$ under High Pressure. *Phys. Rev. B* **2011**, 84, 184110.

(13) Aroyo, M. I.; Orobengoa, D.; de la Flor, G.; Tasci, E. S.; Perez-Mato, J. M.; Wondratschek, H. Brillouin-zone database on the Bilbao Crystallographic Server. *Acta Cryst.* **2014**, A70.

(14) Zallen, R.; Slade, M.; Ward, A.T. Lattice vibrations and Interlayer Interactions in Crystalline $As_2S_3$ and $As_2Se_3$. *Phys. Rev. B* **1971**, 3, 4257.

(15) Forneris, R. The infrared and Raman spectra of realgar and orpiment. *Am. Miner.* **1969**, 54, 1062.

(16)  Liu, K. X.; Dai, L. D.; Li, H. P.; Hu, H. Y.; Yang, L. F.; Pu, C.; Hong, M. L.; Liu, P.F. Phase Transition and Metallization of Orpiment by Raman Spectroscopy, Electrical Conductivity and Theoretical Calculation under High Pressure. *Materials* **2019**, 12, 784.